\documentclass[showpacs,preprintnumbers,amsmath,amssymb]{revtex4}
\usepackage{amsmath,amssymb,graphics,epsfig,subfigure}
\usepackage{color}

\begin{document}
\renewcommand{\baselinestretch}{1.3}

\title{Ruppeiner Geometry, Phase Transitions, and the Microstructure of Charged AdS Black Holes}

\author{Shao-Wen Wei$^{1,2}$  \footnote{weishw@lzu.edu.cn},
        Yu-Xiao Liu$^{1}$  \footnote{liuyx@lzu.edu.cn},
        Robert B. Mann$^{2}$ \footnote{rbmann@uwaterloo.ca}}

\affiliation{ $^{1}$Institute of Theoretical Physics $\&$ Research Center of Gravitation, Lanzhou University, Lanzhou 730000, People's Republic of China,\\
$^{2}$Department of Physics and Astronomy, University of Waterloo, Waterloo, Ontario, Canada, N2L 3G1}

\begin{abstract}
We present a novel approach for probing the microstructure of a thermodynamic system that combines thermodynamic phase transitions with the Ruppeiner scalar curvature. Originally considered for van der Waals fluids and charged black holes [Phys. Rev. Lett. 123, 071103 (2019)], we  extend and generalize our approach to higher-dimensional charged AdS black holes. Beginning with thermodynamic fluctuations, we construct the line element of the Ruppeiner geometry and obtain a universal formula for the scalar curvature $R$. We first review the thermodynamics of a van der Waals fluid and calculate the coexistence and spinodal curves. From this we are able to clearly display the phase diagram. Notwithstanding the invalidity of the equation of state in the coexistence phase regions, we find that the scalar curvature is always negative for the van der Waals fluid,  indicating that attractive interactions dominate amongst the fluid microstructures. Along the coexistence curve, the scalar curvature $R$ decreases with  temperature, and goes to negative infinity at a critical temperature. We then numerically study the critical phenomena associated with the scalar curvature, and find that the critical exponent is 2, and that $R(1-\tilde{T})^{2}C_{v}\approx1/8$, where $\tilde{T}$ and $C_{v}$ are the respective reduced temperature and heat capacity. We next consider four-dimensional charged AdS black holes. Vanishing of the heat capacity at constant volume yields a divergent scalar curvature. In order to extract the corresponding information, we define a new scalar curvature that has  behaviour similar  to that of a van der Waals fluid. We analytically confirm that at the critical point of the small/large black hole phase transition, the scalar curvature has a critical exponent 2, and $R(1-\tilde{T})^{2}C_{v}=1/8$, the same as that of a van der Waals fluid. However we also find a significant distinction:  the scalar curvature can be positive for the small charged AdS black hole, implying that repulsive interactions  dominate among the black hole microstructures. We then generalize our study to higher-dimensional charged AdS black holes, and investigate the influence of the dimensionality on the black hole microstructures and the scalar curvature. Our novel approach provides a universal way for probing the microstructure of charged AdS black holes from a geometric construction.
\end{abstract}



\maketitle

\section{Introduction}
\label{aaa}

Geometric methods have proven to be quite useful and interesting in furthering investigations of black hole thermodynamics. The scalar curvature of the corresponding parameter space geometry can provide   information about black hole phase transitions and is an important tool for investigating the microstructure of a black hole.

In equilibrium thermodynamics Weinhold \cite{Weinhold} introduced the first thermodynamic geometry, where the internal energy is chosen to be the thermodynamic potential. Motivated by this Ruppeiner \cite{Ruppeiner}, starting from the Boltzmann entropy formula, introduced another thermodynamic geometry in which the entropy was chosen to be the thermodynamic potential. The Riemann scalar curvature associated with the phase space geometry can be worked out, and  it was found that it provides information about phase transitions.

This treatment has subsequently been applied to different fluid systems including, for example,   ideal fluids, van der Waals (VdW) fluids, one-dimensional Ising models, quantum gases and so on \cite{Ruppeiner79,Ingarden79,Janyszek90,Ruppeiner81,Janyszek89b,Janyszek90a}. These results, taken in conjunction, clarify the  physical meaning of the geometry. The line element of the geometry measures the distance between two neighbouring fluctuation states. The less probable a fluctuation between two neighbouring states, the longer  the distance between them. For the Fermi or Bose ideal gas, the corresponding scalar curvature is positive or negative, and vanishes for the classical ideal gas. This led to the suggestion  \cite{Oshima} that the sign of the scalar curvature can be used to test the type of interaction amongst the micromolecules of a thermodynamic system. Positive and negative scalar curvatures are respectively related to repulsive and attractive interactions. This property has also been used to test the microscopic properties of other fluid systems \cite{Ruppeiner2010,Mausbach}.

It has also been suggested that the scalar curvature $R$  is related to the correlation length $\xi$ of the thermodynamic system via \cite{Ruppeiner}
\begin{equation}
 R\sim \kappa  \xi^{\bar{d}},
\end{equation}
where $\kappa$ is a dimensionless constant, and $\bar{d}$ denotes the  physical dimensionality of the system. Since near the critical point the correlation length $\xi$ diverges, the scalar curvature $R$ therefore must go to infinity at the critical point.

These results indicate that the Ruppeiner geometry can be exploited to probe the microstructure of a thermodynamic system from its macroscopic thermodynamic quantities. In some sense this is just the inverse process of statistical physics, where one starts with the microstructure of a thermodynamic system, and then obtains the macroscopic quantities.

This is of particular interest in the physics of black holes, which have long been understood as thermodynamic systems. After identifying the black hole horizon area as the entropy and the surface gravity as the temperature, four black hole thermodynamic laws were established \cite{Hawking,Bekensteina,Bekensteinb,Bardeen}, which are strikingly similar to the laws for an ordinary thermodynamic system. A significant difference between them is that the black hole entropy is proportional to the black hole area  rather than its volume. The macroscopic quantities of  a black hole, such as its mass, charge, angular momentum, and so on, can be obtained through the action. However, the microstructure of black holes is completely unclear, and its elucidation remains a huge challenge.

It is at this juncture that the Ruppeiner geometry can play a useful role.  Since it has proved to be useful in understanding the microstructure of a thermodynamic system from its macroscopic quantities, it provides us with a natural approach for probing the microstructure of a black hole. Since entropy is chosen as the thermodynamic potential in the Ruppeiner geometry, it is straightforwardly generalized to black hole systems \cite{CaiCho}; for a recent review of the Ruppeiner geometry, see Ref. \cite{Ruppeinera} and references therein.

An early application came
for a charged Reissner-Nordstrom (RN) black hole -- the scalar curvature of the Ruppeiner geometry vanishes \cite{Aman}, indicating that no non-trivial interaction exists amongst the microstructures of the black hole.
However we still expect there exists some microstate interaction for a charged RN black hole because  the corresponding spacetime is curved. This led to the suggestion  \cite{Mirza}   that   the scalar curvature should be computed for a complete set of thermodynamic variables, where the angular momentum and the cosmological constant are included; the finite scalar curvature for the charged RN black hole is then obtained by taking limits of the parameters.

Another interesting behaviour of the scalar curvature was observed  for a VdW fluid \cite{Sahay}. Taking the temperature and fluid density as parameter space variables, the Ruppeiner scalar curvature, plotted as  a function of   pressure for fixed temperature was found to have  two divergent points and a self-crossing behaviour. This behaviour was later observed in other black hole systems \cite{Pankaj,Wei,Chaturvedi}. Although it was suggested that  similar self-crossing behaviour can be used to determine the phase transition point \cite{Sengupta} for normal hydrogen (or even a black hole), this idea is still speculative. Some studies also show that the scalar curvature does not diverge at the singular points of the heat capacity. In order to  solve this problem, some other geometries were proposed; see \cite{Quevedo,Lu,Tian,Liu,Roychowdhury,Hosseini,Zengd,Fazel,Naderi,Dolanb,Mansooria,Mansoorib} and references therein.

Recently, there has been great interest in studying black hole thermodynamics and phase transitions in  extended phase space, where the cosmological constant  is interpreted as  thermodynamic pressure \cite{Kastor,Gunasekaran}
\begin{eqnarray}
 P=-\frac{\Lambda}{8\pi}=\frac{(d-1)(d-2)}{16\pi l^{2}},\label{pL}
\end{eqnarray}
 where $\Lambda$ is the cosmological constant, $l$ is the AdS radius, and $d$ is the dimension of spacetime.
Adopting this interpretation, the conjugate variable of the pressure is the thermodynamic volume.
One of the first applications of this idea was for charged AdS black holes, which were found to exhibit $P$-$v$ criticality (with $v$ being the specific volume) \cite{Kubiznak}. The results demonstrate that the phase transition and the critical exponents are just like that of  a VdW fluid. This approach has been  extended to a very large variety of AdS black holes, and phenomena such as   reentrant phase transitions, triple points, isolated critical points, and $\lambda$-line phase transitions have all been observed in different black hole systems \cite{Altamirano,AltamiranoKubiznak,Altamirano3,Wei2,Frassino,Cai,XuZhao,Kostouki,Hennigar,Hennigar2,Tjoa2,Ruihong}. All these results imply that the black hole systems are quite similar to ordinary thermodynamic systems, and the subject has come to be known as {\it Black Hole Chemistry}.

Combining Ruppeiner geometry and this new understanding of black hole phase transitions has led to a novel way to probe black hole microstructure. By taking the mass and the pressure as the parameter space coordinates for a charged black hole, the Ruppeiner scalar curvature can be expressed as the number density of the microstates \cite{WeiLiu} for two distinct phases. For a large black hole  an  attractive interaction was shown to exist between two `black hole molecules', whereas for a small black hole a repulsive interaction was obtained in some parameter range. This yielded the first insight into black hole microstructure. This approach has since been generalized to other black hole systems \cite{Dehyadegari,Moumni,Deng,Sheykhi,Miao,Miao2,Miao3,Li,Chen,Guo,Du}.

This substantive progress in understanding black hole microstructure has presented us with a puzzle. At the critical point, the scalar curvature has a finite constant value rather than diverging, which is obviously different from the previous findings of the Ruppeiner geometry. In order to clarify this problematic point, we recently  reexamined the Ruppeiner geometry for the black hole, and proposed a new normalized scalar curvature \cite{LiuLiu}. We empolyed it to explore the microstructures of a four-dimensional charged AdS black hole, and found that while an attractive interaction dominates for most parameter ranges, a weak repulsive interaction dominates for small black holes at high temperature. This property distinguishes the black hole system from that of a VdW fluid. Two more novel universal properties of the Ruppeiner geometry for charged black holes were also obtained.

The purpose of this paper is to provide a detailed study of our proposed approach \cite{LiuLiu}, extending our work to van der Waals fluids and higher dimensional charged AdS black holes. In so doing, we obtain more detailed information concerning the microstructures of charged AdS black holes.

The outline of our paper is as follows. In Sec. \ref{ggg}, we give a brief introduction to Ruppeiner geometry.  Then, by employing the first law of black hole thermodynamics, we obtain  a universal line element for the Ruppeiner geometry with different parameter coordinates. In Sec. \ref{ppp}, we obtain  a universal scalar curvature for the Ruppeiner geometry. In Sec. \ref{vvv}, we  apply our approach to a VdW fluid. First, we review the phase transition for the VdW fluid, and exhibit its phase structures. Then the scalar curvature is calculated and analyzed along the coexistence curve,  and we investigate critical phenomena at the critical point. This novel approach is also extended to four-dimensional charged AdS black holes in Sec. \ref{fff}, where some analytical results are obtained. Finally in Sec. \ref{aaa} we also carry out the calculation for higher-dimensional charged AdS black holes to see how spacetime dimension affects these results.

\section{First laws and Ruppeiner geometry}
\label{ggg}

In this section we provide a brief review of Ruppeiner geometry from the thermodynamic fluctuation approach, and construct the associated line element.

\subsection{Thermodynamic fluctuations}

One of the mainstays of statistical mechanics is Boltzmann's entropy formula
\begin{equation}
 S=k_{\rm B}\ln\Omega,
\end{equation}
where the Boltzmann constant $k_{\rm B}\approx 1.38\times 10^{-23} J/K$, and $\Omega$ denotes the number of the microscopic states of the corresponding thermodynamic system. Its inversion is
\begin{equation}
 \Omega=e^{\frac{S}{k_{\rm B}}},\label{Os}
\end{equation}
which is the starting point for the thermodynamic fluctuation theory.

Here we would like to consider that a thermodynamic system has two independent variables, $x^{0}$ and $x^{1}$. Then the probability of finding the system in the internals
$x^{0}+dx^{0}$ and $x^{1}+dx^{1}$ is proportional to the number of the microstates
\begin{equation}
 P(x^{0},x^{1}) dx^{0}dx^{1}=C\, \Omega(x^{0},x^{1}) dx^{0}dx^{1}
 \end{equation}
where the parameter $C$ is a normalization constant. Substituting (\ref{Os})  into the above equation, we have
\begin{equation}
  P(x^{0}, x^{1})  \propto e^{\frac{S}{k_{\rm B}}}.\label{pp}
\end{equation}
Here we consider a thermodynamic system  partitioned into  a small sub-system plus its remainder, which can be regarded as the environment. Therefore the total entropy can be expressed as the sum of the entropy of the system $S$ and environment $E$
\begin{equation}
 S(x^{0},x^{1})=S_{\rm S}(x^{0},x^{1})+S_{\rm E}(x^{0},x^{1}),
\end{equation}
with $S_{\rm S}\ll S_{\rm E}\sim S$.

Further, near the local maximum of the entropy at $x^{\mu} = x^{\mu}_0$, one can expand the total entropy in the following form
\begin{eqnarray}
 S&=&S_{0} + \left. \frac{\partial S_{\rm S}}{\partial x^{\mu}}  \right|_{0} \Delta x^{\mu}_{\rm S}
      + \left.\frac{\partial S_{\rm E}}{\partial x^{\mu}}   \right|_{0}   \Delta x^{\mu}_{\rm E}\nonumber\\
 &&   + \left. \frac{1}{2}\frac{\partial^{2}S_{\rm S}}{\partial x^{\mu}\partial x^{\nu}}
        \right|_{0}  \Delta x^{\mu}_{\rm S}\Delta x^{\nu}_{\rm S}
      + \left. \frac{1}{2}\frac{\partial^{2}S_{\rm E}}{\partial x^{\mu}\partial x^{\nu}}
        \right|_{0}  \Delta x^{\mu}_{\rm E}\Delta x^{\nu}_{\rm E}
   +\cdots, \quad (\mu, \nu=0,1),
\end{eqnarray}
where $S_{0}$ denotes the local maximum of the entropy $S(x^{\mu}_0)$ and ``$|_{0}$'' is short for $|_{x^{\mu} = x^{\mu}_0}$. For a closed system, we suppose that the fluctuating parameters are conservative and additive i.e., $x^{\mu}_{\rm S}+x^{\mu}_{\rm E}=x^{\mu}_{\rm total}=constant$; hence
\begin{equation}
  \left.\frac{\partial S_{\rm S}}{\partial x^{\mu}} \right|_{0} \Delta x^{\mu}_{\rm S}
   =-\left.\frac{\partial S_{\rm E}}{\partial x^{\mu}} \right|_{0} \Delta x^{\mu}_{\rm E}.\end{equation}
Therefore,
\begin{equation}
 \Delta S=    \left. \frac{1}{2}\frac{\partial^{2}S_{\rm S}}{\partial x^{\mu} \partial x^{\nu}}
              \right|_{0} \Delta x^{\mu}_{\rm S} \Delta x^{\nu}_{\rm S}
          +   \left. \frac{1}{2}\frac{\partial^{2}S_{\rm E}}{\partial x^{\mu} \partial x^{\nu}}
              \right|_{0} \Delta x^{\mu}_{\rm E} \Delta x^{\nu}_{\rm E}
          +   \cdots\;.
 \end{equation}
Since $S_{\rm E}$ is of the same scale as that of the total system, the second term is much smaller than the first term, so we ignore it. Then by truncating the  series at   second order and absorbing $S_{0}$ into the normalization constant, we   have the probability
\begin{equation}
 P(x^{0},x^{1}) \propto e^{-\frac{1}{2}\Delta l^{2}}.
\end{equation}
In the thermodynamic information geometry, $\Delta l^{2}$ measures the distance between two neighbouring fluctuation states, and its expression is given by
\begin{eqnarray}
 \Delta l^{2}&=&-\frac{1}{k_{\rm B}}g_{\mu\nu} \Delta x^{\mu}\Delta x^{\nu},\label{Ds}\\
 g_{\mu\nu}&=&\frac{\partial^{2}S_{\rm S}}{\partial x^{\mu}\partial x^{\nu}}.\label{gmunu}
 \end{eqnarray}
For simplicity, we set $k_{\rm B}=1$ and drop the  index S in the entropy $S_{\text{S}}$.

\subsection{Riemannian geometry}

Under the transformation
\begin{equation}
 x^{\mu}\rightarrow x'^{\mu},
\end{equation}
$g_{\mu\nu}$ transforms as follows \cite{Landau}
\begin{equation}
 g_{\mu\nu}'=\frac{\partial x^{\rho}}{\partial x'^{\mu}}\frac{\partial x^{\sigma}}{\partial x'^{\nu}}g_{\rho\sigma}
\end{equation}
which is just like that of the metric tensor in Riemannian geometry. In other words, we can  treat $\Delta l^2$ in (\ref{Ds}) as the distance between two neighbouring states in the state space. As we know information in Riemannian geometry is contained in the scalar curvature. We shall compute the scalar curvature in this parameter space following the approach in Riemannian geometry, where we employ
the conventions
\begin{eqnarray}
 \Gamma^{\sigma}_{\;\;\mu\nu}&=&\frac{1}{2}g^{\sigma\rho}
 (\partial_\nu g_{\rho\mu} + \partial_\mu g_{\rho\nu} - \partial_\rho g_{\mu\nu}) \\
 R^{\sigma}_{\;\;\rho\mu\nu}&=& \partial_\nu \Gamma^{\sigma}_{\;\;\rho\mu}
 - \partial_\mu \Gamma^{\sigma}_{\;\;\rho\nu}+\Gamma^{\delta}_{\;\;\rho\mu}\Gamma^{\sigma}_{\;\;\delta\nu}-\Gamma^{\delta}_{\;\;\rho\nu}\Gamma^{\sigma}_{\;\;\delta\mu} \\
 R_{\mu\nu}&=&R^{\sigma}_{\;\;\mu\sigma\nu} \qquad
 R=g^{\mu\nu}R_{\mu\nu}
\end{eqnarray}
with $g^{\mu\nu}$  the inverse of the metric tensor $g_{\mu\nu}$. In a 2-dimensional space, the scalar curvature is \cite{Sokolnikoff}
\begin{eqnarray}
 R=&-&\frac{1}{\sqrt{g}}\bigg[\frac{\partial}{\partial x^{0}}\left(\frac{g_{01}}{g_{00}\sqrt{g}}\frac{\partial g_{00}}{\partial x^{1}}-\frac{1}{\sqrt{g}}\frac{\partial g_{11}}{\partial x^{0}}\right)\nonumber\\
 &+&\frac{\partial}{\partial x^{1}}\left(\frac{2}{\sqrt{g}}\frac{\partial g_{01}}{\partial x^{1}}-\frac{1}{\sqrt{g}}\frac{\partial g_{00}}{\partial x^{1}}-\frac{g_{01}}{g_{00}\sqrt{g}}\frac{\partial g_{00}}{\partial x^{0}}\right)\bigg],
\end{eqnarray}
where the determinant $g=g_{00}g_{11}-g_{01}g_{10}$   with $g_{01}=g_{10}$. If the metric is diagonal, the scalar curvature can be simplified to
\begin{equation}
 R=\frac{1}{\sqrt{g}}\left[\frac{\partial}{\partial x^{0}}\left(\frac{1}{\sqrt{g}}\frac{\partial g_{11}}{\partial x^{0}}\right)+\frac{\partial}{\partial x^{1}}\left(\frac{1}{\sqrt{g}}\frac{\partial g_{00}}{\partial x^{1}}\right)\right].\label{scalar}
\end{equation}
We shall subsequently show that the scalar curvature $R$ plays an important role in exploring the microscopic properties of a thermodynamic system.

\subsection{First law and Ruppeiner geometry}

The Ruppeiner geometry is based on the line element given in (\ref{Ds}). In this  subsection we   consider the Ruppeiner geometry from the perspective of  the first law of thermodynamics.

For a thermodynamic system, the first law reads
\begin{equation}
 dU=TdS-PdV+\sum_{i}y_{i}dx^{i},
\end{equation}
 where $x^{i}$ are the thermodynamic variables,  and $y_{i}$ are the corresponding chemical potentials. For simplicity, we absorb the $PdV$ term into the last term, writing
\begin{equation}
 dU=TdS+\sum_{i}y_{i}dx^{i}.
\end{equation}
Since the thermodynamic potential of the Ruppeiner geometry is the entropy, we can change this law to
\begin{equation}
 dS=\frac{1}{T}dU-\sum_{i}\frac{y_{i}}{T}dx^{i}.\label{firstenlaw}
\end{equation}
Here we denote the indices $\mu$=0,1,2,$\cdots$, and $i$=1,2,3,$\cdots$. Then we have the respective extensive and intensive quantities
\begin{equation}
 x^{\mu} = (U,V,\cdots), \qquad
 y_{i}= (-P,\cdots),
\end{equation}
and from (\ref{firstenlaw}), we find the quantity $z_{\mu}=\frac{\partial S}{\partial x^{\mu}}$, which is
\begin{equation}
 z_{\mu}= \left(\frac{1}{T},-\frac{y_{i}}{T}\right). \label{zmu}
\end{equation}
With these symbols, (\ref{firstenlaw}) can be expressed as
\begin{equation}
 dS=z_{\mu}dx^{\mu}.
\end{equation}
Moreover, the line element (\ref{Ds}) can be rewritten as
\begin{equation}
 \Delta l^{2}=-\Delta z_{\mu}\Delta x^{\mu}.\label{eles}
\end{equation}
From (\ref{zmu}), we have
\begin{eqnarray}
 \Delta z_{0}&=&\Delta\left(\frac{1}{T}\right)=-\frac{1}{T^{2}}\Delta T,\\
 \Delta z_{i}&=&\Delta\left(-\frac{y_{i}}{T}\right)=\frac{y_{i}}{T^{2}}\Delta T-\frac{1}{T}\Delta y_{i}.
\end{eqnarray}
Then the line element (\ref{eles}) will be of the following form
\begin{equation}
 \Delta l^{2}=\frac{1}{T}\Delta T\Delta S+\frac{1}{T}\Delta y_{i}\Delta x^{i}.\label{dl2}
\end{equation}

We now consider two choices for the parameter space coordinates:  the ($T$, $x^{i}$) coordinate and the ($T$, $y_{i}$) coordinate. We shall henceforth refer to these as {\it fluctuation coordinates}, since
it is these coordinates whose fluctuations induce the change $\Delta l$.
\\

\par
\noindent \textbf{Case I:} ($T$, $x^{i}$) fluctuation coordinate.

In this case the most important coordinate is ($T$, $V$) or ($T$, $\rho$) with $\rho$ the density of the system. Before pursuing the specific form of the line element, let us first look at the corresponding thermodynamic potential of the geometry.

Since the independent variables are $T$ and $x^{i}$, the potential must be the Helmholtz free energy $F=U-TS$. The differential law is
\begin{equation}
 dF=-SdT+y_{i}dx^{i}.\label{Helmf}
\end{equation}
It is clear that the Helmholtz free energy $F$ is  a function of $T$ and $x^{i}$. Interestingly, one has the relation
\begin{equation}
 \frac{\partial S}{\partial x^{i}}=-\frac{\partial y_{i}}{\partial T}.\label{Mexr}
\end{equation}
Next, we expand $\Delta S$ and $\Delta y_{i}$ by using the independent variables,
\begin{eqnarray}
 \Delta S&=&\left(\frac{\partial S}{\partial T}\right)\Delta T+\left(\frac{\partial S}{\partial x^{i}}\right)\Delta x^{i},\\
 \Delta y_{i}&=&\left(\frac{\partial y_{i}}{\partial T}\right)\Delta T+\left(\frac{\partial y_{i}}{\partial x^{j}}\right)\Delta x^{j}.
\end{eqnarray}
Plunging them into (\ref{dl2}) and using the relation (\ref{Mexr}), we have the line element of the Ruppeiner geometry under the $T$ and $x^{i}$ fluctuations
\begin{equation}
 \Delta l^{2}=\frac{1}{T}\left(\frac{\partial S}{\partial T}\right) \Delta T^{2}
 +\frac{1}{T}\left(\frac{\partial y_{i}}{\partial x^{j}}\right)\Delta x^{i}\Delta x^{j}.
\end{equation}
According to (\ref{Helmf}), one has $S=-\partial_{T}F$ and $y_{i}=\partial_{x^{i}}F$. Therefore, the line element can be further expressed as
\begin{equation}
 \Delta l^{2}=-\frac{1}{T}\left(\frac{\partial^{2}F}{\partial T^{2}}\right)\Delta T^{2}
   +\frac{1}{T}\left(\frac{\partial^{2}F}{\partial x^{i}\partial x^{j}}\right)\Delta x^{i}\Delta x^{j}. \label{ddl2}
\end{equation}
Here the thermodynamic potential is the Helmholtz free energy, which is the same with that proposed in Ref. \cite{Lu}. However, these two geometries are rather different, because that there is no  $\Delta T\Delta x^{i}$ term. Moreover, if the indices $i$ and $j$ have only one value, this metric is diagonal, and the scalar curvature can be easily calculated. The geometry of the fluctuation coordinates ($T$, $V$) and ($T$, $\rho$) belong to this case.\\

\par
\noindent \textbf{Case II:} ($T$, $y^{i}$) fluctuation coordinate.

This choice is very useful for investigating   ($T$, $P$) fluctuations. The corresponding thermodynamic potential is
\begin{equation}
 W=U-TS-y_{i}x^{i},
\end{equation}
and thus
\begin{equation}
 dW=-SdT-x^{i}dy_{i}.
\end{equation}
Employing this differential form, the entropy $S$ and $x^{i}$ can be obtained by
\begin{equation}
 S=-\frac{\partial W}{\partial T},\quad
 x^{i}=-\frac{\partial W}{\partial y_{i}}.
\end{equation}
Expanding $\Delta S$ and $\Delta x^{i}$ in terms of the independent variables $T$ and $y_{i}$ yields
\begin{eqnarray}
 \Delta S&=&\left(\frac{\partial S}{\partial T}\right)\Delta T+\left(\frac{\partial S}{\partial y_{i}}\right)\Delta y_{i},\\
 \Delta x^{i}&=&\left(\frac{\partial x^{i}}{\partial T}\right)\Delta T+\left(\frac{\partial x^{i}}{\partial y_{j}}\right)\Delta y_{j}.
\end{eqnarray}
Combining them together, the line element will be in the following form
\begin{equation}
 \Delta l^{2}=-\frac{1}{T}\left(\frac{\partial^{2}W}{\partial p_{\mu}\partial p_{\nu}}\right) \Delta p_{\mu}\Delta p_{\nu},\quad p_{\mu}=(T, y_{i}).
\end{equation}

\section{Scalar curvature and phase transition}
\label{ppp}

In applying Ruppeiner geometry to black hole thermodynamics,  the first thing to  consider is phase transitions.  The scalar curvature should diverge at the critical point of a phase transition as  fluid systems do. This property can be used to investigate the information contained in  black hole phase transitions from the thermodynamic geometric perspective. The sign of the scalar curvature can be used as a diagnostic of the interaction between two microscopic constituents (or molecules) of the system. A positive scalar curvature indicates a repulsive interaction, whereas a  negative one signifies an attractive interaction. In this section we consider the relation between  phase transitions and the scalar curvature.

Consider the scalar curvature associated with (\ref{ddl2}), which actually is extensively used   for  fluid systems. For simplicity, we take the fluctuation coordinates $T$ and $x$, where $x$ can be any one of the volume $V$,   charge $Q$,  angular momentum $J$, or other extensive quantities.
For this case, the metric is  two-dimensional and diagonal
\begin{equation}
 g_{\mu\nu}=\frac{1}{T}\left(
  \begin{array}{cc}
     -\left(\frac{\partial^{2}F}{\partial T^{2}}\right)_{x} & 0\\
    0 & \left(\frac{\partial^{2}F}{\partial x^{2}}\right)_{T}\\
  \end{array}
\right)
=
\frac{1}{T}\left(
  \begin{array}{cc}
     \left(\frac{\partial S}{\partial T}\right)_{x} & 0\\
    0 & \left(\frac{\partial y}{\partial x}\right)_{T}\\
  \end{array}
\right) .
\end{equation}
Noting that the heat capacity at constant $x$ is
\begin{equation}
 C_{x}=T\left(\frac{\partial S}{\partial T}\right)_{x},
\end{equation}
the line element can be expressed as
\begin{equation}
 dl^{2}=\frac{C_{x}}{T^{2}}dT^{2}+\frac{(\partial_{x}y)_{T}}{T}dx^{2},\label{xxy}
\end{equation}
and  the scalar curvature is easily worked out
\begin{eqnarray}
 R&=&\frac{1}{2C_{x}^{2}(\partial_{x}y)^{2}}
 \bigg\{
 T(\partial_{x}y)\bigg[(\partial_{x}C_{x})^{2}+(\partial_{T}C_{x})(\partial_{x}y-T\partial_{T,x}y)\bigg]\nonumber\\
 &+&C_{x}\bigg[(\partial_{x}y)^{2}+T\left((\partial_{x}C_{x})(\partial_{x,x}y)-T(\partial_{T,x}y)^{2}\right)
 +2T(\partial_{x}y)(-(\partial_{x,x}C_{x})+T(\partial_{T,T,x}y))\bigg]
 \bigg\} \nonumber\\\label{RR}
\end{eqnarray}
using (\ref{scalar}). Note that $R$ can potentially diverge at $C_{x}=0$ or $(\partial_{x}y)_{T} =0$. Indeed, for a VdW-like phase transition, the condition determining the critical point is
\begin{equation}\label{critpt1}
 \left(\partial_{x}y\right)_{T}=\left(\partial_{x,x}y\right)_{T}=0
\end{equation}
and so it is quite obvious that the scalar curvature of the Ruppeiner geometry has  divergent behaviour at the critical point of the phase transition. This property provides a possible link between the scalar curvature and the correlation length, which approaches infinity at the critical point according to   phase transition theory.
For constant heat capacity $C_{x}$  the scalar curvature (\ref{RR})   reduces to
\begin{equation}
 R=\frac{(\partial_{x}y)^{2}-T^{2}(\partial_{T,x}y)^{2}+2T^{2}(\partial_{x}y)(\partial_{T,T,x}y)}{2C_{x}(\partial_{x}y)^{2}}
\end{equation}
which diverges when \eqref{critpt1} holds provided $\partial_{T,T,x}y \neq 0$.

A previous examination of the    relationship between  phase transitions and the Ruppeiner geometry,  indicated that scalar curvature does not go to infinity but rather has a finite negative value at the critical point  \cite{WeiLiu}. The reason for this is that  the mass $M$ and pressure $P$ were taken to be the fluctuation variables. In the following sections, we will instead take the temperature $T$ and other extensive quantities as independent variables to investigate black hole phase transitions.

\section{Van der Waals fluid}
\label{vvv}

A VdW fluid is the first and simplest model of an interacting thermodynamic system exhibiting a liquid-gas phase transition. Its thermodynamic geometry was constructed by Ruppeiner \cite{Ruppeiner}, and the scalar curvature was obtained in terms of $(T, \rho)$ parameter coordinates. A novel result was also revealed in Ref. \cite{Sahay}: the isothermal line in an $R$-$P$ diagram shows a self-crossing behaviour, leading some authors to suggest that such a crossing point can be approximately treated as the phase transition point. In this section we   give a brief view of the thermodynamics of the VdW fluid, examine the relationship between the Ruppeiner geometry and the liquid-gas phase transition, and reveal the critical phenomena associated with the scalar curvature.

\subsection{Thermodynamics and phase structures}
\label{svdw}

Compared with an ideal fluid, two more characteristic quantities, $a$ and $b$ are introduced for the VdW fluid. The quantity $a$ measures the interaction between two fluid molecules, and $b$ denotes the size of the molecule.  The specific Helmholtz free energy of the VdW fluid is \cite{Landau}
\begin{equation}
  F=-\frac{3}{2}T\ln T-\xi T+\epsilon-T\ln(e(v-b))-\frac{a}{v},\label{fffd}
\end{equation}
where $e$, $\xi$ and $\epsilon$ are constants. One should note that the specific volume $v>b$. The last term is a correction term that originates from the interaction of the fluid molecules, and  from which we can clearly see dependence on the parameter $a$. Note that here $v$ is the specific volume of the fluid. The relation between  the specific volume and the total volume is $V=Nv$  with $N$ the total number of all the microscopic molecules. In  what follows,  thermodynamic quantities shall generally refer to   specific ones.

The entropy and energy are
\begin{eqnarray}
 S&=&-\left(\frac{\partial F}{\partial T}\right)_{v}
  =\frac{3}{2}\left(1 +\frac{2}{3}\xi+\ln T\right) +\ln(e(v-b)),\\
 U&=&F+TS
  =\frac{3}{2}T-\frac{a}{v} +\epsilon.
\end{eqnarray}
The heat capacity at constant volume is
\begin{equation}
 C_{v}=\frac{3}{2}k_{\rm B}
\end{equation}
where we have restored the Boltzmann constant $k_{\rm B}$. So $C_{v}$ is an extremely small quantity. We note  that the heat capacity is the same as that of the ideal fluid.
%%%%%%%%%%%%%%%
\begin{figure*}
\begin{center}
\subfigure[]{\label{ISOa}
\includegraphics[width=7cm]{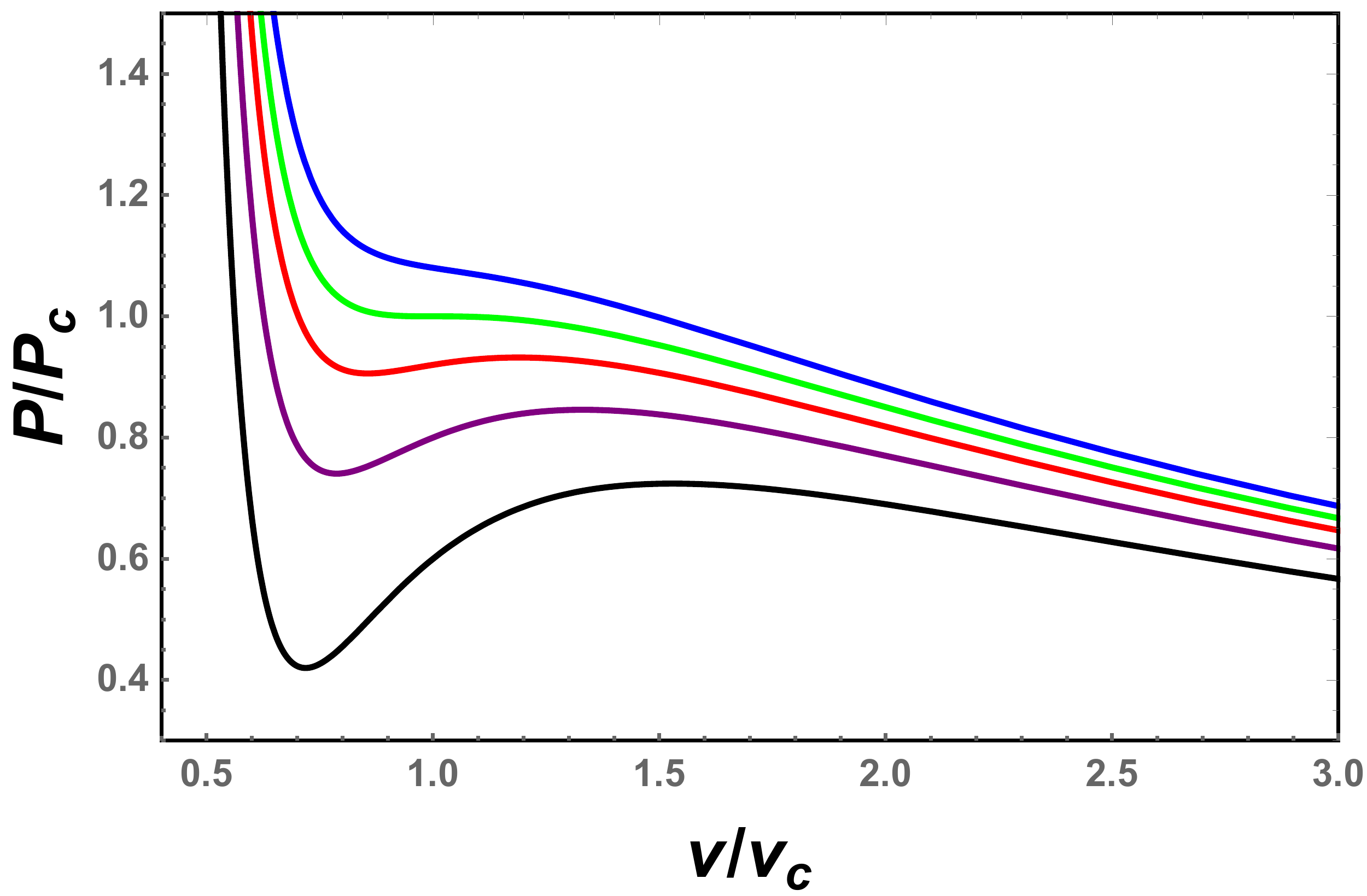}}
\subfigure[]{\label{ISPb}
\includegraphics[width=7cm]{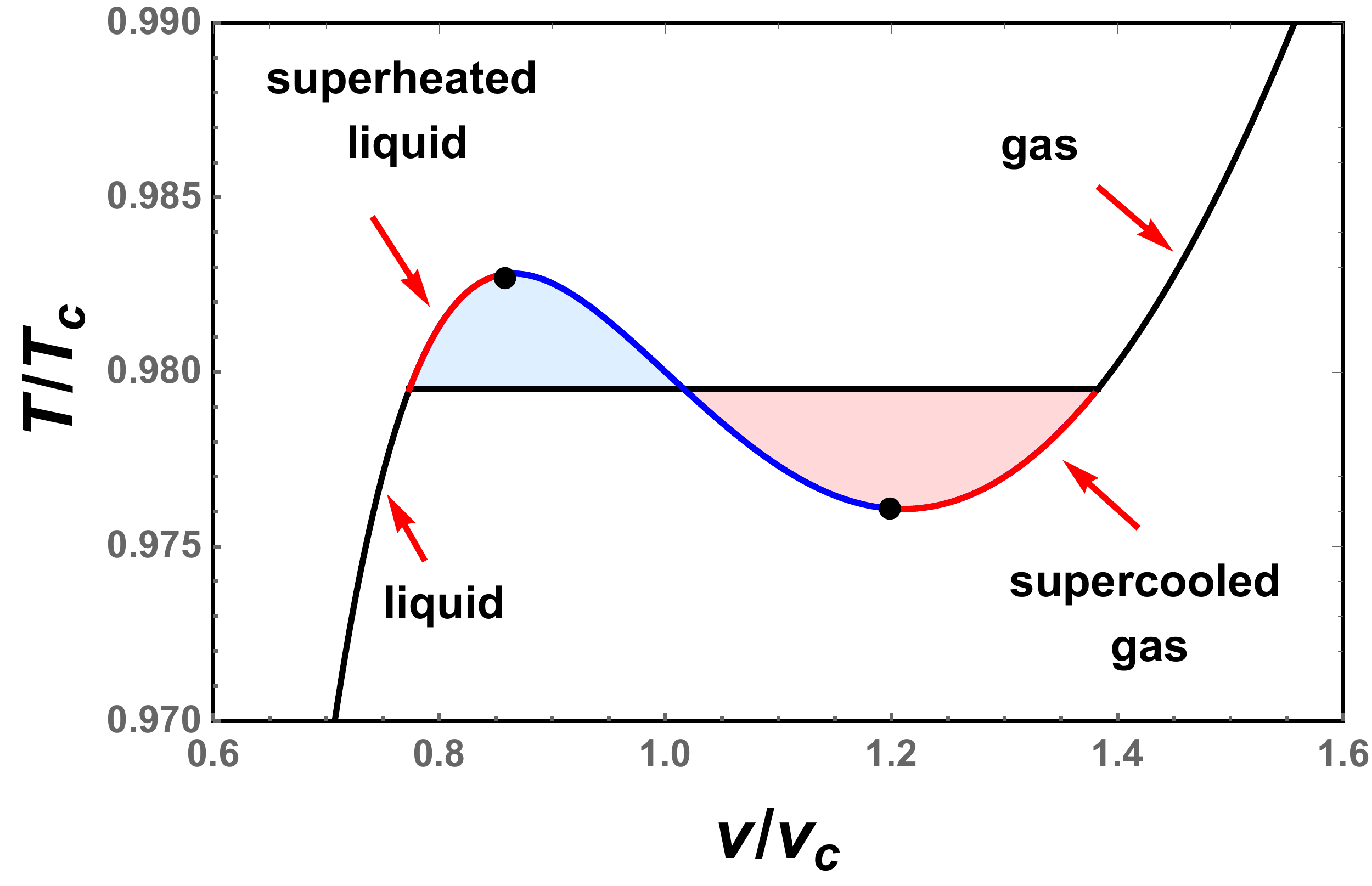}}
\end{center}
\caption{(a) Isothermal curve with $\tilde{T}=$0.9, 0.95, 0.98, 1.00, and 1.02 from bottom to top in $\tilde{P}$-$\tilde{v}$ diagram. (b) Isobaric curve with $\tilde{P}=0.92$ in $\tilde{T}$-$\tilde{v}$ diagram. The black horizontal line of $T/T_{\rm c}=0.98$ corresponds the phase transition temperature calculated from the free energy. The black curves are the liquid and gas branches, respectively. The red curves are two metastable branches, the superheated liquid branch and the supercooled gas branch. The blue curve of negative slope is an unstable branch, and it is replaced by the horizontal line when the phase transition is considered. These two black dots are the spinodal points, which separate the metastable branches from the unstable branch. Note that under this case, the equal area law does not hold.}\label{pISPb}
\end{figure*}
%%%%%%%%%%%%%%%

The pressure or the equation of state can be calculated as
\begin{equation}
 P=-\left(\frac{\partial F}{\partial v}\right)_{T}
   =\frac{T}{v-b}-\frac{a}{v^{2}}.\label{stateeq}
\end{equation}
Employing this result, the enthalpy for one molar VdW fluid is
\begin{equation}
 H=U+Pv=T\left(\frac{v}{v-b}+\frac{3}{2}\right)-\frac{2a}{v} +\epsilon.
\end{equation}
The VdW fluid admits a liquid-gas phase transition.
The first-order phase transition point can be determined from the swallowtail behaviour of the free energy or the Maxwell equal area law. The critical point is a second-order phase transition point, at which the latent heat vanishes, and   is determined by the condition
\begin{equation}
 (\partial_{v}P)_{T}=(\partial_{v,v}P)_{T}=0.
\end{equation}
With the help of the equation of state  (\ref{stateeq}), the critical point is given by
\begin{equation}
 P_{\rm c}=\frac{a}{27b^{2}},\quad v_{\rm c}=3b,\quad T_{\rm c}=\frac{8a}{27b}.
\end{equation}
In the reduced parameter space, the equation of state (\ref{stateeq}) has the following form
\begin{equation}\label{VdWeos}
 \tilde{P}=\frac{8\tilde{T}}{3\tilde{v}-1}-\frac{3}{\tilde{v}^{2}},
\end{equation}
where the reduced pressure, temperature, and specific volume are defined by $\tilde{P}=P/P_{\rm c}$, $\tilde{T}=T/T_{\rm c}$, and $\tilde{v}=v/v_{\rm c}$. The reduced volume $\tilde{v}$ must be larger than 1/3. More interestingly, this reduced state function does not depend on the parameters $a$ and $b$.

We depict the isothermal curves in Fig. \ref{ISOa} with the reduced temperature $\tilde{T}=$0.9, 0.95, 0.98, 1.00, and 1.02 from bottom to top. It is clear that, for $\tilde{T}<1$, there are two extremal points. These two extremal points divide one isothermal curve into three branches. Two of them, liquid and gas phases, with negative slopes are stable, while another one is unstable. One can construct equal areas to obtain the phase transition point at each isothermal curve. When $\tilde{T}>1$, no extremal point exists. In order to show more phases, we plot the isobaric curve with $\tilde{P}=0.92$ in a $\tilde{T}$-$\tilde{v}$ diagram in Fig. \ref{ISPb}. The black horizontal line of $T/T_{\rm c}=0.98$ corresponds the phase transition temperature. We can find that below the critical point, there are two extremal points. Note that the equal area law does not hold for this case due to the use of specific volume and so the  light blue and red shaded regions are not required to be equal. However if we depict the isobaric line in $T$-$S$ diagram, the equal area law will be applicable -- a discussion about this issue can be found in Ref. \cite{Wei5}.   Note that two metastable branches plotted in red color are shown. One is the superheated liquid branch, and another is the supercooled gas branch. The blue curve of negative slope is the unstable branch. The extremal points (black dots) separate the two metastable branches from the unstable branch, and thus we call them the spinodal points.

\begin{figure*}
\begin{center}
\subfigure[]{\label{VdWPTP}
\includegraphics[width=7cm]{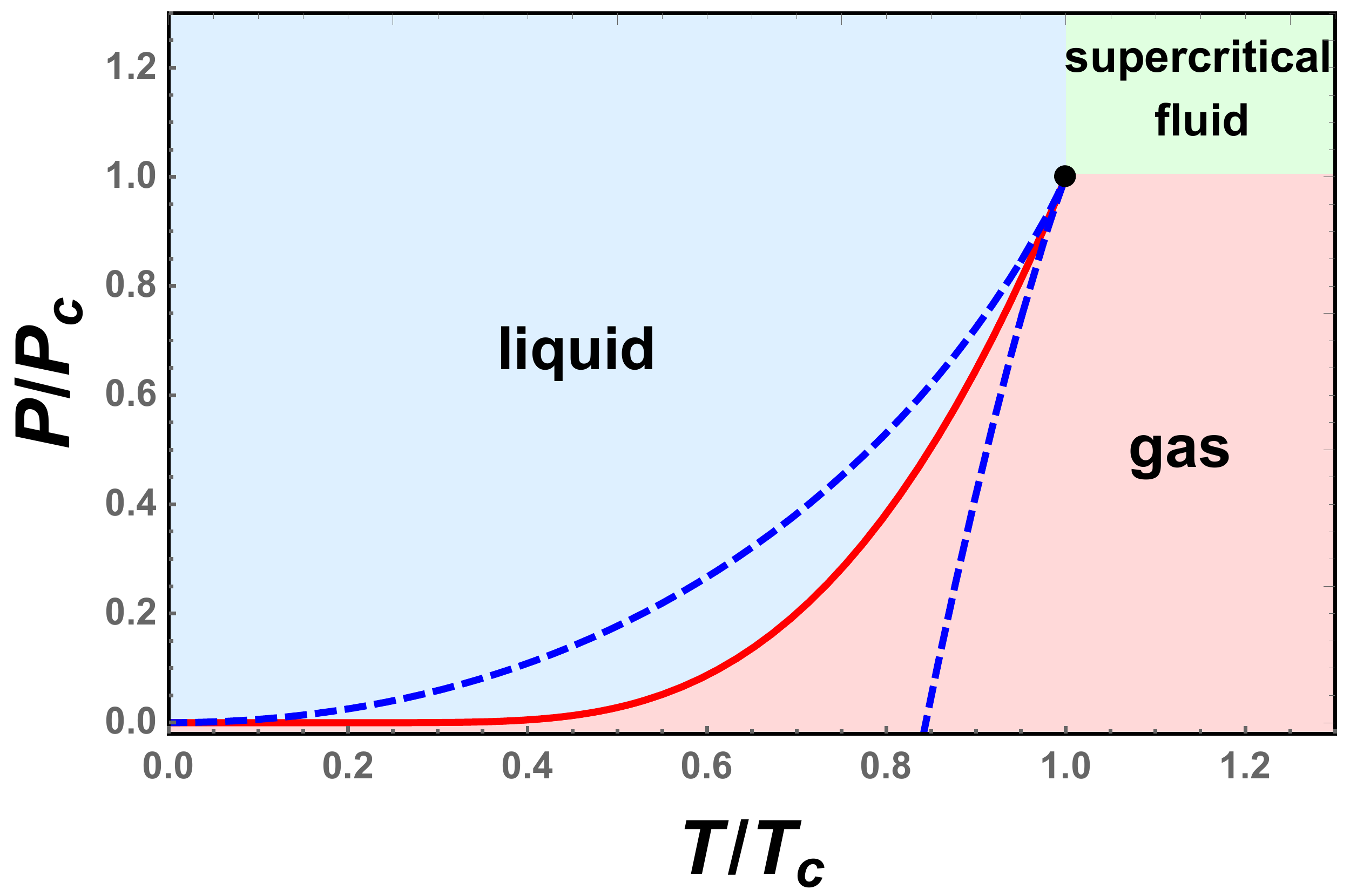}}
\subfigure[]{\label{VdWPTPb}
\includegraphics[width=7cm]{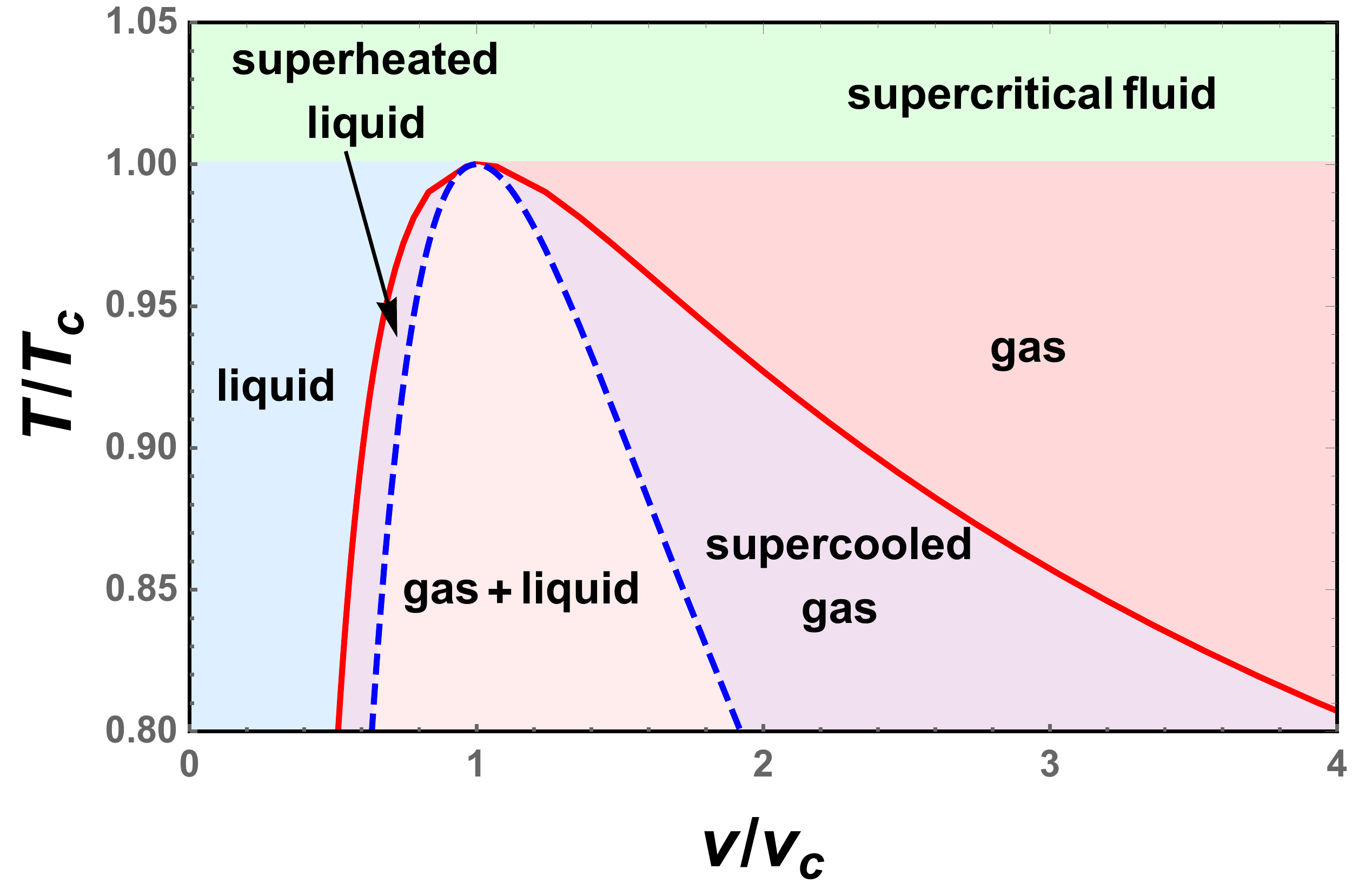}}
\end{center}
\caption{Phase diagram and spinodal curves for VdW fluid. The red solid curves and blue dashed curves are the coexistence curves and spinodal curves, respectively. (a) $\tilde{P}$-$\tilde{T}$ phase diagram. Black dot denotes the critical point. The regions for the gas, liquid, and supercritical fluid phases are displayed. The bottom blue dashed curve starts at $\tilde{T}$=27/32. (b) $\tilde{T}$-$\tilde{v}$ phase diagram. The regions for the liquid phase, gas phase, coexistence phase, superfluid phase, metastable superheated liquid phase, and metastable supercooled gas phase are clearly displayed.}\label{pVdWPTP}
\end{figure*}

In fact, the phase transition point is a coexistence phase of liquid and gas. All these points form one coexistence curve. Although there is no the analytic form for the curve, it can be described by a fitting formula \cite{Johnstonq}
\begin{equation}
 \ln\tilde{P}=\sum_{i=0}^{9}c_{i}\left(\frac{1}{\tilde{T}}\right)^{i},
\end{equation}
where the reduced pressure and temperature are $\tilde{P}\in(0,1)$ and $\tilde{T}\in(0,1)$. The ten fitting coefficients are   \cite{Johnstonq}
\begin{eqnarray}
 c_{0}&=&5.66403835,\; c_{1}=-8.73724257,\; c_{2}=5.14022974,\;
 c_{3}=-2.92538942,\nonumber\\
 c_{4}&=&1.09108819,\; c_{5}=-2.741948\times10^{-1},\;
 c_{6}=4.5992265\times10^{-2},\nonumber\\
 c_{7}&=&-4.92809927\times10^{-3},\;
 c_{8}=3.04520105\times10^{-4},\; c_{9}=-8.24218733\times10^{-6}
\end{eqnarray}
from which we plot the phase diagram in Fig. \ref{VdWPTP}. The red solid line is the coexistence curve of the liquid and gas phases. Above and below the curve are the liquid phase and gas phase, respectively. The black dot denotes the critical point. The region on the top right corner marked in light green color is the supercritical fluid phase, at which the gas and liquid cannot be clearly distinguished.

The spinodal curves are the two blue dashed curves, and are determined by the condition
\begin{equation}
 (\partial_{v}P)_{T}=0,\quad \text{or}\quad
 (\partial_{v}T)_{P}=0.
\end{equation}
Solving them, we obtain the respective gas and liquid spinodal curves
\begin{equation}
 \tilde{P}_{\rm gs,ls}=\frac{8\tilde{T}\tilde{v}^{2}_{\rm gs,ls}-9\tilde{v}_{\rm gs,ls}+3}
   {\tilde{v}_{\rm gs,ls}^{2}(3\tilde{v}_{\rm gs,ls}-1)},
\end{equation}
where
\begin{eqnarray}
 \tilde{v}_{\rm gs}&=&\frac{3+2\sqrt{9-8\tilde{T}}}{4\tilde{T}}\cos\left(\frac{1}{3}\arccos\left(\frac{8\tilde{T}^{2}-36\tilde{T}+27}{(9-8\tilde{T})^{3/2}}\right)\right),\\
 \tilde{v}_{\rm ls}&=&\frac{3-2\sqrt{9-8\tilde{T}}}{4\tilde{T}}\cos\left(\frac{1}{3}\left(\pi+\arccos\left(\frac{8\tilde{T}^{2}-36\tilde{T}+27}{(9-8\tilde{T})^{3/2}}\right)\right)\right).
\end{eqnarray}
Alternatively, in the $\tilde{T}$-$\tilde{v}$ diagram, the spinodal curve has the following compact form
\begin{equation}
 \tilde{T}_{\rm sp}=\frac{(3\tilde{v}-1)^{2}}{4\tilde{v}^{3}},\label{spcurvevdw}
\end{equation}
where $1/3<\tilde{v}<1$ is for the liquid spinodal curve and $\tilde{v}>1$ is for the gas spinodal curve (see the blue dashed curve in Fig. \ref{VdWPTPb}).

In Fig. \ref{VdWPTP}, the top blue dashed curve is the gas spinodal curve starting at $\tilde{T}$=0 and the bottom one is the liquid spinodal curve starting at $\tilde{T}$=27/32. It is clear that these spinodal curves and the coexistence curve meet each other at the critical point.

%%%%%%%%%%%%%%%
%%%%%%%%%%%%%%%

In Fig. \ref{VdWPTPb}, we show the phase structure in the $\tilde{T}$-$\tilde{v}$ diagram. The regions for the liquid phase, gas phase, coexistence phase, supercritical fluid phase, metastable superheated liquid phase, and metastable supercooled gas phase are clearly displayed. It is obvious that the blue dashed spinodal curve separates the metastable superheated liquid phase and the supercooled gas phase from the coexistence phase. Note that in the  coexistence phase region  the  equation of state (\ref{stateeq}) is invalid.

In the Ruppeiner geometry approach, we take the parameter coordinates $T$ and $x=v$. Then the parameter $y=-P$. Thus the condition $(\partial_{x}y)_{T}=0$, where the scalar curvature diverges, will reduce to $(\partial_{v}P)_{T}=0$. Therefore the scalar curvature diverges at the spinodal curve.

\subsection{Ruppeiner geometry}

In this subsection, we would like to examine the Ruppeiner geometry for the VdW fluid. Here, we take the fluctuation coordinates $T$ and $v$. Note that  Ruppeiner chose the number density $\rho$ rather the volume $v$ as one fluctuation \cite{Ruppeiner}. In order to investigate the relationship between the scalar curvature and the thermodynamic phase transition, we adopt the volume as a fluctuation quantity. For simplicity, we take the parameters $e=\xi=\epsilon=1$ in (\ref{fffd}).

In this approach, the line element of the Ruppeiner geometry is
\begin{eqnarray}
 dl^{2}&=&\frac{C_{v}}{T^{2}}dT^{2}-\frac{(\partial_{v}P)_{T}}{T}dv^{2}\nonumber\\
 &=&\frac{3}{2T^{2}}dT^{2}+\frac{Tv^{3}-2a(v-b)^{2}}{Tv^{3}(v-b)^{2}}dv^{2}.
\end{eqnarray}
Plunging it into (\ref{RR}), we obtain the scalar curvature
\begin{eqnarray}
 R=\frac{4a(v-b)^{2}\left(a(v-b)^{2}-Tv^{3}\right)}{3\left(2a(v-b)^{2}-Tv^{3}\right)^{2}}.
\end{eqnarray}
It is easy to check that the denominator of the scalar curvature $R$ vanishes at the critical point. Obviously, $R$=0 when $v=b$ -- at this point all the volume is occupied by the fluid molecules, so no extra space remains. Thus the total fluid becomes a rigid body, and no interaction can exist between the fluid molecules -- the distance between any two of its constituents is always  constant. This is consistent with the result that $R$=0 corresponds to vanishing interaction \cite{Oshima}.

Expressing the scalar curvature in the reduced parameter space
\begin{eqnarray}
 R=\frac{(3\tilde{v}-1)^{2}\big((3\tilde{v}-1)^{2}-8\tilde{T}\tilde{v}^{3}\big)}{3\big((3\tilde{v}-1)^{2}-4\tilde{T}\tilde{v}^{3}\big)^{2}},
\end{eqnarray}
we see that it does not depend on the parameters $a$ and $b$ of the VdW fluid. This property is similar to the reduced equation of state. It implies that  fluids modelled as VdW fluids share the same scalar curvature in the reduced parameter space. Therefore the scalar curvature can reveal   universal properties of different fluid systems.

%%%%%%%%%%%%%%%
\begin{figure*}
\begin{center}
\includegraphics[width=9cm]{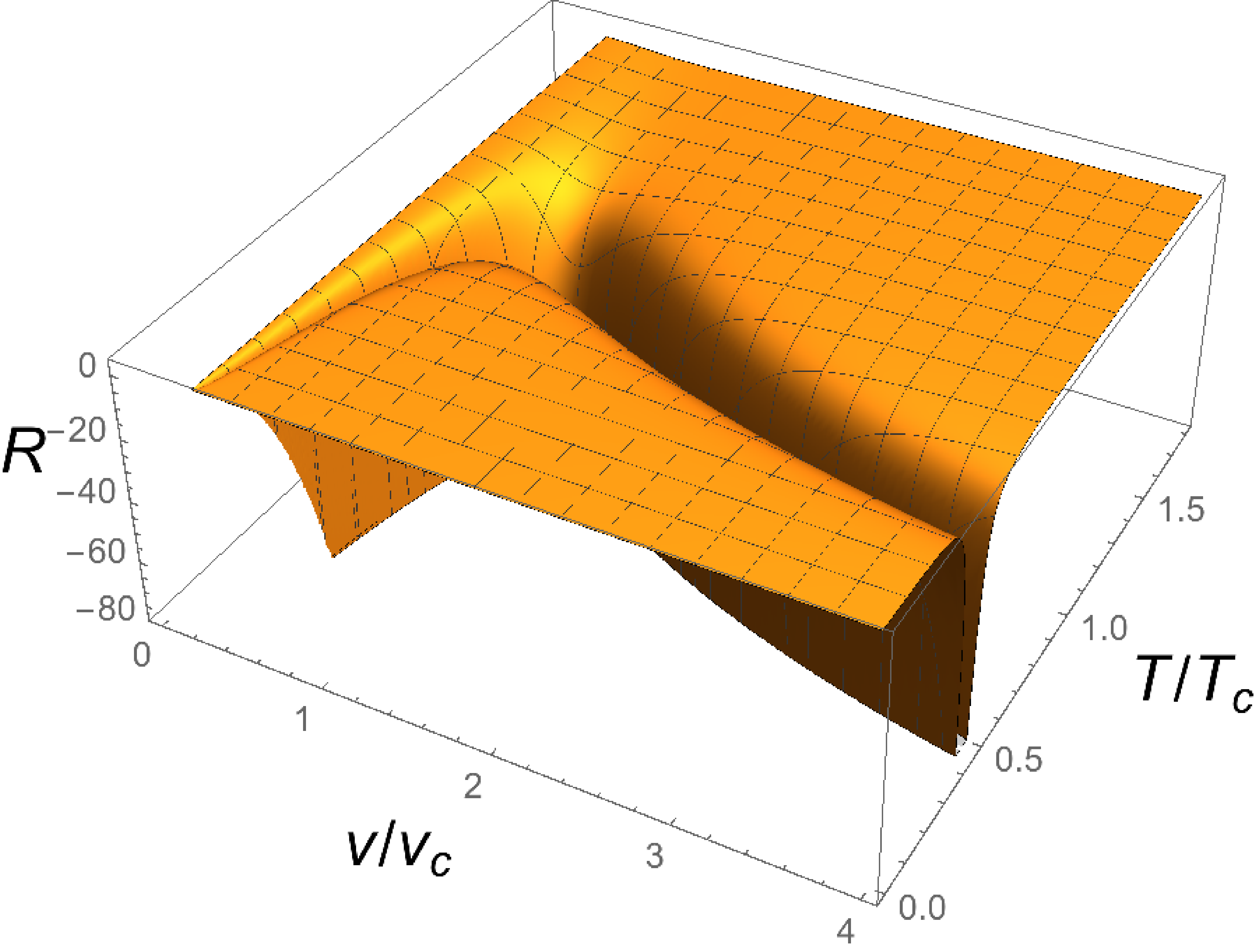}
\end{center}
\caption{Behavior of the scalar curvature as a function of $\tilde{v}$ and $\tilde{T}$.}\label{pRcruv}
\end{figure*}
%%%%%%%%%%%%%%%

%%%%%%%%%%%%%%%
\begin{figure*}
\begin{center}
\subfigure[]{\label{Rva}
\includegraphics[width=7cm]{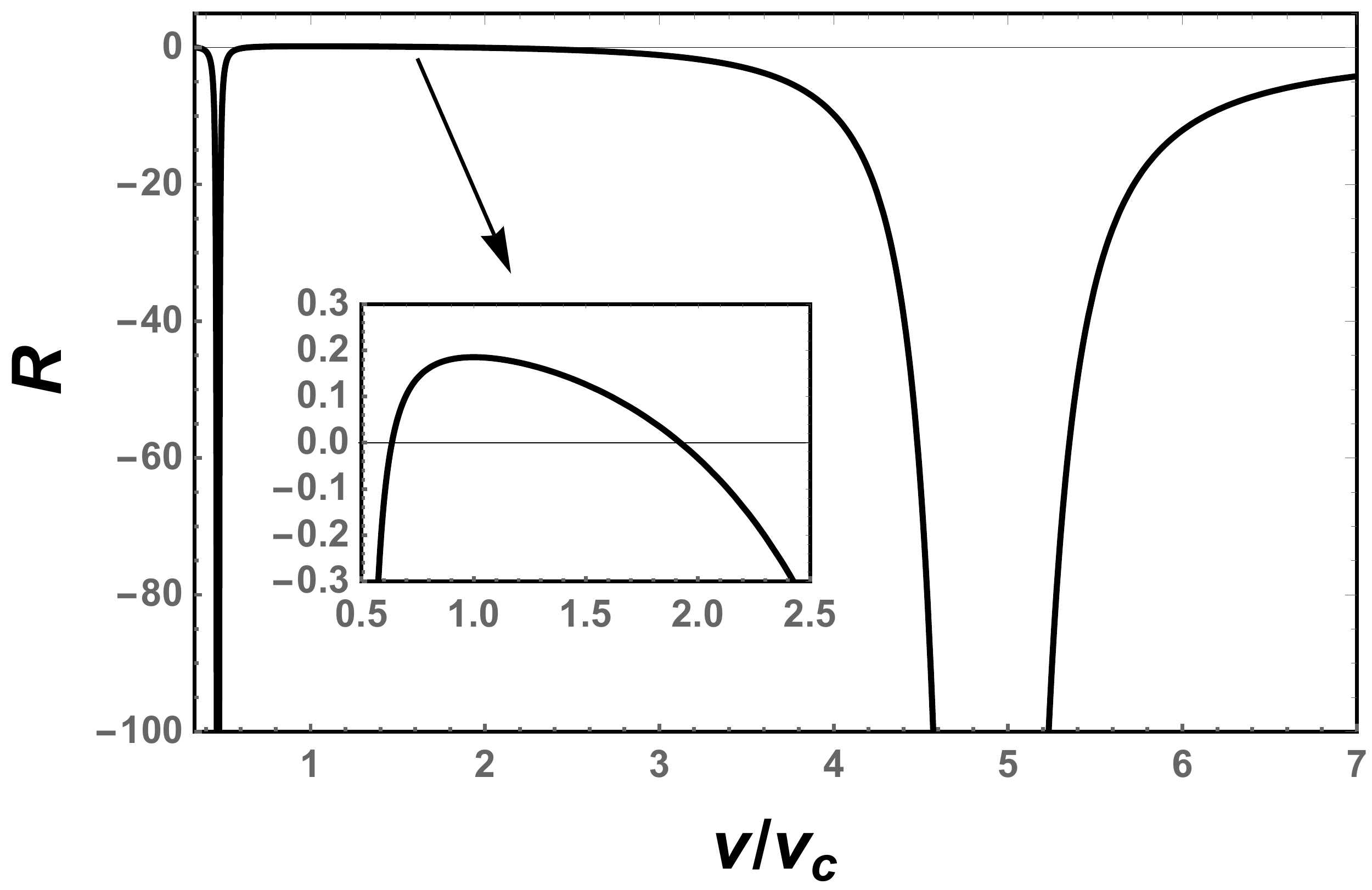}}
\subfigure[]{\label{Rvab}
\includegraphics[width=7cm]{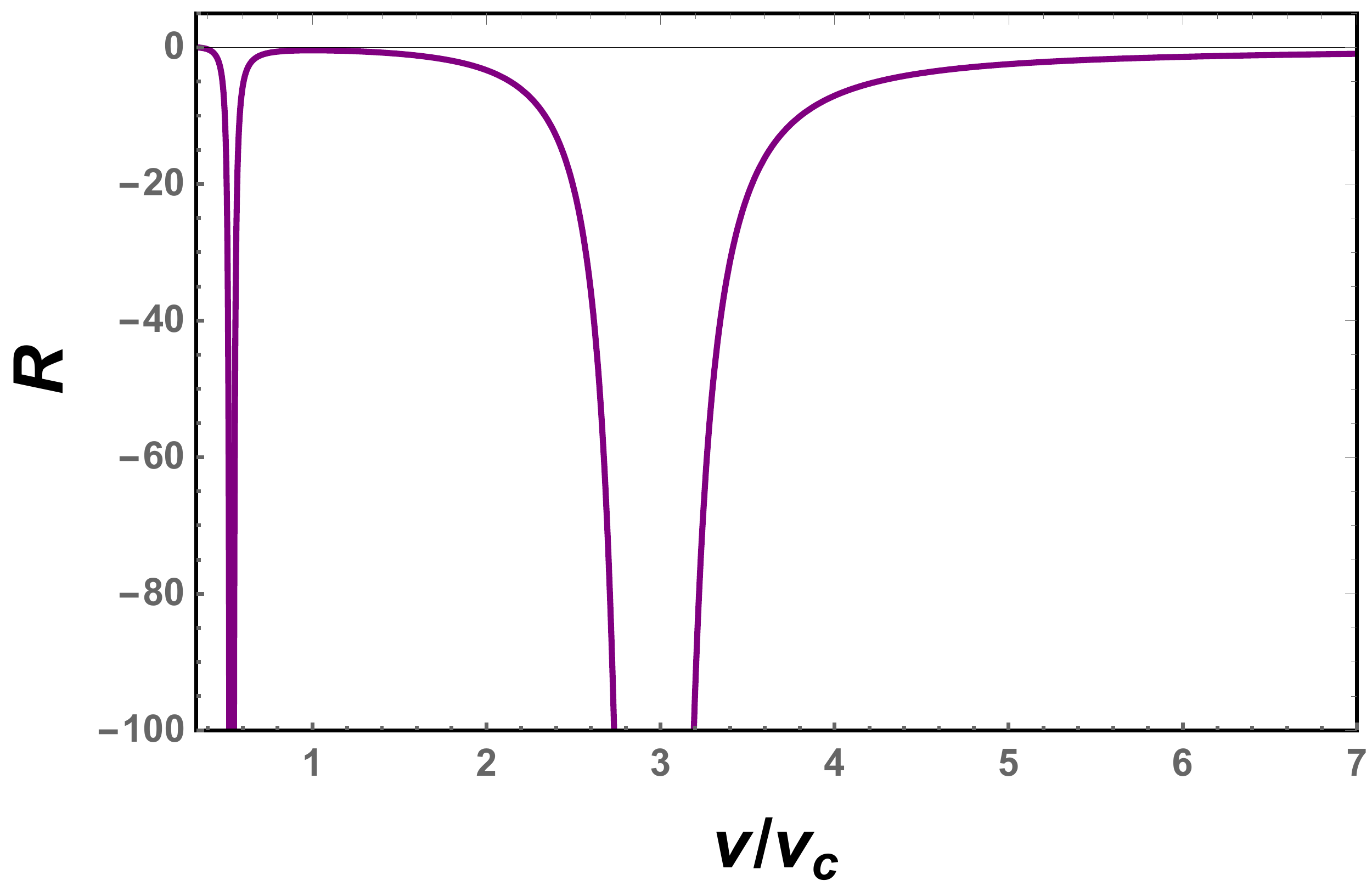}}\\
\subfigure[]{\label{Rvac}
\includegraphics[width=7cm]{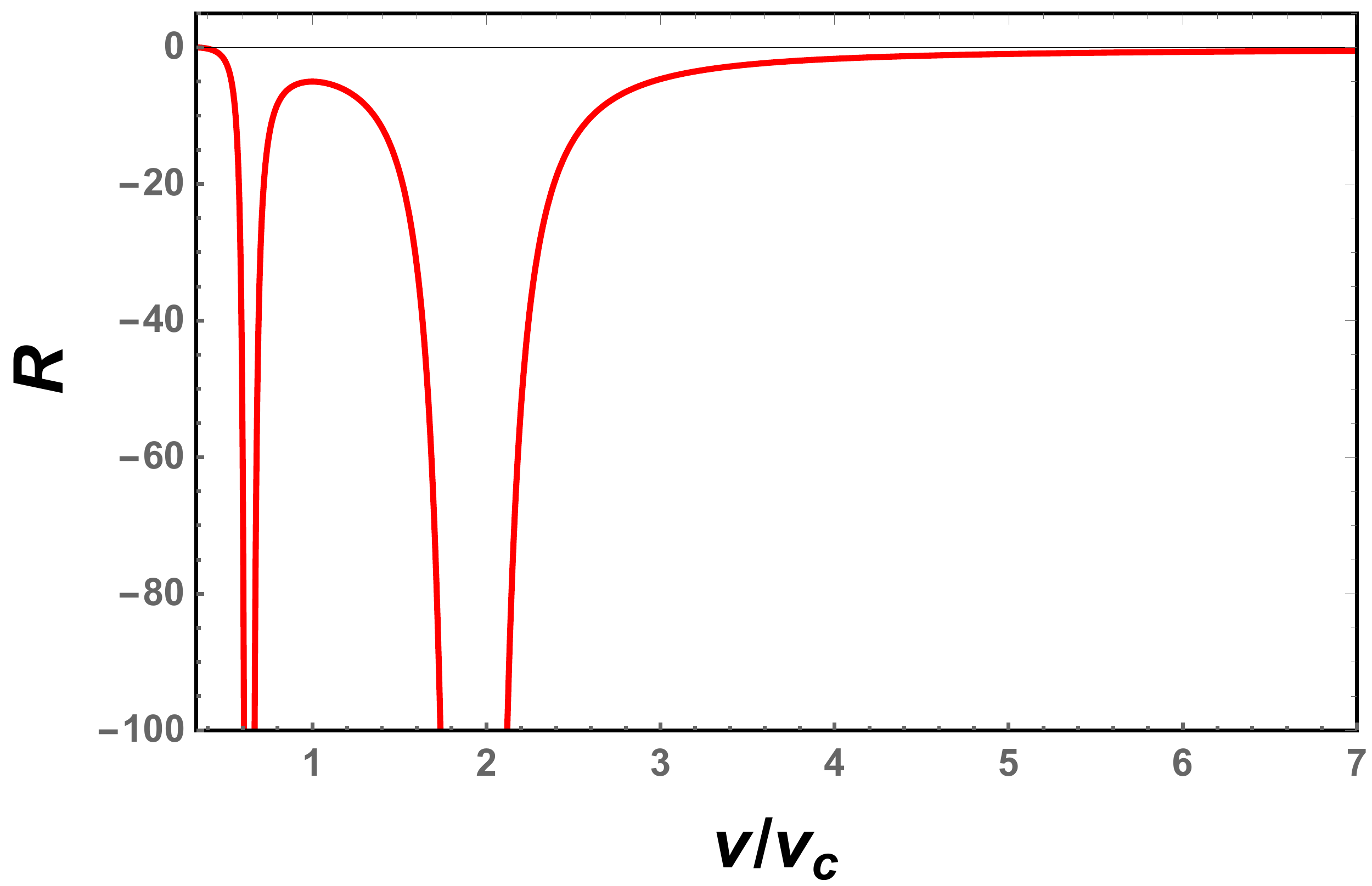}}
\subfigure[]{\label{Rvad}
\includegraphics[width=7cm]{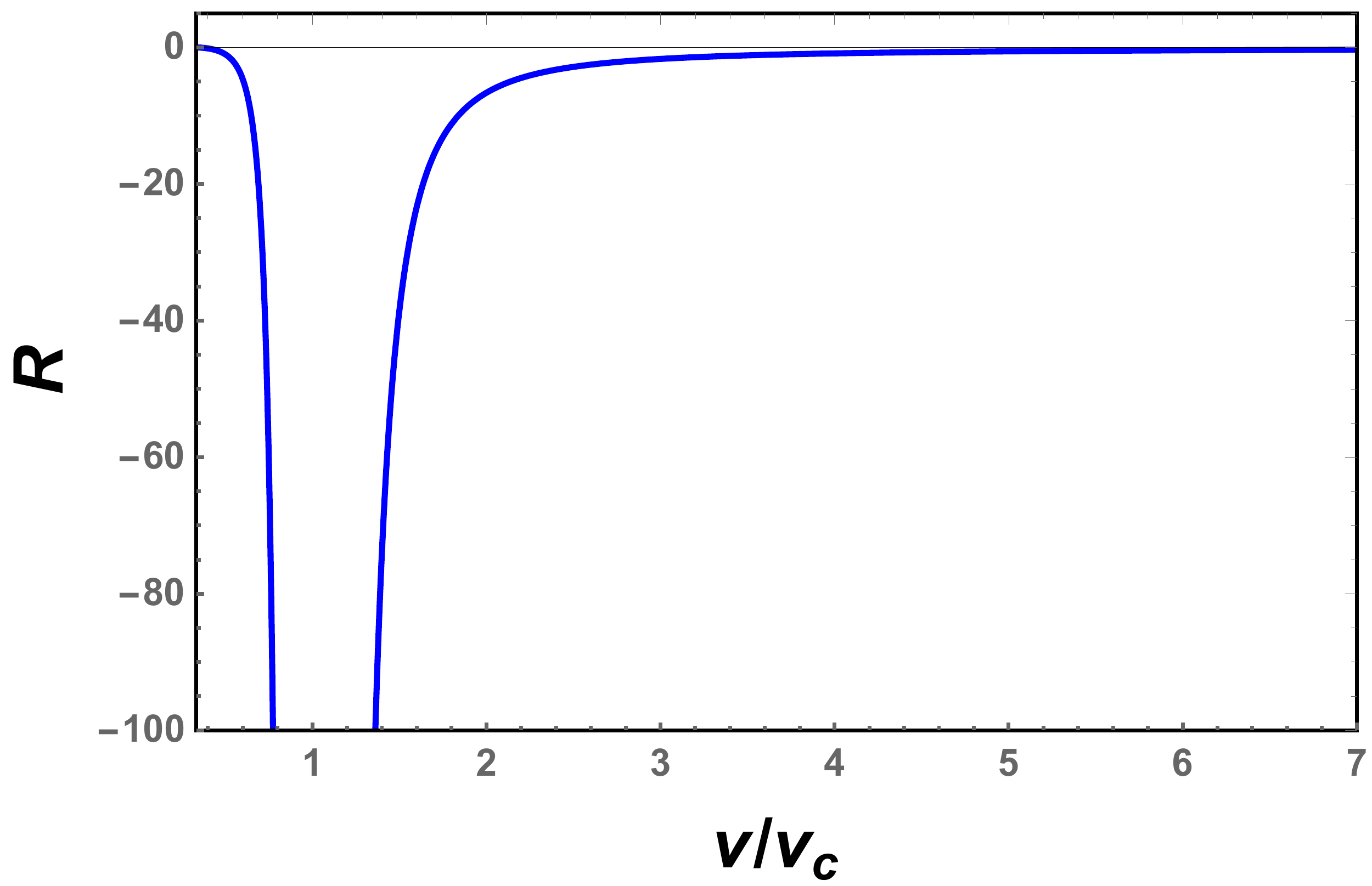}}\
\end{center}
\caption{$R$ vs. $\tilde{v}$ at constant temperature. (a) $\tilde{T}$=0.4. (b) $\tilde{T}$=0.6. (c) $\tilde{T}$=0.8. (d) $\tilde{T}$=1.0.}\label{pRva}
\end{figure*}
%%%%%%%%%%%%%%%

We plot the scalar curvature $R$ in Fig. \ref{pRcruv} as {a function of} $\tilde{v}$ and $\tilde{T}$. For low reduced temperature $\tilde{T}<1$, there are two divergent points, whereas for higher temperature $\tilde{T}$, no divergent point exists. As discussed above, the scalar curvature diverges at the spinodal curve given in (\ref{spcurvevdw}), and vanishes and changes sign at
\begin{equation}
 T_{0}=\frac{\tilde{T}_{\rm sp}}{2}=\frac{(3\tilde{v}-1)^{2}}{8\tilde{v}^{3}}.\label{to}
\end{equation}
For the sake of clarity, we show $R$ as a function of $\tilde{v}$ for fixed temperature $\tilde{T}$=0.4, 0.6, 0.8, and 1.0 in Fig. \ref{pRva}. From it, we can see that for $\tilde{T}<1$, there are two divergent points, where the scalar curvature goes to negative infinity. With an increase in temperature these two divergent points get close, and coincide at $\tilde{T}=1$. Another novel behaviour is shown in Fig. \ref{Rva}, where the scalar curvature is positive near $\tilde{v}=1$, implying a repulsive interaction between the microscopic molecules. When $\tilde{T}>1$, the divergent point disappears. However a negative well is present at $\tilde{v}=1$.

\begin{figure*}
\begin{center}
\includegraphics[width=9cm]{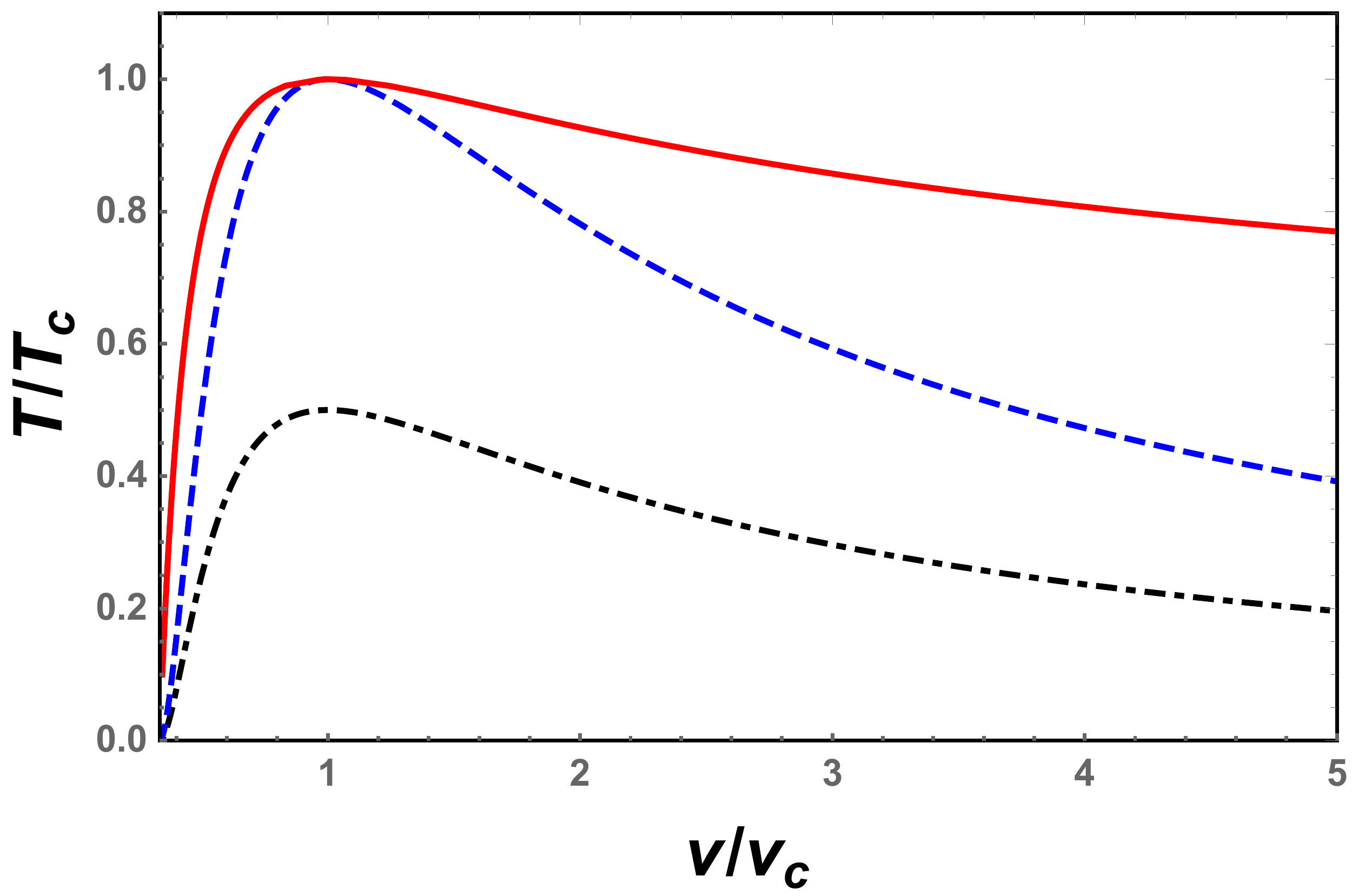}
\end{center}
\caption{The coexistence curve (red solid line), spinodal curve (blue dashed line), and the sign-changing curve where the scalar curvature $R$ changes its sign (black dot dashed line). The sign-changing curve has a maximum value 1/2 at $\tilde{v}=1$.}\label{pRnegpos}
\end{figure*}

Moreover, ignoring the metastable superheated liquid phase and supercooled gas phase, the equation of state is invalid in the coexistence phase. After considering this, we find that the divergent and positive behaviours disappear. This result can also be found in Fig. \ref{pRnegpos}, where the coexistence curve, spinodal curve,  and sign-changing curve of $R$ are described by the red solid line, blue dashed line,  and black dot-dashed line, respectively. Below the sign-changing curve, $R$ is positive, while above it, $R$ is negative. Since the positive value region of $R$ and the spinodal curve are in the coexistence phase, the positive and divergent behaviours will be excluded due to the fact that the equation of state is invalid in the coexistence phase. So there is only an attractive interaction between the microscopic molecules of the VdW fluid in the liquid and gas phases.

%%%%%%%%%%%%%%%
%%%%%%%%%%%%%%%

\subsection{Coexistence curves and critical phenomena}
\label{VdWcccp}

In this subsection, we consider the behaviour of the scalar curvature $R$ along the coexistence curve, and its critical phenomena at the critical point.

As shown in Fig. \ref{VdWPTP}, the coexistence curve in $\tilde{P}$-$\tilde{T}$ diagram is a monotonically increasing curve, and it ends at the critical point. However in the $\tilde{T}$-$\tilde{v}$ diagram, see Fig. \ref{VdWPTPb}, the coexistence curve is not monotonically increasing but instead reaches a maximum at the critical point (1, 1).  On either side of this point the coexistence curve is monotonic.  We call these respective curves the coexistence saturated liquid curve for $\tilde{v}\in (0, 1)$ and the coexistence saturated gas curve for $\tilde{v}\in (1, \infty)$.

\begin{figure*}
\begin{center}
\includegraphics[width=9cm]{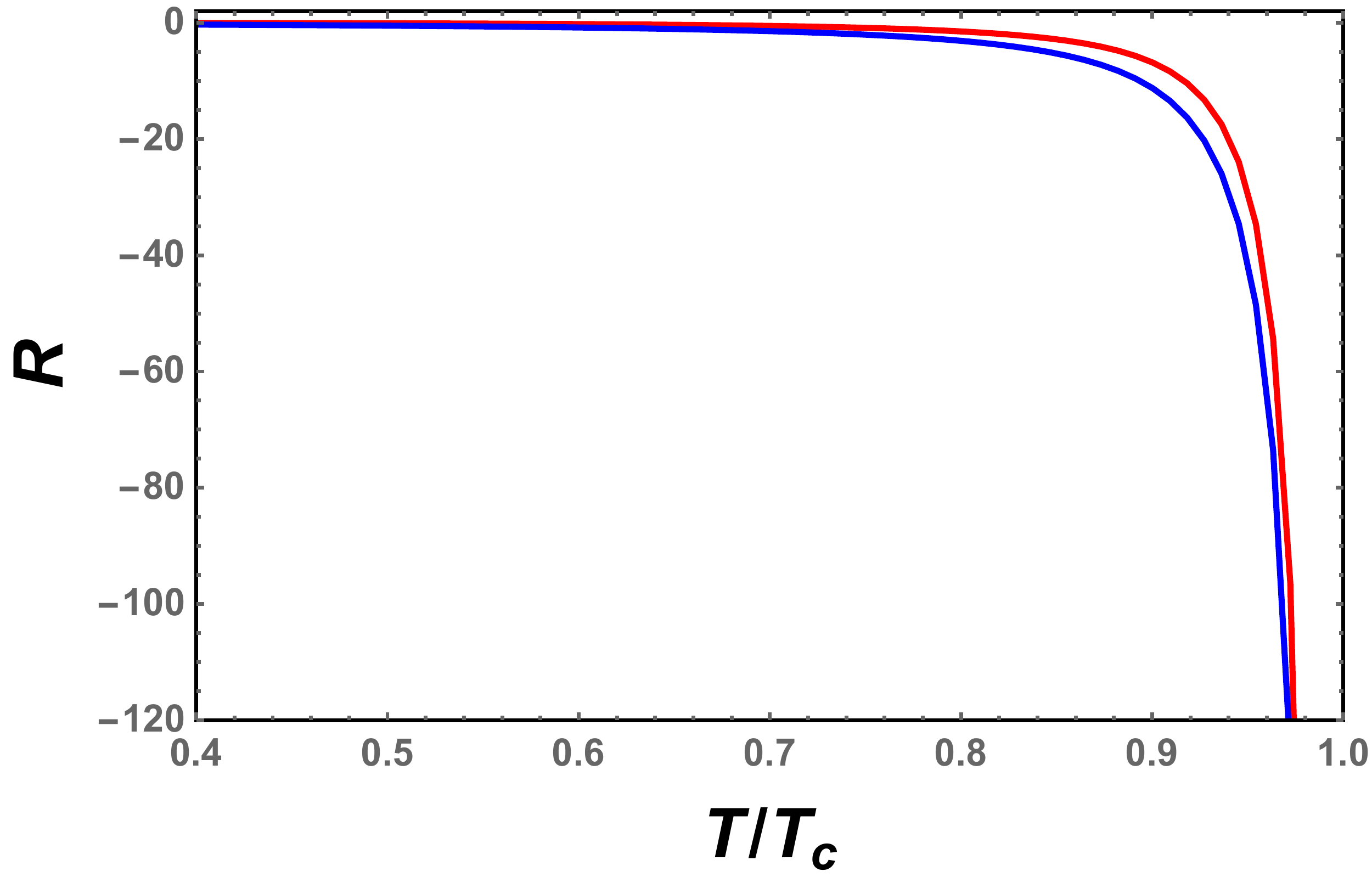}
\end{center}
\caption{The behaviour of the scalar curvature along the coexistence gas curve (top red line) and liquid curve (bottom blue line).}\label{pCoexis}
\end{figure*}

Along the coexistence gas and liquid curve, the scalar curvature is depicted in Fig. \ref{pCoexis}. The scalar curvature is always negative and decreases with the temperature. At low temperatures, $R\rightarrow 0^{-}$. At the critical point, $R$ goes to negative infinity, which is consistent with the fact that the correlation length diverges. Moreover, for fixed phase transition temperature, the absolute value of the scalar curvature of the coexistence liquid curve is larger than that of the coexistence gas curve. If we conjecture that the absolute value of $R$ is proportional to the intensity of the interaction, we can arrive the conclusion that the attractive interaction of the liquid is stronger than the gas one.

%%%%%%%%%%%%%%%
%%%%%%%%%%%%%%%

\begin{figure*}
\begin{center}
\subfigure[]{\label{Liquidp}
\includegraphics[width=7cm]{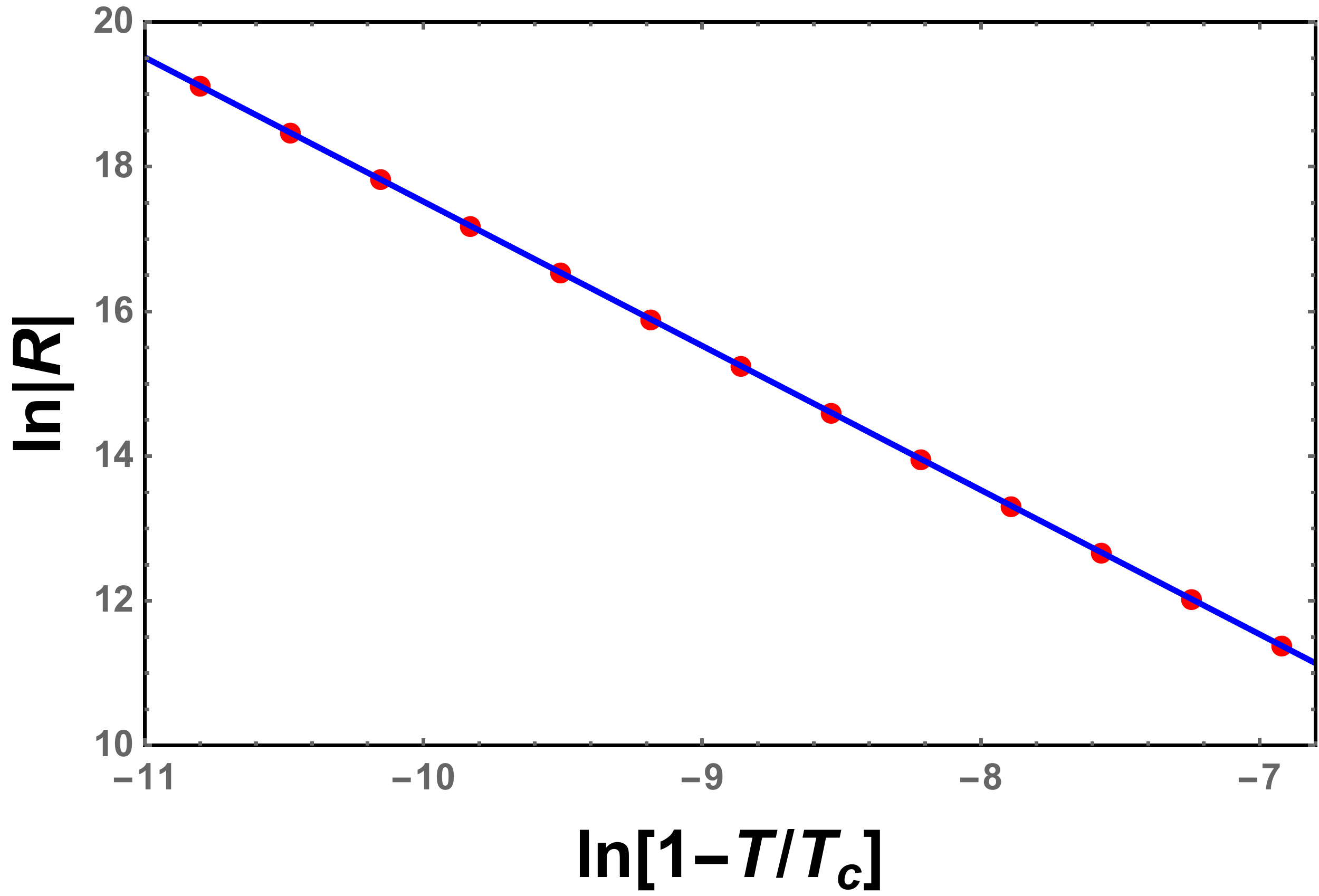}}
\subfigure[]{\label{gas}
\includegraphics[width=7cm]{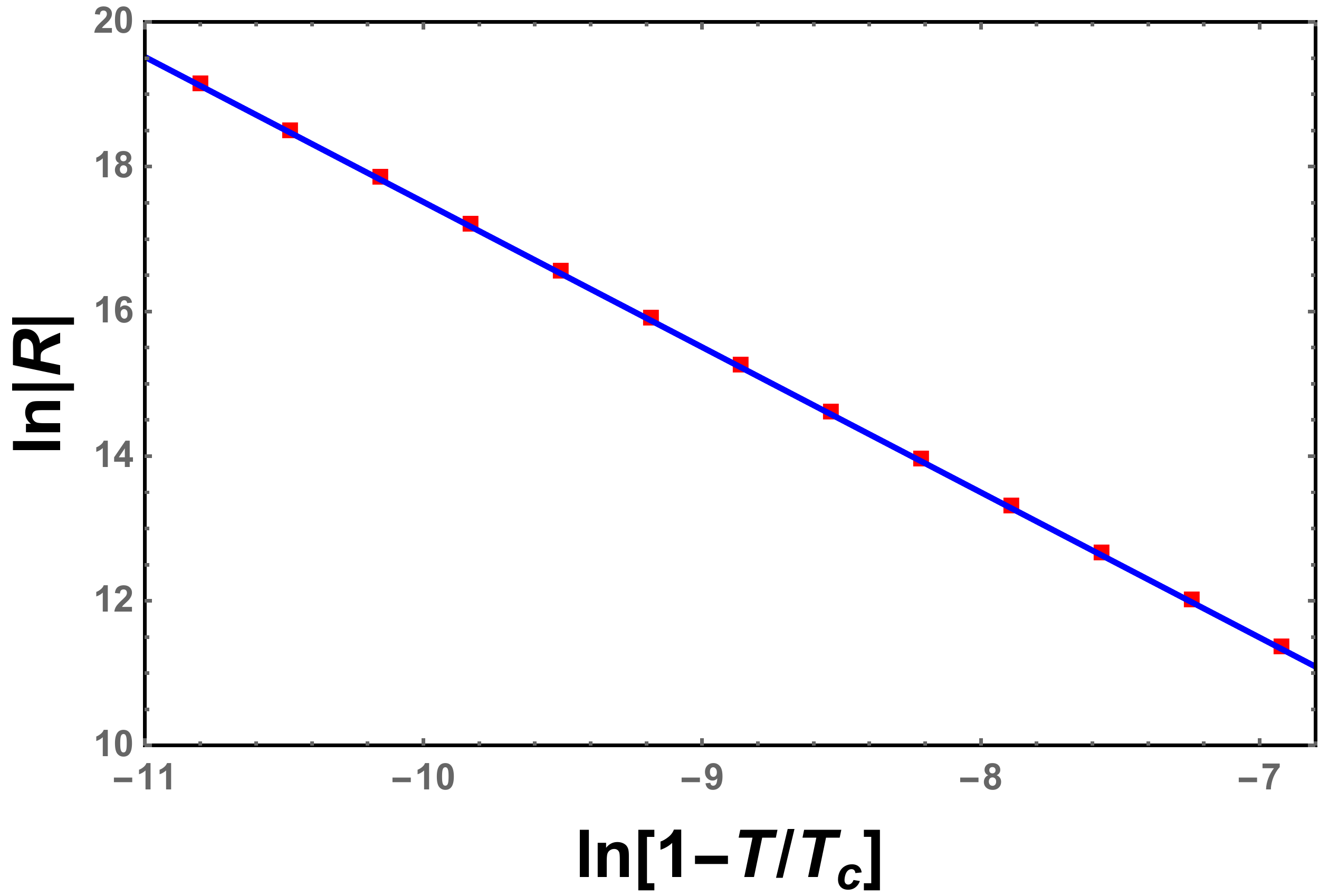}}
\end{center}
\caption{The scalar curvature $\ln |R|$ vs. $\ln(1-\tilde{T})$.  The numerical data are described by the red markers and the fitting formulas are described by blue solid lines. (a) The coexistence saturated liquid curve -- the slope is -1.99323. (b) The coexistence saturated gas curve -- the slope is -2.00628.}\label{pgas}
\end{figure*}

Next, we would like to examine the critical exponent of $R$ near the critical point, which can give us some universal properties of the fluid systems.
Although the coexistence curve of the VdW fluid has no analytical form, we can numerically fit the formula by
assuming that the scalar curvature $R$ has the following form
\begin{equation}
 R\sim-(1-\tilde{T})^{-\alpha},
\end{equation}
or, equivalently,
\begin{equation}
 \ln|R|=-\alpha\ln(1-\tilde{T})+\beta.\label{423}
\end{equation}
near the critical point.
Along the coexistence saturated liquid and gas curves, we get the following fitting results
\begin{eqnarray}
 \ln|R|&=&-1.99323\ln(1-\tilde{T})-2.41428, \quad \text{coexistence saturated liquid curve},\label{lic1}\\
 \ln|R|&=&-2.00628\ln(1-\tilde{T})-2.55089, \quad \text{coexistence saturated gas curve}, \label{lic2}
\end{eqnarray}
which we illustrate in  Fig. \ref{pgas}. The numerical data and fitting formulas are described by the red markers and the blue solid lines, respectively; clearly they are highly consistent.  Taking numerical
error into account, the slopes indicate that the critical exponent must be $\alpha=2$, which is consistent with Refs. \cite{tpo1,tpo2}.

Moreover, from (\ref{lic1}) and (\ref{lic2}) we can construct
\begin{equation}
 R(1-\tilde{T})^{2}C_{v}=-\frac{3}{2}e^{-\frac{2.41428 + 2.55089}{2}}=-0.125291\approx-\frac{1}{8} \label{oneeight}
\end{equation}
which is independent of the parameters $a$ and $b$ and is therefore  a universal result. More interestingly, we will show in the next section that the constant  $1/8$ exactly agrees with the analytic one from a charged AdS black hole.

%%%%%%%%%%%%%%%
%%%%%%%%%%%%%%%

\section{Charged AdS black holes}
\label{fff}

In this section, we investigate the thermodynamics and Ruppeiner geometry for a charged AdS black hole in  extended phase space, where the cosmological constant (or  AdS radius) is
identified with the thermodynamic pressure
\begin{equation}
 P=\frac{3}{8\pi l^{2}}.\label{ppl}
\end{equation}
There are a number of reasons for considering this \cite{Teo}. In a `more fundamental' theory, the cosmological constant is not fixed a priori, but arises as a vacuum expectation value of the stress energy, and thus its value can vary.  Scaling relations imply that the Smarr
formula relating the various thermodynamic parameters of a black hole  only holds when the variation of the cosmological constant is included  \cite{Kastor}. Furthermore, the precise analogy between a  black hole  and a VdW fluid becomes complete \cite{Kubiznak}.

The conjugate quantity of the pressure is the thermodynamic volume, yielding a $PdV$ term in the first law of  black hole thermodynamics.  A phase transition between small and large black hole phases results and is found to be the similar to the liquid-gas phase transition of the VdW fluid.
It is in this context that  we examine the behaviour of the scalar curvature for the charged AdS black hole.

\subsection{Thermodynamics and phase structures}

Let us start with a brief review of the thermodynamics and phase structure for the charged AdS black hole. More importantly, we shall clearly show the spinodal curve, which is ignored in previous works.

The bulk action  describing a charged AdS black hole is
\begin{equation}
 S=\frac{1}{16\pi}\int\sqrt{-g}dx^{4}\left(\mathcal{R} -F^{\mu\nu}F_{\mu\nu}+\frac{6}{l^{2}}\right),
\end{equation}
and the solution to its equations of motion is
\begin{eqnarray}
 ds^{2} &=&-f(r)dt^{2}+f^{-1}(r)dr^{2}+r^{2}d\Omega^{2}_{2},\\
 F&=& dA,~~~~\quad A_{\mu}=\left(-\frac{Q}{r},0,0,0\right),
\end{eqnarray}
where $d\Omega^{2}_{2}$ is the standard element on $S^{2}$. The metric function $f(r)$ is given by
\begin{equation}
 f(r)=1-\frac{2M}{r}+\frac{Q^{2}}{r^{2}}+\frac{r^{2}}{l^{2}}.
\end{equation}
The parameters $M$ and $Q$ are the black hole mass and charge. The temperature can be calculated with the `Euclidean trick'
and the entropy obeys the Bekenstein-Hawking area entropy relation,
\begin{eqnarray}
 T&=&\frac{1}{4\pi r_{\rm h}}\left(1+\frac{3r_{\rm h}^{2}}{l^{2}}-\frac{Q^{2}}{r_{\rm h}^{2}}\right),\label{ct}\\
 S&=&\frac{A}{4}=\pi r_{\rm h}^{2},
\end{eqnarray}
where $r_{\rm h}$ is the radius of the event horizon, and it satisfies $f(r_{\rm h})=0$. Solving the equation, the black hole mass can be expressed as
\begin{equation}
 M=\frac{r_{\rm h}}{2}+\frac{Q^{2}}{2r_{\rm h}}+\frac{r_{\rm h}^{3}}{2l^{2}}.
\end{equation}
The electric potential on the horizon measured at infinity is $\Phi=\frac{Q}{r_{\rm h}}$. Rewriting  the AdS radius $l$ in terms of the pressure $P$ using (\ref{ppl}), the first law and Smarr formula are
\begin{eqnarray}
 dM&=&TdS+\Phi dQ+VdP,\label{fflaw}\\
 M&=&2TS+Q\Phi-2PV.
\end{eqnarray}
According to the first law, the thermodynamic volume is
\begin{equation}
 V=\left(\frac{\partial M}{\partial P}\right)_{Q,S}=\frac{4}{3}\pi r_{\rm h}^{3}.
\end{equation}
Note that although the thermodynamic volume and the geometric volume share the same expression, they have different meanings. Another thing worth pointing out is that, from (\ref{fflaw}), we can clearly see that the black hole mass $M$ should be treated as the enthalpy, i.e., $M\equiv H$ rather than the internal energy $E$ of the system. The internal energy $E$ reads
\begin{equation}
 E=H-PV=\frac{r_{\rm h}}{2}+\frac{Q^{2}}{2r_{\rm h}}.
\end{equation}
Obviously, the energy is independent of the pressure $P$.

Since the heat capacity has important impact on the local thermodynamic stability, we would like to examine its behaviour before pursuing the phase structure. The heat capacity  at constant pressure is
\begin{equation}
 C_{P}=T\left(\frac{\partial S}{\partial T}\right)_{P}=\frac{2S(8PS^{2}+S-\pi Q^{2})}{8PS-S+3\pi Q^{2}}
\end{equation}
where local stability requires $C_{P}>0$. The heat capacity   at constant volume is
\begin{equation}
 C_{V}=T\left(\frac{\partial S}{\partial T}\right)_{V}
\end{equation}
where constant volume means $dr_{\rm h}=0$ since $V\sim r_{\rm h}^{3}$.  Furthermore, since $S=\pi r_{\rm h}^{2}$ we have $dS=0$. Hence
\begin{equation}
 C_{V}=0,\label{cc0}
\end{equation}
which (as we shall see) will be somewhat problematic for the charged AdS black hole.

Turning next to  phase behaviour, from (\ref{ct}) we obtain  the equation of state  \cite{Kubiznak}
\begin{equation}
 P=\frac{T}{v}-\frac{1}{2\pi v^{2}}+\frac{2Q^{2}}{\pi v^{4}}
\end{equation}
where $v=2r_{\rm h}$ denotes the specific volume. From (\ref{cc0}), we also have the specific heat capacity
\begin{equation}
 C_{v}=0.\label{cc1}
\end{equation}
Comparing with that of the VdW fluid, $C_{v}^{\text{VdW}}=\frac{3}{2}k_{\rm B}$ ($k_{\rm B}\approx 1.38\times 10^{-23} J/K$), the specific heat capacity of the charged AdS black hole can be treated as a limit of the VdW fluid, i.e.,
\begin{equation}
 k_{\rm B}\rightarrow 0^{+}.
\end{equation}
We shall treat these problematic quantities by normalizing them with $C_{v}$ or $C_{V}$.

For the charged AdS black hole there exists a stable small-large black hole phase transition, with a critical point determined via $(\partial_{v}P)_{T}=(\partial_{v,v}P)_{T}=0$,  which gives \cite{Kubiznak}
\begin{equation}
 P_{\rm c}=\frac{1}{96\pi Q^{2}},\quad T_{\rm c}=\frac{\sqrt{6}}{18\pi Q},\quad V_{\rm c}=8\sqrt{6}\pi Q^{3},\quad v_{\rm c}=2\sqrt{6}Q.
\end{equation}
In the reduced parameter space, the equation of state reads
\begin{equation}
 \tilde{P}=\frac{8 \tilde{T}}{3\tilde{v}}-\frac{2}{\tilde{v}^{2}}+\frac{1}{3\tilde{v}^{4}}.\label{crstate}
\end{equation}
This reduced equation of state is similar to that of the VdW fluid \eqref{VdWeos} insofar as it depends only on $(\tilde{P},\tilde{T},\tilde{v})$ because the charge $Q$ scales out. Furthermore the reduced temperature and Gibbs free energy in terms of $\tilde{P}$ and $\tilde{V}$ are \cite{Wei5}
\begin{eqnarray}
 \tilde{T}&=&\frac{3 \tilde{P} \tilde{V}^{4/3}+6
   \tilde{V}^{2/3}-1}{8 \tilde{V}},\\
 \tilde{G}&=&\frac{1}{8} \left(-\tilde{P}\tilde{V}+6
   \sqrt[3]{\tilde{V}}+\frac{3}{\sqrt[3]{\tilde{V}}}\right),
\end{eqnarray}
where $\tilde{G}=G/G_{\rm c}$ with $G_{\rm c}=\frac{\sqrt{6}}{3}Q$. In a $\tilde{G}$-$\tilde{T}$ diagram there will be  swallowtail behaviour when $\tilde{P}<1$, which signifies a phase transition. Alternatively one can determine the phase transition point by constructing the equal area law on the isothermal curve  in the $P$-$V$ diagram. However, one must be very careful in constructing the equal area law because it does not hold in a $P$-$v$ diagram \cite{Wei5}.

\begin{figure*}
\begin{center}
\subfigure[]{\label{Pvp}
\includegraphics[width=7cm]{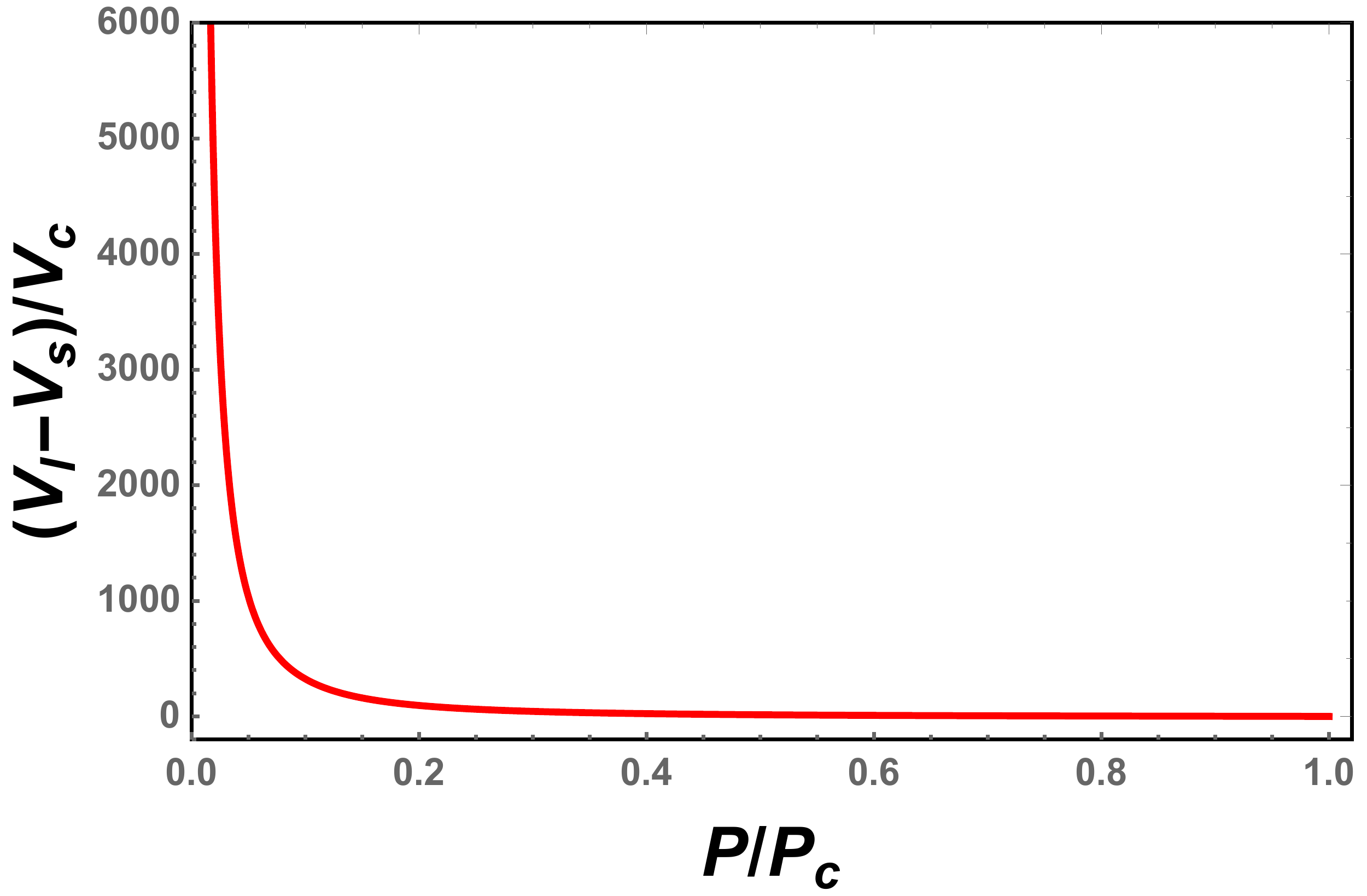}}
\subfigure[]{\label{Pvt}
\includegraphics[width=7cm]{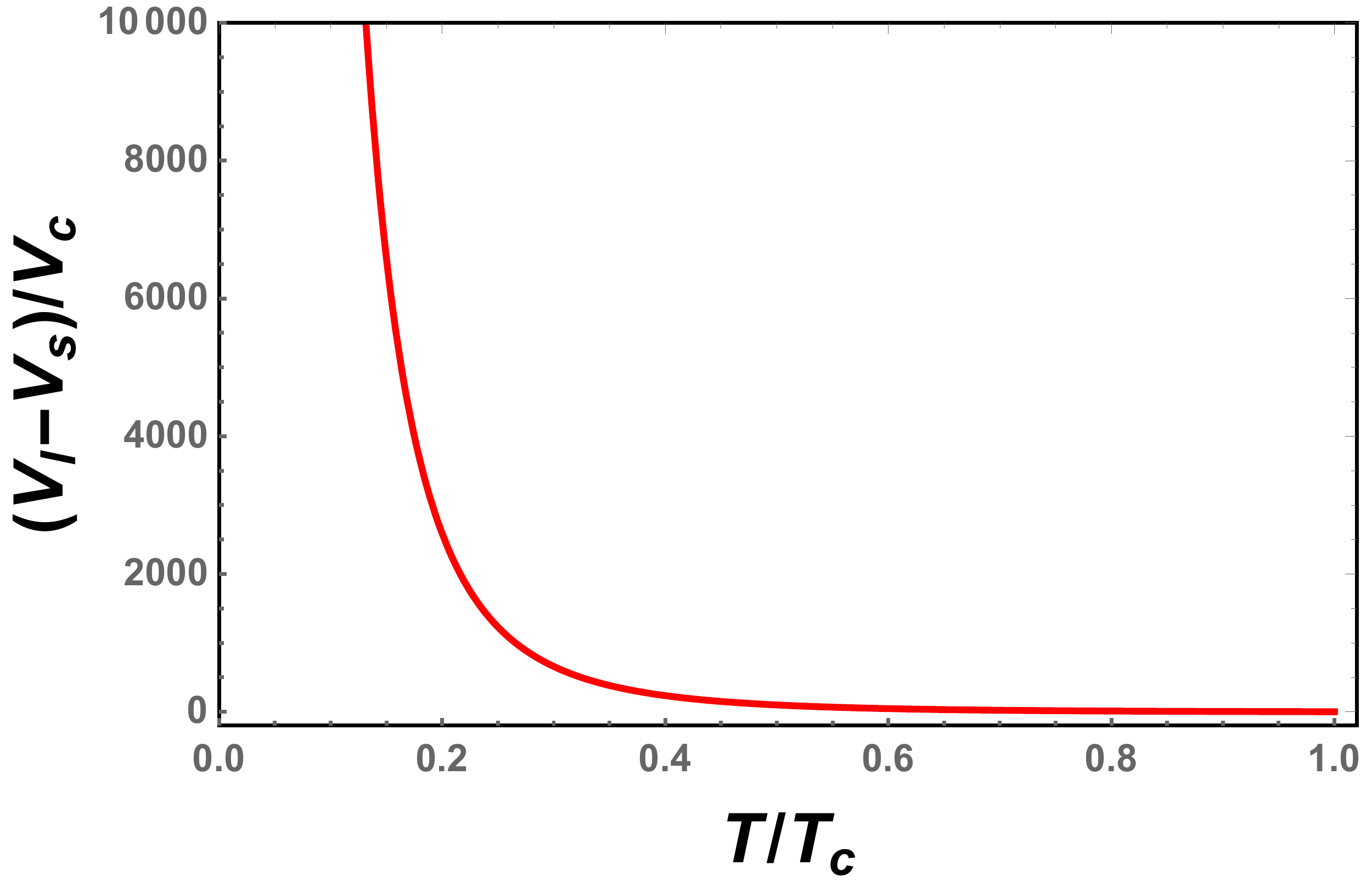}}
\end{center}
\caption{The behaviour of the change of the volume at the black hole phase transition. (a) $\Delta\tilde{V}$ vs. $\tilde{P}$. (b) $\Delta\tilde{V}$ vs. $\tilde{T}$. The critical point is at $\tilde{P}$=1 and $\tilde{T}$=1.}\label{pPvt}
\end{figure*}

It is amazing that there exists an analytic form of the coexistence curve of small and large black hole phases for the charged AdS black hole system. The first formula was given  by constructing the equal area law on an isobaric curve  in the $T$-$S$ diagram  \cite{Spallucci}
\begin{eqnarray}
 \tilde{T}^{2}=\tilde{P}(3-\sqrt{\tilde{P}})/2,\label{tt}
\end{eqnarray}
or, equivalently
\begin{eqnarray}
 \sqrt{\tilde{P}}=1-2\cos\left(\frac{\arccos(1-\tilde{T}^{2})+\pi}{3}\right).\label{cppp}
\end{eqnarray}
Employing this formula we can obtain the reduced thermodynamic volumes for small and large black holes along the coexistence curve
\begin{eqnarray}
 \tilde{V}_{\rm s}&=&\left(\frac{\sqrt{3-\sqrt{\tilde{P}}}-\sqrt{3-3\sqrt{\tilde{P}}}}{\sqrt{2\tilde{P}}}\right)^{3},\label{cvs}\\
 \tilde{V}_{\rm l}&=&\left(\frac{\sqrt{3-\sqrt{\tilde{P}}}+\sqrt{3-3\sqrt{\tilde{P}}}}{\sqrt{2\tilde{P}}}\right)^{3}.\label{cvl}
\end{eqnarray}
From this result, we can obtain the following compact relations
\begin{eqnarray}
 \tilde{V}^{\frac{1}{3}}_{\rm s}+\tilde{V}^{\frac{1}{3}}_{\rm l} &=&\sqrt{\frac{6}{\sqrt{\tilde{P}}}-2},\\
 \tilde{V}_{\rm s}\tilde{V}_{\rm l} &=& \frac{1}{\sqrt{\tilde{P}^{3}}}.
\end{eqnarray}
It is also interesting to investigate the change $\Delta\tilde{V}=\tilde{V}_{\rm l}-\tilde{V}_{\rm s}$ of the volume at the phase transition. Employing (\ref{cvs}), (\ref{cvl}) and combining with (\ref{cppp}), we easily obtain $\Delta\tilde{V}$, and illustrate the result  as a function of the respective reduced phase transition pressure and temperature in Fig. \ref{pPvt}. From the figures it is quite obvious that an increase of $\tilde{P}$ or $\tilde{T}$ yields a monotonic decrease in $\Delta\tilde{V}$, and vanishes when the critical point is approached.  Examining the critical exponent of $\Delta\tilde{V}$ near the critical point we obtain
\begin{eqnarray}
 \Delta\tilde{V}&=&3\sqrt{3}(1-\tilde{P})^{\frac{1}{2}}+\frac{51\sqrt{3}}{8}(1-\tilde{P})^{\frac{3}{2}}+\mathcal{O}(1-\tilde{P})^{\frac{5}{2}},\\
 \Delta\tilde{V}&=&6\sqrt{2}(1-\tilde{T})^{\frac{1}{2}}+\frac{61\sqrt{2}}{2}(1-\tilde{T})^{\frac{3}{2}}+\mathcal{O}(1-\tilde{T})^{\frac{5}{2}}
\end{eqnarray}
implying that $\Delta\tilde{V}$ has an exact exponent of $\frac{1}{2}$. Furthermore, it monotonically decreases as a function of $\tilde{P}$ and of $\tilde{T}$ and vanishes at the critical point.  Taken in conjunction, these properties imply  that
$\Delta\tilde{V}$ can be treated as an order parameter to describe this small-large black hole phase transition.

%%%%%%%%%%%%%%%
%%%%%%%%%%%%%%%

Next, let us turn to the spinodal curves for the charged AdS black hole. Here we would like to carry out our study in the reduced parameter space.  Because $\tilde{V} =  \tilde{v}^3$,
the spinodal curve determined by $(\partial_{\tilde{V}}\tilde{P})=0$ is the same as that of  $(\partial_{\tilde{v}}\tilde{P})=0$. By using the reduced equation of state (\ref{crstate}), we obtain
\begin{eqnarray}
 \tilde{V}_{\rm l}&=&\left(\frac{1+2\cos\left(\frac{1}{3}\arccos(1-2\tilde{T}^{2})\right)}{2\tilde{T}}\right)^{3},\\
 \tilde{V}_{\rm s}&=&\left(\frac{1-2\cos\left(\frac{1}{3}(\pi+\arccos(1-2\tilde{T}^{2}))\right)}{2\tilde{T}}\right)^{3}
\end{eqnarray}
and insertion of these expressions into the reduced equation of state (\ref{crstate}) yields the small and large black hole spinodal curves in the $P$-$T$ diagram. Note that when $\tilde{T}$=0, $\tilde{V}_{\rm s}$ attains its smallest value of $1/3\sqrt{3}$.
Alternatively, in the $\tilde{T}$-$\tilde{V}$ diagram, the spinodal curve has a compact form
\begin{equation}
 \tilde{T}_{\rm sp}=\frac{3\tilde{V}^{\frac{2}{3}}-1}{2\tilde{V}},\label{spcurve}
\end{equation}
where $1/3\sqrt{3}<\tilde{V}<1$ is for the small black hole spinodal curve, while $\tilde{V}>1$ is for the large black hole spinodal curve.

%%%%%%%%%%%%%%%
\begin{figure*}
\begin{center}
\subfigure[]{\label{Cbhpt}
\includegraphics[width=7cm]{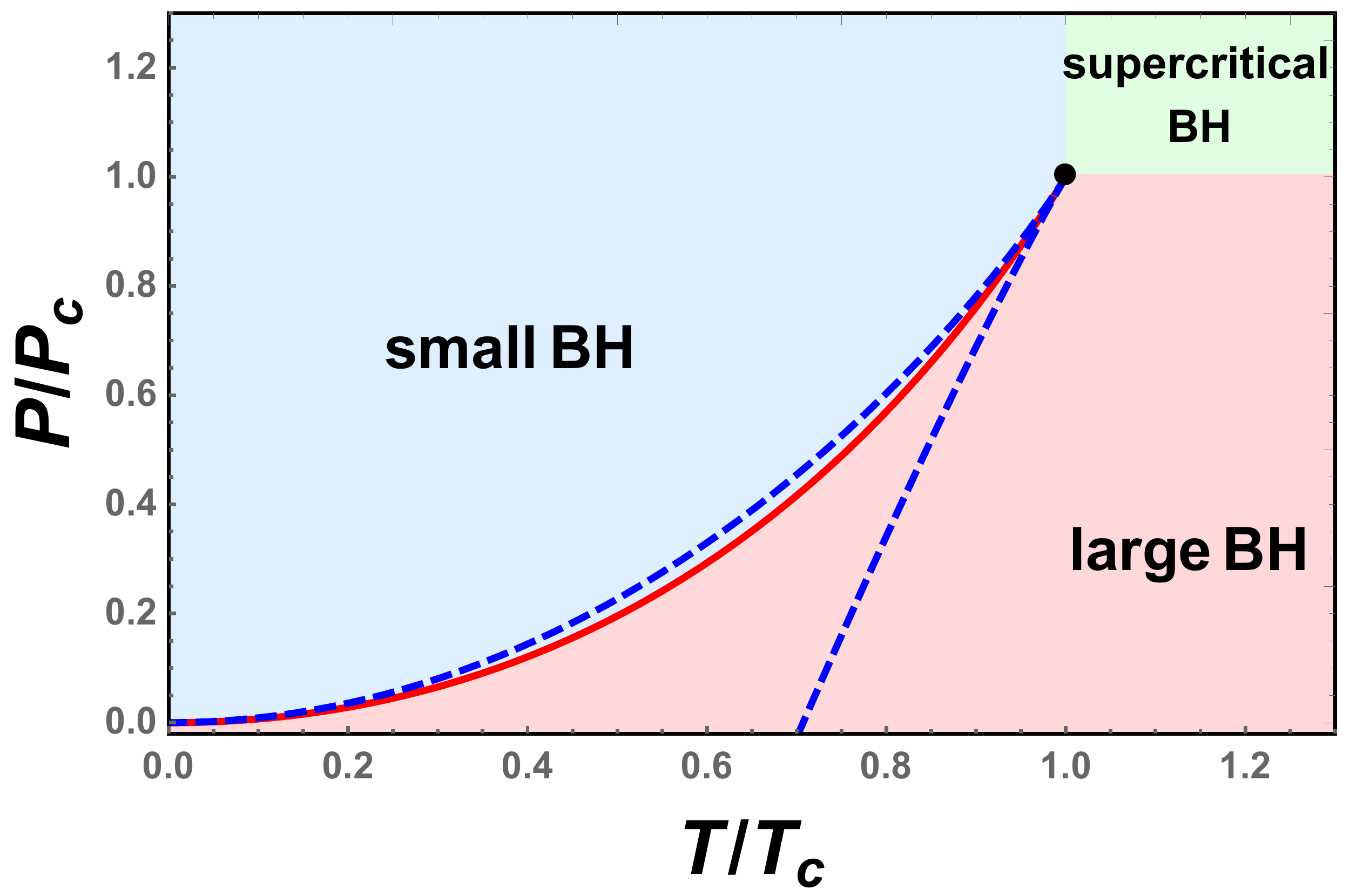}}
\subfigure[]{\label{Cbhtv}
\includegraphics[width=7cm]{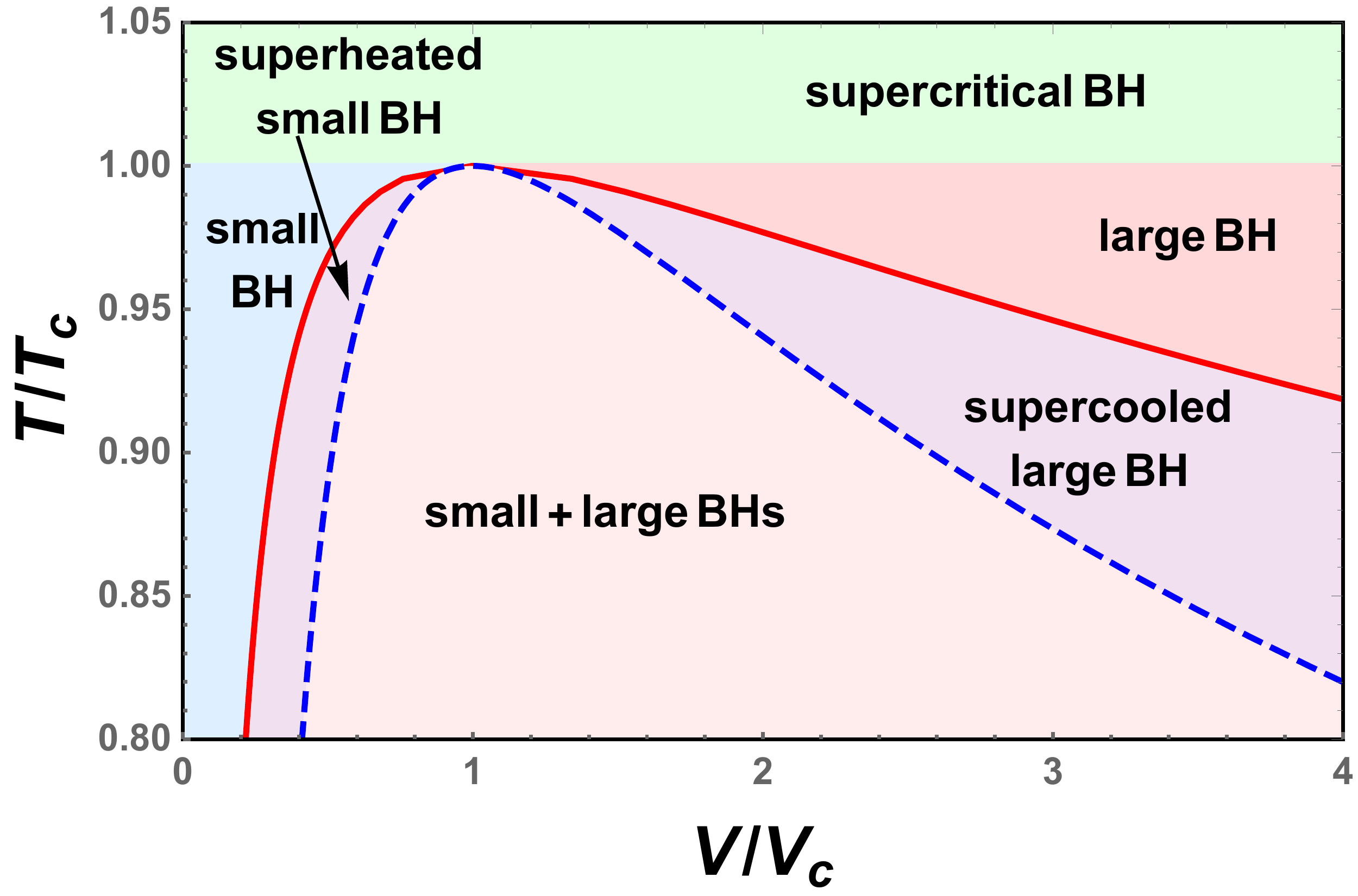}}
\end{center}
\caption{Phase diagram and spinodal curve for the charged AdS black hole. The red solid curve and blue dashed curve are the coexistence curves and spinodal curves, respectively. (a) $\tilde{P}$-$\tilde{T}$ phase diagram. Black dot denotes the critical point. The regions for the small black hole, large black hole, and supercritical black hole are displayed. The bottom blue dashed curve starts at $\tilde{T}=\sqrt{2}/2$, which is smaller than 27/32 of the VdW fluid. (b) $\tilde{T}$-$\tilde{v}$ phase diagram. The regions of the small black hole, larger black hole, coexistence black hole, supercritical black hole, metastable superheated small black hole, and metastable supercooled larger black hole are clearly displayed. One can check that the small black hole spinodal curve has a reduced volume $\tilde{V}=1/3\sqrt{3}$ when $\tilde{T}=0$, while for the large black hole spinodal curve, $\tilde{V}$ approaches infinity.}\label{pCbhpt}
\end{figure*}
%%%%%%%%%%%%%%%

In Fig. \ref{pCbhpt} we show the phase structures and spinodal curves  in the $\tilde{P}$-$\tilde{T}$ and $\tilde{T}$-$\tilde{V}$ diagrams respectively. The red solid lines denote the coexistence curves for small and large black holes, and the blue dashed lines are   the spinodal curves. In Fig. \ref{Cbhpt}, we can see that the small/large black hole phases are above/below the coexistence curve. The top right corner is  the supercritical black hole phase, where   small and large black holes cannot be distinguished. These phase diagrams are qualitatively identical to those of the VdW fluid in Fig. \ref{VdWPTP}.  Note that
the spinodal curves are always under the coexistence curve.
A small quantitative distinction is that  the spinodal curve of the small black hole starts at $\tilde{T}=\sqrt{2}/2$, which is smaller than the value 27/32 for the VdW fluid. Analogous to the VdW fluid, in the $\tilde{T}$-$\tilde{V}$ phase diagram Fig. \ref{Cbhtv}, two metastable phases, the superheated small black hole and supercooled large black hole, are clearly shown. A detailed study of the superheated small black hole and supercooled large black hole can be found in \cite{Wei6}.

\subsection{Ruppeiner geometry}
\label{rgr}

We now apply Ruppeiner geometry to help discern the microstructure of a charged AdS black hole,  taking the temperature $T$ and thermodynamic volume $V$ as the fluctuation coordinates.

Setting $x=V$ and $y=-P$, the line element of the Ruppeiner geometry (\ref{xxy}) reduces to\begin{equation}
 dl^{2}=\frac{C_{V}}{T^{2}}dT^{2}-\frac{(\partial_{V}P)_{T}}{T}dV^{2}
\end{equation}
where we see that   its vanishing heat capacity   $C_{V}=0$ (see (\ref{cc0})) yields $g_{TT}$=0 or $g^{TT}=\infty$ and so the metric is not invertible. We shall deal with this by treating the vanishing heat capacity as the limit $k_{\rm B}\rightarrow0^{+}$. In other words we shall treat $C_{V}$ as a constant with its value infinitely close to zero
\begin{equation}
 C_{V}\rightarrow0^{+}
\end{equation}
and normalize the scalar curvature $R$ such that
\begin{equation}
 R_{\rm N}=R C_{V}
\end{equation}
and then use $R_{\rm N}$ to probe the microstructure of the charged AdS black hole.

Combining these ideas, from (\ref{RR})  the normalized scalar curvature $R_{\rm N}$ is
\begin{equation}
 R_{\rm N}=\frac{(\partial_{V}P)^{2}-T^{2}(\partial_{V,T}P)^{2}+2T^{2}(\partial_{V}P)(\partial_{V,T,T}P)}{2(\partial_{V}P)^{2}},
\end{equation}
which becomes
\begin{equation}
 R_{\rm N}=\frac{(3\tilde{V}^{\frac{2}{3}}-1)(3\tilde{V}^{\frac{2}{3}}-4\tilde{T}\tilde{V}-1)}{2(3\tilde{V}^{\frac{2}{3}}-2\tilde{T}\tilde{V}-1)^{2}},\label{crnn}
\end{equation}
in terms of the reduced volume and temperature. Note that $R_{\rm N}$ is independent of the charge in terms of these reduced variables.

\begin{figure*}
\begin{center}
\includegraphics[width=9cm]{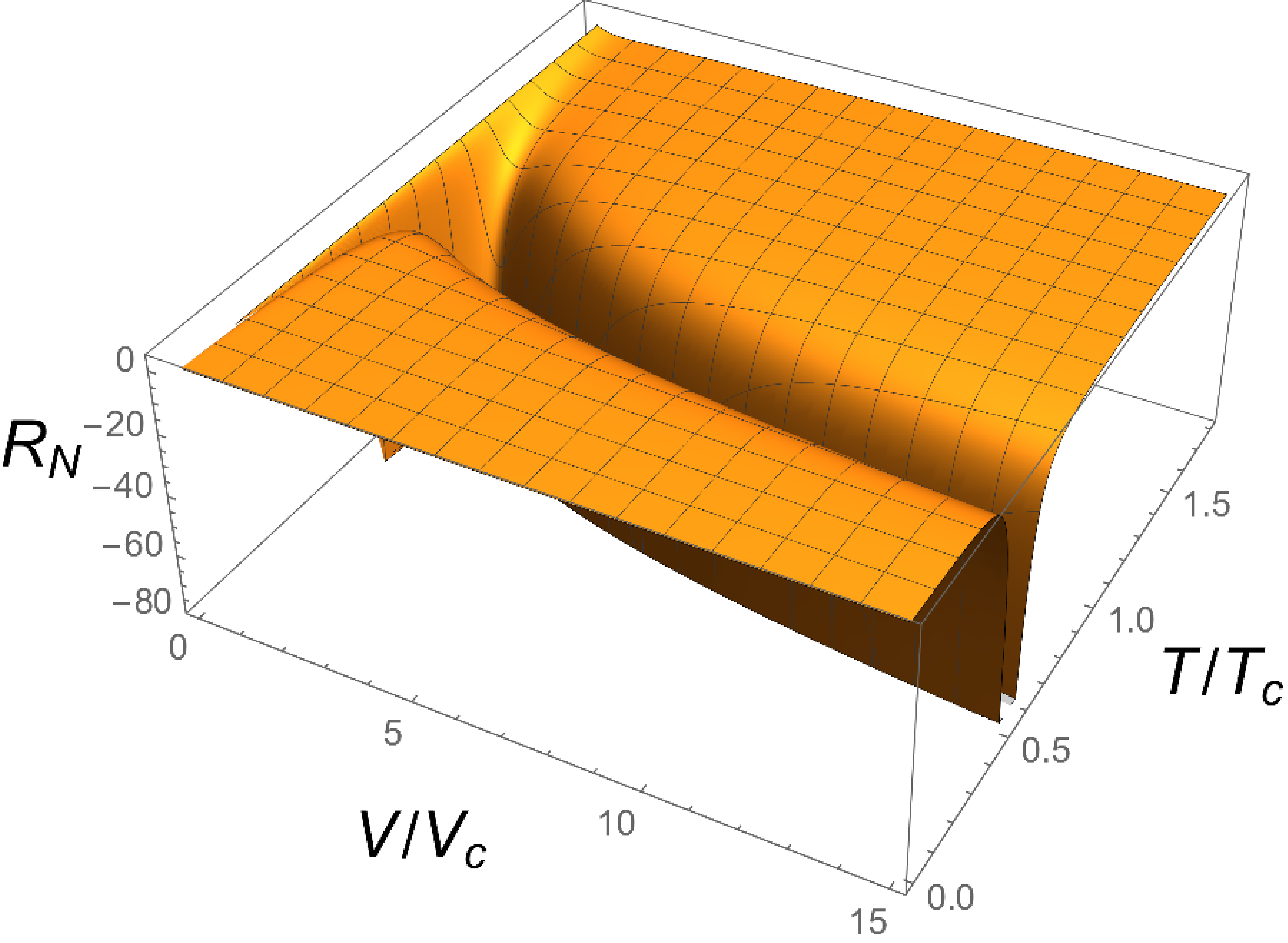}
\end{center}
\caption{Behavior of the normalized scalar curvature $R_{\rm N}$ as a function of $\tilde{V}$ and $\tilde{T}$ for the charged AdS black hole.}\label{pCBHR}
\end{figure*}

\begin{figure*}
\begin{center}
\subfigure[]{\label{RNa}
\includegraphics[width=7cm]{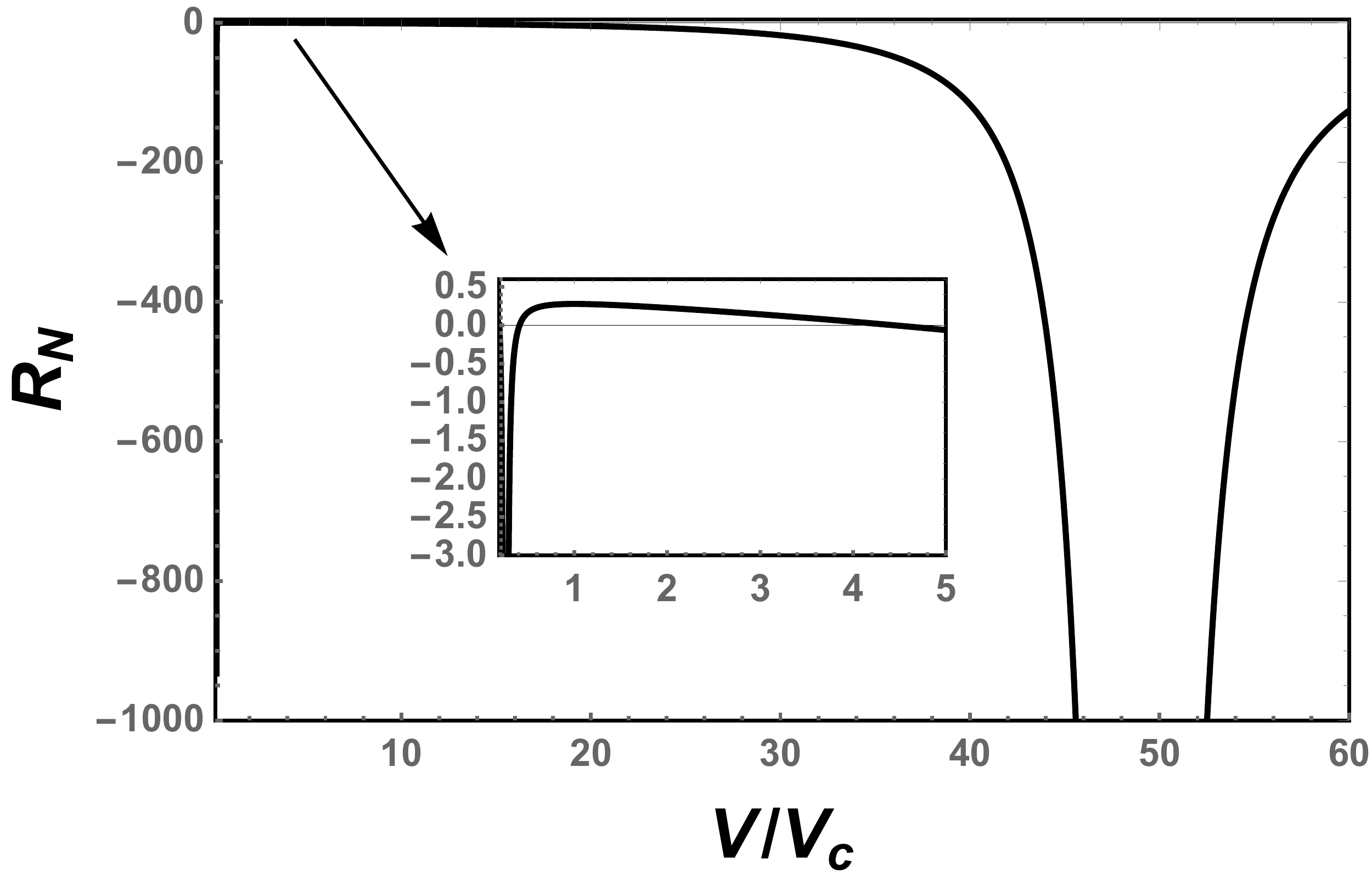}}
\subfigure[]{\label{RNab}
\includegraphics[width=7cm]{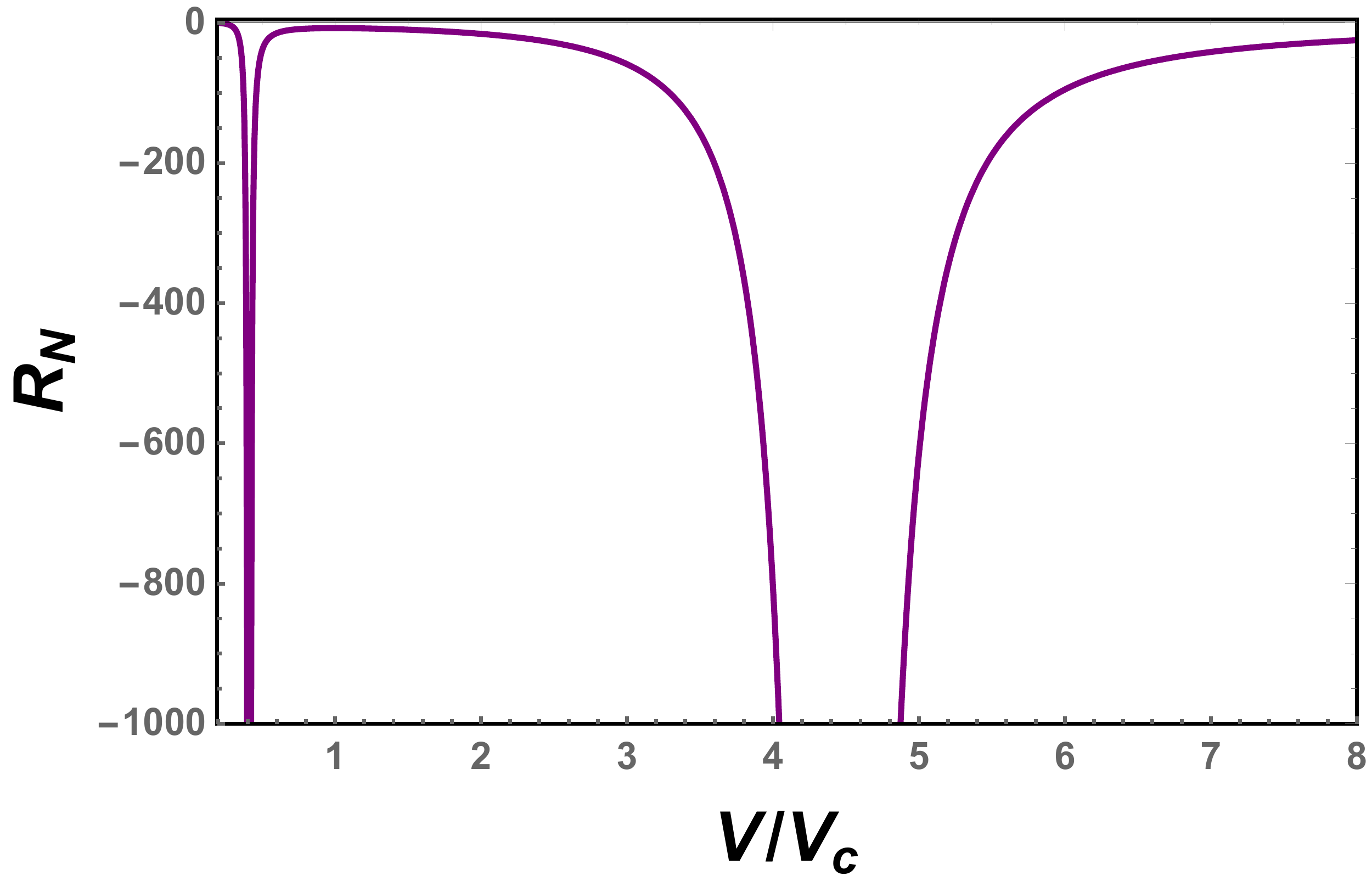}}\\
\subfigure[]{\label{RNac}
\includegraphics[width=7cm]{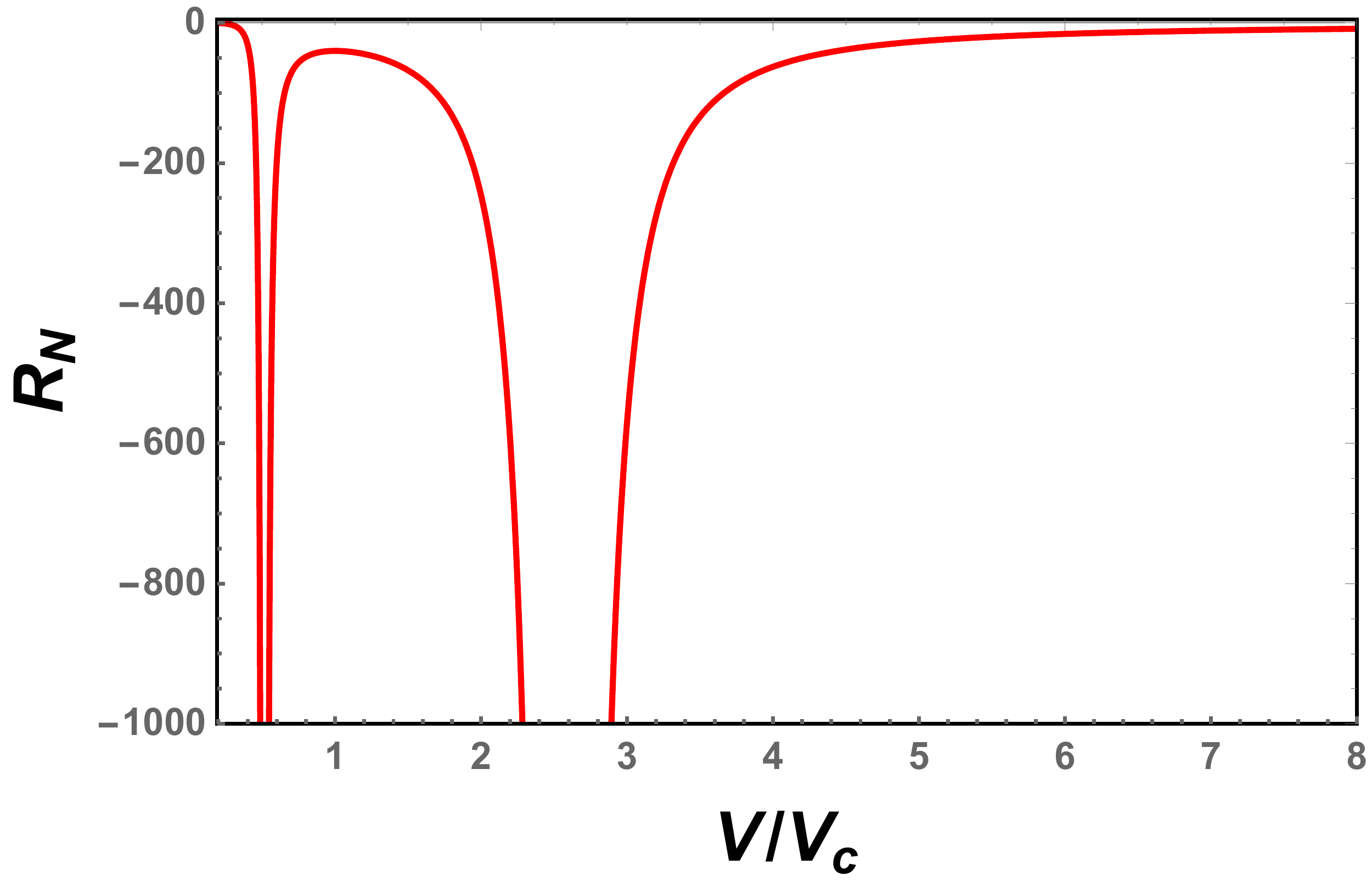}}
\subfigure[]{\label{RNad}
\includegraphics[width=7cm]{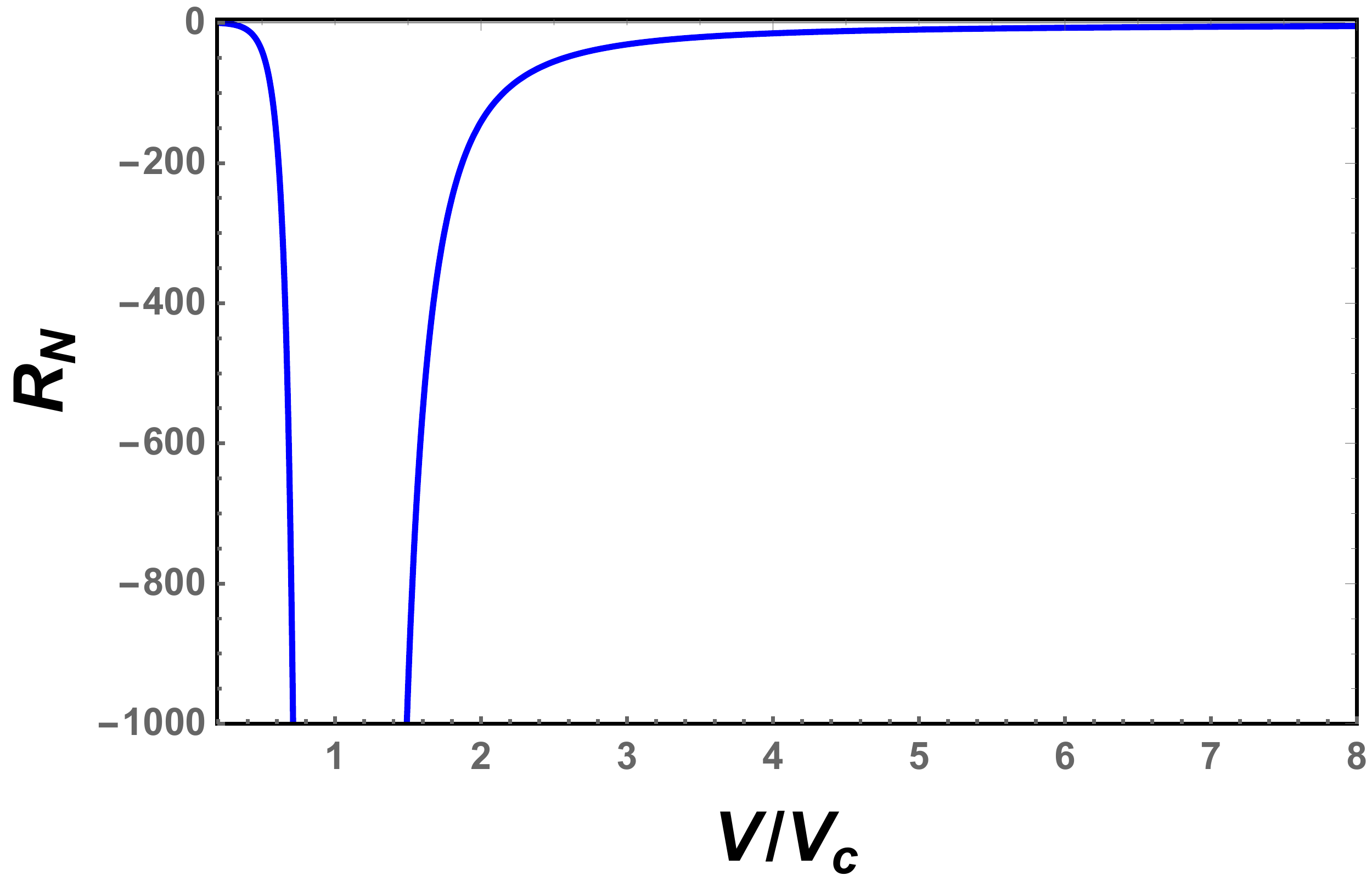}}
\end{center}
\caption{$R_{\rm N}$ vs. $\tilde{V}$ at constant temperature. (a) $\tilde{T}$=0.4. (b) $\tilde{T}$=0.8. (c) $\tilde{T}$=0.9. (d) $\tilde{T}$=1.0. For $\tilde{T}<1$, there are two negative divergent points, and one point for $\tilde{T}=1$. When $\tilde{T}$=0.4, $R_{\rm N}$ is positive in the range $\tilde{V}\in(0.41, 4.44)$.}\label{pRNa}
\end{figure*}

We plot the normalized scalar curvature $R_{\rm N}$  as a function of $\tilde{V}$ and $\tilde{T}$ in Fig. \ref{pCBHR}. For fixed low temperature $\tilde{T}<1$ there are two negative divergent points, more easily
visible in Fig.~\ref{pRNa}. One is located at small $\tilde{V}$, and another is at large $\tilde{V}$. At the critical temperature $\tilde{T}=1$, these two points coincide  at $\tilde{V}=1$. Above the critical temperature, no divergent point exists. Note that $R_{\rm N}$ always has an extremal point at $\tilde{V}$=1.

The behaviour of this normalized scalar curvature is very similar to the scalar curvature of the VdW fluid and so can be expected to indicate similar microstructure. For a range of low temperatures $R_{\rm N}$ can be positive (see Fig. \ref{pRNa}(a)), which according to the physical interpretation of the Ruppeiner geometry, implies  repulsive interaction between  the microstructures of a charged AdS black hole. The full divergent and positive behaviour of $R_{\rm N}$ is displayed in Fig. \ref{pRNa}.
%%%%%%%%%%%%%%%
%%%%%%%%%%%%%%%

%%%%%%%%%%%%%%%
%%%%%%%%%%%%%%%

\begin{figure*}
\begin{center}
\includegraphics[width=9cm]{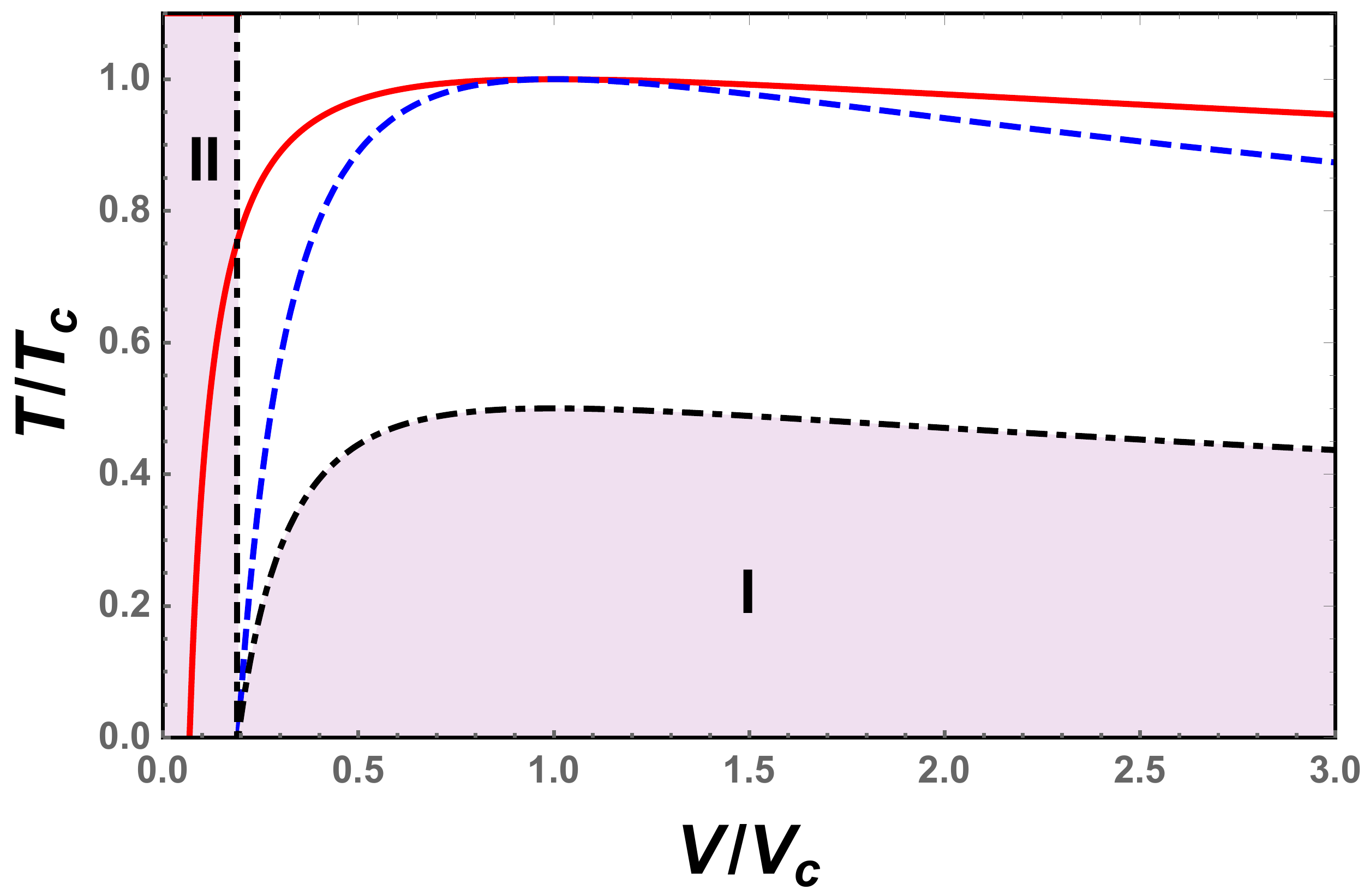}
\end{center}
\caption{The coexistence curve (red solid line), spinodal curve (blue dashed line), and the sign-changing curve of $R_{\rm N}$ (black dot dashed line). The normalized scalar curvature $R_{\rm N}$ diverges at the spinodal curve. Both the spinodal and sign-changing curves start at $\tilde{V}=1/3\sqrt{3}$, and the coexistence curve starts at $1/6\sqrt{6}$. In the shadow region, $R_{\rm N}$ is positive, otherwise, it is negative.}\label{pCsign}
\end{figure*}

Although $R_{\rm N}$ is positive in some regions, we need to check whether these regions are
thermodynamically
allowed. To this end  we  set $R_{\rm N}$=0 to obtain the sign-changing curve
\begin{equation}
 T_{0}=\frac{\tilde{T}_{\rm sp}}{2}=\frac{3\tilde{V}^{\frac{2}{3}}-1}{4\tilde{V}},
\end{equation}
indicating that  the sign-changing temperature is half of the spinodal curve temperature, which is the same relation as that of the VdW fluid (\ref{to}). Moreover, there is one more solution
\begin{equation}
 \tilde{V}=\frac{1}{3\sqrt{3}}.
\end{equation}
We depict the coexistence curve (red solid line), spinodal curve (blue dashed line), and   sign-changing curve (black dot dashed line) in Fig. \ref{pCsign}. The normalized scalar curvature $R_{\rm N}$ diverges at the spinodal curve. The regions I and II in shadow have positive $R_{\rm N}$. Similar to the VdW fluid, region I is completely in the coexistence region of small and large black holes (below the red solid line). However, different from the VdW fluid, there is another region II where $R_{\rm N}$ can be positive.  Part of it is under the coexistence region, which can be excluded from the thermodynamic
viewpoint like region I. However the top part of region II has a positive $R_{\rm N}$, so according to the physical interpretation of the Ruppeiner geometry the  `molecules' of the charged AdS black hole system have a repulsive interaction.

%%%%%%%%%%%%%%%
%%%%%%%%%%%%%%%

\subsection{Coexistence curves and critical phenomena}

Following the study in Sec. \ref{VdWcccp} for the VdW fluid, we here  examine the corresponding behaviour of $R_{\rm N}$ for a charged AdS black hole. One significant difference between them is that   an analytic expression for the coexistence curve exists for the charged AdS black hole, so we can explore these properties analytically.

\begin{figure*}
\begin{center}
\includegraphics[width=9cm]{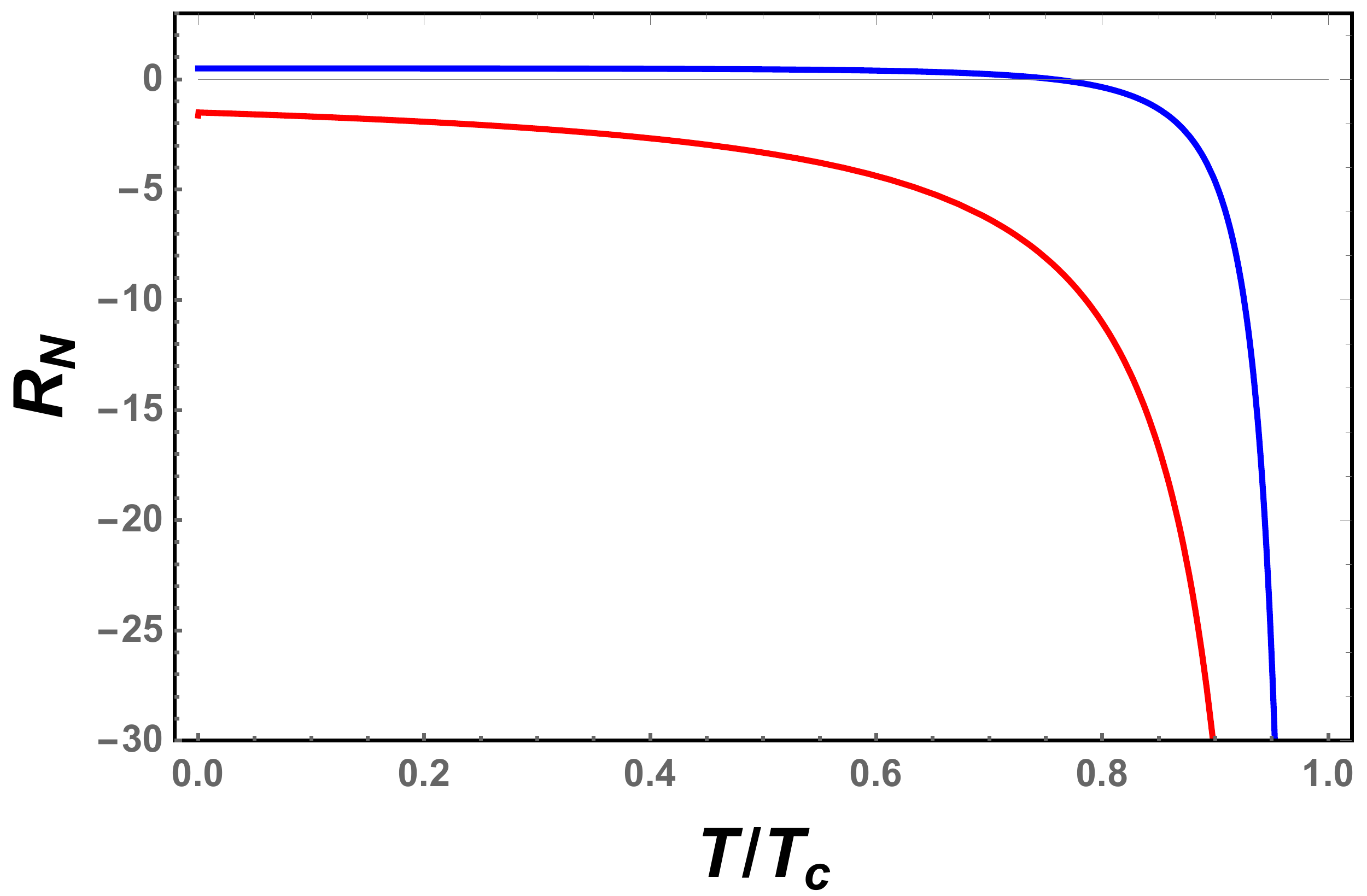}
\end{center}
\caption{The behaviour of the normalized scalar curvature $R_{\rm N}$ along the coexistence saturated large black hole (bottom red line) and small black hole (top blue line) curves for the charged AdS black hole. $R_{\rm N}$ of the coexistence saturated small black hole vanishes at $\tilde{T}$=0.7581.}\label{pCbhcorn}
\end{figure*}

Using phase transition parameters (\ref{cppp})-(\ref{cvl}) and the normalized scalar curvature (\ref{crnn}), we plot $R_{\rm N}$ along the coexistence saturated small (top blue line) and large black hole (bottom red line) curves in Fig. \ref{pCbhcorn}. At the critical point, both $R_{\rm N}$ go to negative infinity. Along the coexistence saturated large black hole curve, $R_{\rm N}$ is always negative, whereas along the coexistence saturated small black hole we observe that $R_{\rm N}$ is positive for low $\tilde{T}$ and negative for high $\tilde{T}$. This phenomenon implies that at low temperature there are repulsive interactions for the small black hole molecules and attractive interactions for the large black hole molecules. Therefore when the phase transition occurs the interaction between the microscopic molecules changes its type, leading to a huge change in the microstructure.

Since we have an analytic coexistence curve for the charged AdS black hole, we can examine the critical exponent of the normalized scalar curvature by expanding at the critical point. Along the coexistence curves for small and large black holes we  find
\begin{eqnarray}
 R_{\rm N}({\rm{SBH}})&=&-\frac{1}{8}(1-\tilde{T})^{-2}+\frac{1}{2\sqrt{2}}(1-\tilde{T})^{-\frac{3}{2}}-\frac{13}{24}(1-\tilde{T})^{-1}+\mathcal{O}(1-\tilde{T})^{-\frac{1}{2}},\\
 R_{\rm N}({\rm{LBH}})&=&-\frac{1}{8}(1-\tilde{T})^{-2}-\frac{1}{2\sqrt{2}}(1-\tilde{T})^{-\frac{3}{2}}-\frac{13}{24}(1-\tilde{T})^{-1}+\mathcal{O}(1-\tilde{T})^{-\frac{1}{2}}
\end{eqnarray}
as a series in $\sqrt{1-\tilde{T}}$.
Thus $R_{\rm N}$ has a critical exponent 2. Note that the coefficients of the odd terms are the same, while for even terms they have opposite values.
We also see that
\begin{equation}
 R_{\rm N}(1-\tilde{T})^{2}=-\frac{1}{8}
\end{equation}
as $\tilde{T}\to 1$, confirming that $R(1-\tilde{T})^{2}C_{v}$ has  a universal value of $1/8$ near the critical point.  Remarkably this exactly agrees with that of the VdW fluid (see \eqref{oneeight}). So the normalized scalar curvature $R_{\rm N}$ can   substitute for the conventional one to explore the microscopic properties of a charged AdS black hole.

Summarizing, an examination of the normalized scalar curvature $R_{\rm N}$ indicates that both charged AdS black holes and VdW fluids have a Ruppeiner scalar curvature with a critical exponent of 2 and constant value of $-1/8$ for $R_{\rm N}(1-\tilde{T})^{2}$ as $\tilde{T}\to 1$.  However for some small charged AdS black holes, their microscopic molecules can have repulsive interaction, in contrast to a VdW fluid.

%%%%%%%%%%%%%%%
%%%%%%%%%%%%%%%

\section{Higher-dimensional charged AdS black holes}
\label{aaa}

We have seen that   Ruppeiner geometry can be used to provide information about the microscopic properties of four-dimensional charged AdS black holes. Since higher-dimensional charged black holes also exhibit   VdW-type phase transitions \cite{Gunasekaran}, we examine in this section their Ruppeiner scalar curvature and to study the effect of the dimension of the spacetime on the black hole microscopic structures.

\subsection{Thermodynamics and phase structures}

We begin with a brief review of the thermodynamics for the $d$-dimensional charged AdS black holes \cite{Gunasekaran}. The metric reads
\begin{eqnarray}
 ds^{2}&=&-f(r)dt^{2}+f^{-1}(r)dr^{2}+r^{2}d\Omega^{2}_{d-2},\\
 f(r)&=&1-\frac{m}{r^{d-3}}+\frac{q^{2}}{r^{2(d-3)}}+\frac{r^{2}}{l^{2}}.
\end{eqnarray}
The parameters $m$ and $q$, respectively, are related to the black hole enthalpy and charge
\begin{eqnarray}
 H&=&\frac{d-2}{16\pi}\omega_{d-2}m,\\
 Q&=&\frac{\sqrt{2(d-2)(d-3)}}{8\pi}\omega_{d-2}q,
\end{eqnarray}
where $\omega_{d}=2\pi^{(d+1)/2}/\Gamma((d+1)/2)$ is the volume of a unit $d$-sphere.  The black hole temperature, entropy, electric potential, and thermodynamic
volume are
\begin{eqnarray}
 T&=&\frac{16\pi r_{\rm h}^2
      \left(P-\frac{2\pi Q^2 r_{\rm h}^{4-2 d}}{\omega_{d-2}^2}\right)
     +(d-5)d+6}{4\pi (d-2) r_{\rm h}},\label{tep}\\
 S&=&\frac{\omega_{d-2}r_{\rm h}^{d-2}}{4},\quad
 \Phi=\frac{4\pi Q r_{\rm h}^{3-d}}{(d-3)\omega_{d-2}},\quad
 V=\frac{\omega_{d-2} r_{\rm h}^{d-1}}{d-1},
\end{eqnarray}
where the AdS radius $l$ has been replaced by the pressure $P$ following (\ref{pL}). The Gibbs free energy $G=H-TS$ is
\begin{equation}
 G=\frac{\omega_{d-2}}{16\pi}\bigg(r_{\rm h}^{d-3}
       -\frac{16\pi Pr_{\rm h}^{d-1}}{(d-1)(d-2)}
       +\frac{32(2d-5)\pi^{2}Q^{2}r_{\rm h}^{3-d}}{(d-2)(d-3)\omega_{d-2}^{2}}\bigg) \label{gbs}
\end{equation}
and from (\ref{tep}), we can obtain the equation of state
\begin{eqnarray}
 P&=&\frac{d-2}{16\pi}\left(d-3+4\pi T \left(\frac{(d-1)V}{\omega_{d-2}}\right)^{\frac{1}{d-1}}\right)\left(\frac{(d-1)V}{\omega_{d-2}}\right)^{-\frac{2}{d-1}}\nonumber\\
 &&\qquad + 2\pi  Q^{2} \left((d-1)V\right)^{-\frac{2(d-2)}{(d-1)}}\omega_{d-2}^{-\frac{2}{d-1}}\label{ppppq}
\end{eqnarray}
for the $d$-dimensional charged AdS black hole.

As in four dimensions,
this equation of state  describes a small-large black hole phase transition that  is also reminiscent of the liquid-gas phase transition of the VdW fluid. This coexistence curve has positive slope everywhere and terminates at a second order phase transition point, the critical point, determined by $(\partial_{V}P)_{T}=(\partial_{V,V}P)_{T}=0$,
\begin{eqnarray}
 T_{\rm c}=\frac{4(d-3)^{2}}{(d-2)(2d-5)\pi v_{\rm c}},\quad
 P_{\rm c}=\frac{(d-3)^{2}}{(d-2)^{2}\pi v_{\rm c}^{2}}, \quad
 V_{\rm c}&=&\frac{\omega_{d-2}}{d-1}\bigg(\frac{d-2}{4}\bigg)^{d-1}v_{\rm c}^{d-1},
\end{eqnarray}
where the critical specific volume $v_{\rm c}=\frac{4}{d-2}\left(q^{2}(d-2)(2d-5)\right)^{1/(2d-6)}$. In the reduced parameter space, the equation of state becomes
\begin{equation}
 \tilde{P}=\frac{\tilde{V}^{-\frac{2(d-2)}{d-1}}+(d-2)\tilde{V}^{\frac{2}{1-d}}\left(5-2d+4(d-3)\tilde{T} \tilde{V}^{\frac{1}{d-1}}\right)}{(d-3)(2d-5)}
\end{equation}
which is free of the black hole charge $Q$.

Unlike the  four-dimensional case there is no   analytic form of the coexistence curve for small-large black holes. However  highly accurate fitting formulas of the coexistence curves for $d$=5-10 have been computed \cite{Wei5}.

We next consider   the spinodal curve. By solving $(\partial_{\tilde{V}}\tilde{P})=0$, we obtain
\begin{equation}
 \tilde{T}_{\rm sp}=\frac{2d-5-\tilde{V}^{-\frac{2(d-3)}{d-1}}}{2(d-3)\tilde{V}^{\frac{1}{d-1}}}
\end{equation}
 in the reduced parameter space.
The small black hole spinodal curve  starts at $\tilde{V}=(2d-5)^{-\frac{d-1}{2(d-3)}}$, which has a maximum of $\tilde{V} = 1/5$ for $d=5$.
 Alternatively the spinodal curve has the following form
\begin{equation}
 \tilde{P}_{\rm sp}=\frac{d-2-\tilde{V}^{-\frac{2(d-3)}{d-1}}}{(d-3)\tilde{V}^{\frac{2}{d-1}}}
\end{equation}
in terms of pressure.

%%%%%%%%%%%%%%%
\begin{figure*}
\begin{center}
\subfigure[]{\label{Cpt5}
\includegraphics[width=4.5cm]{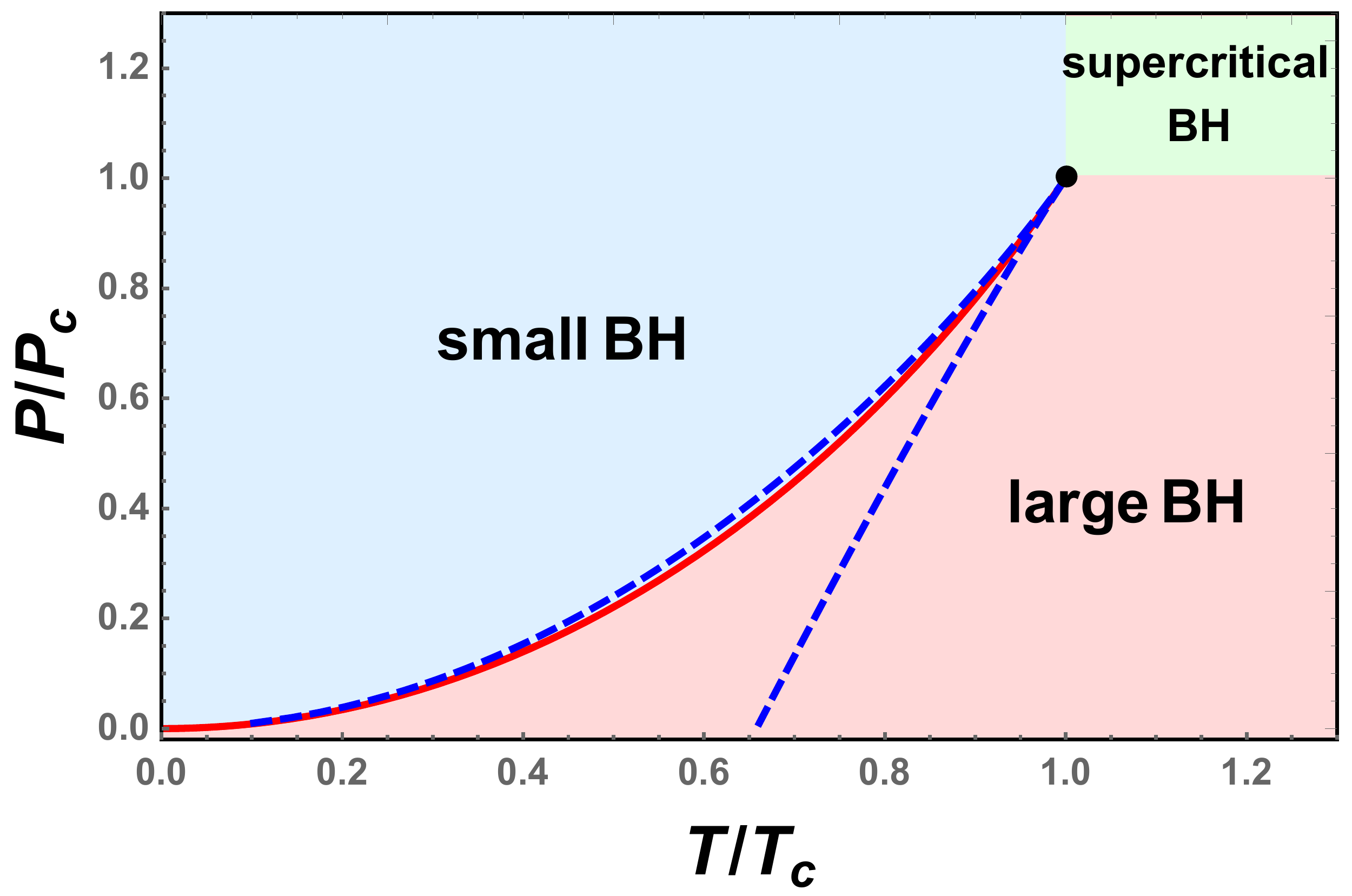}}
\subfigure[]{\label{Cpt6}
\includegraphics[width=4.5cm]{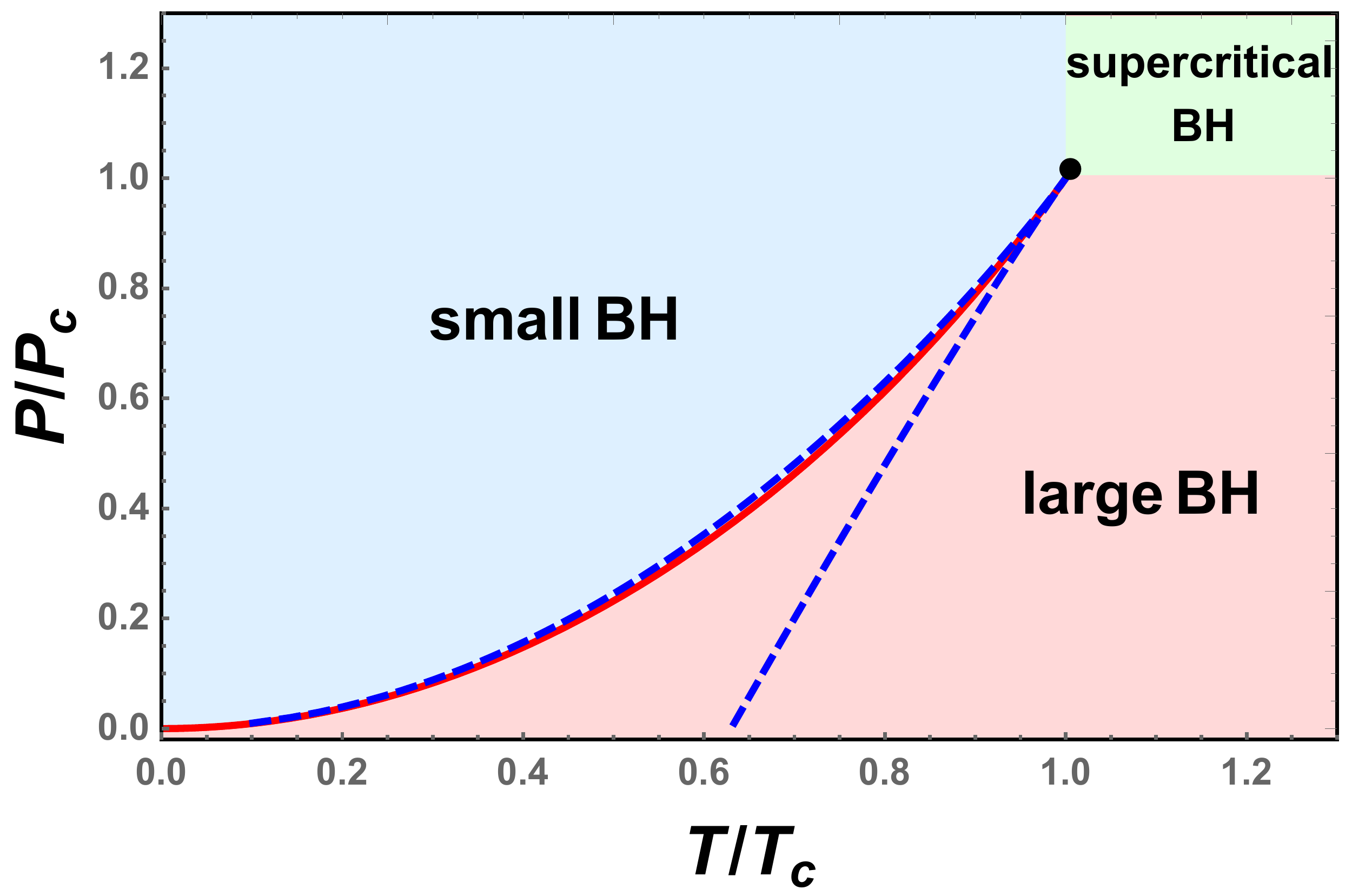}}
\subfigure[]{\label{Cpt7}
\includegraphics[width=4.5cm]{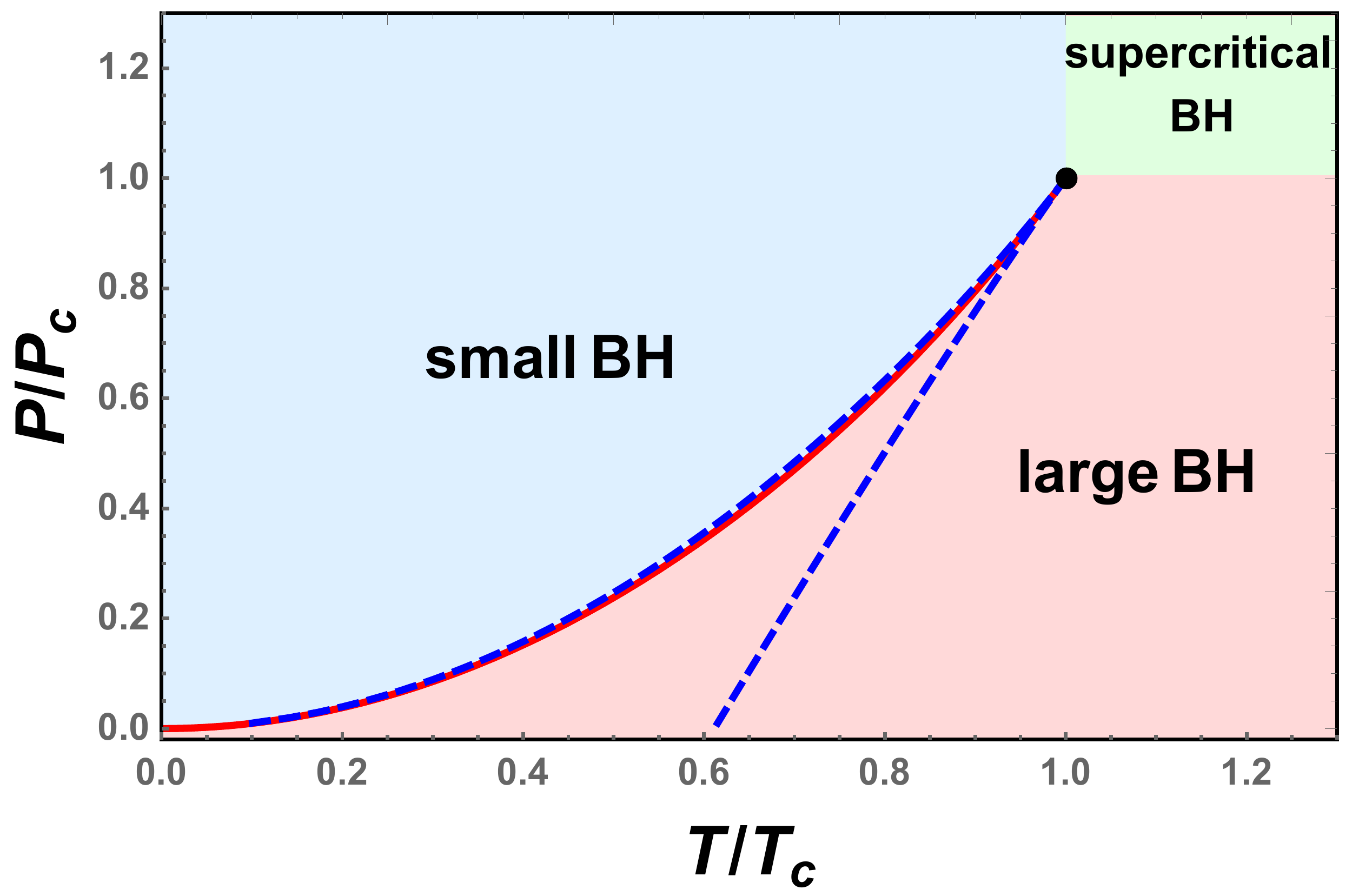}}\\
\subfigure[]{\label{Cpt8}
\includegraphics[width=4.5cm]{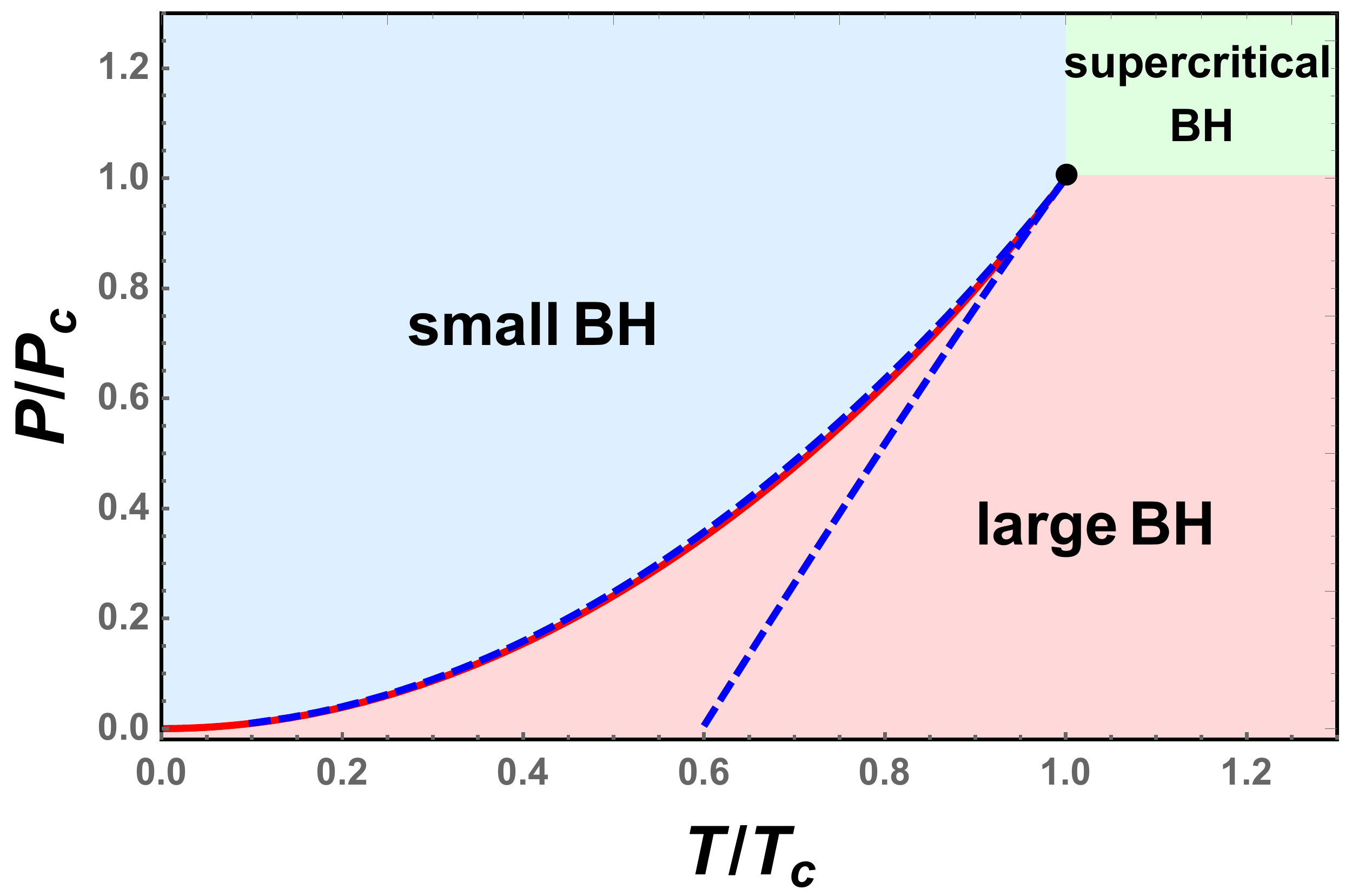}}
\subfigure[]{\label{Cpt9}
\includegraphics[width=4.5cm]{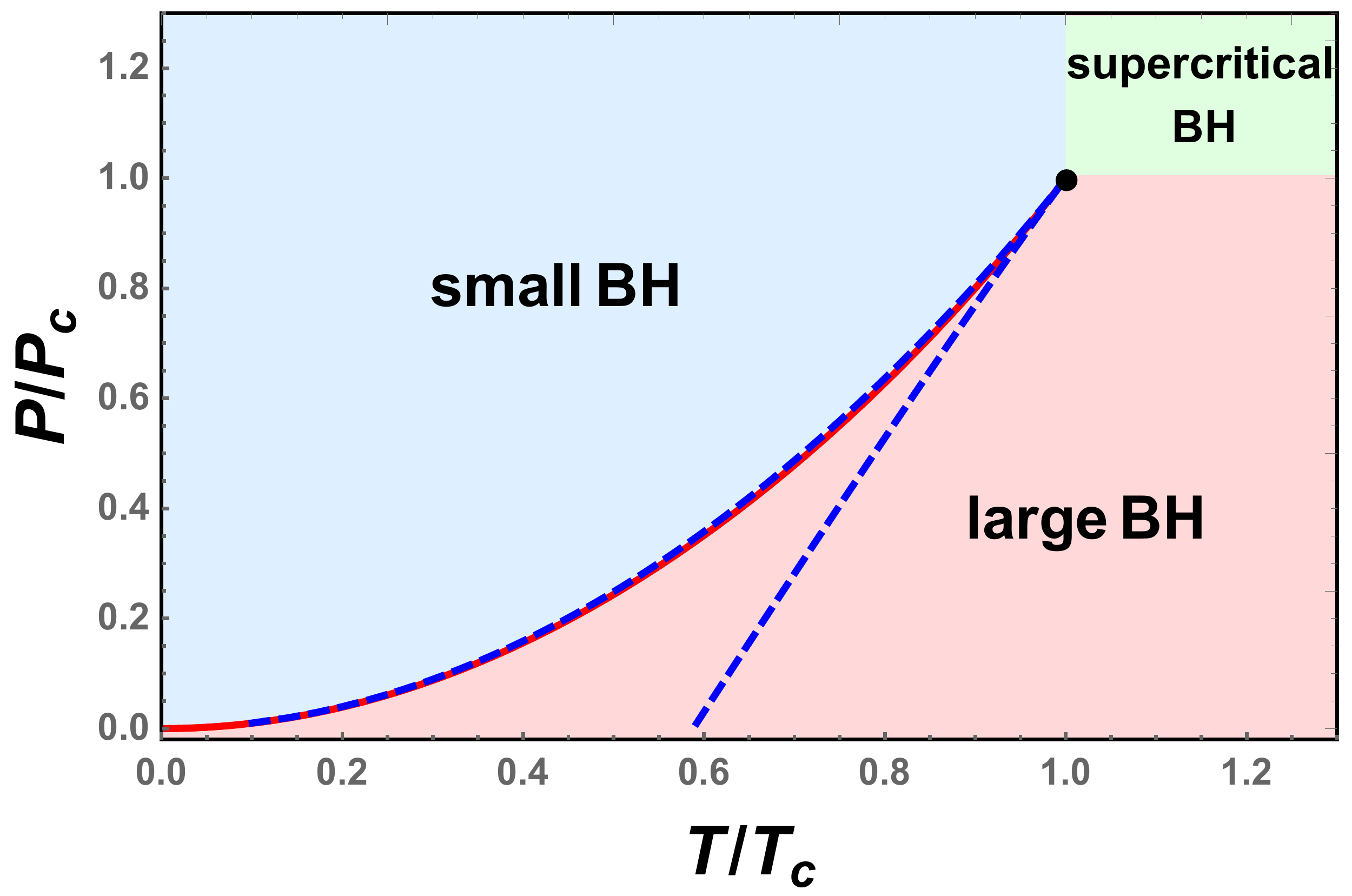}}
\subfigure[]{\label{Cpt10}
\includegraphics[width=4.5cm]{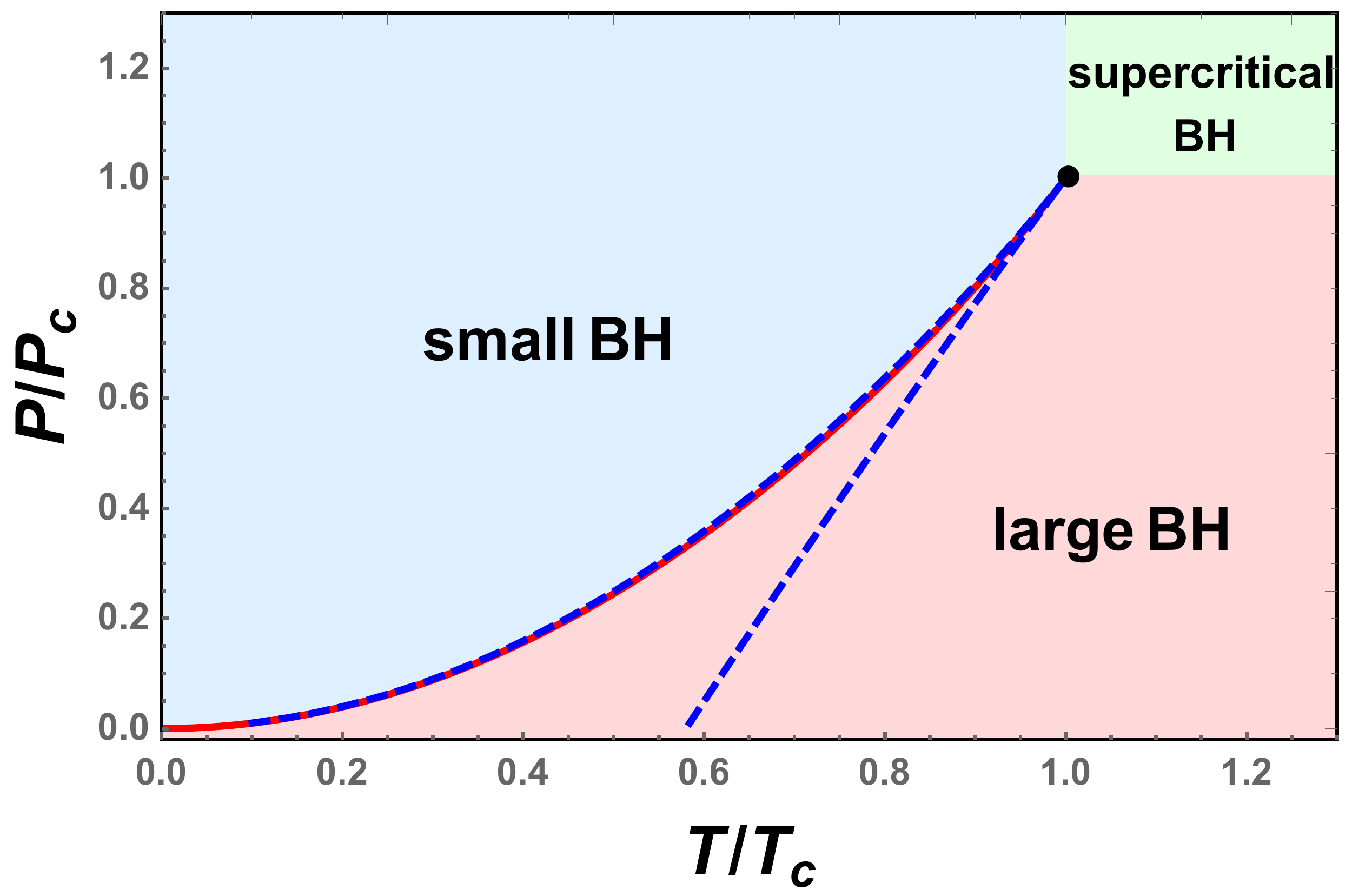}}
\end{center}
\caption{Phase diagrams and spinodal curves for  higher-dimensional charged AdS black  holes in the reduced $\tilde{P}$-$\tilde{T}$ diagram. Black dots denote the critical points. The red solid curves denote the coexistence curves, above and below which are the small black hole phases and large black hole phases, respectively. The supercritical black hole phases are located at right top corners. Top and bottom blue dashed curves are for the large and small black hole spinodal curves, respectively. (a) $d$=5. (b) $d$=6. (c) $d$=7. (d) $d$=8. (e) $d$=9. (f) $d$=10.}\label{pCpt5}
\end{figure*}
%%%%%%%%%%%%%%%

In Fig. \ref{pCpt5}, we show the phase diagrams and spinodal curves for  higher-dimensional charged AdS black holes in the reduced $\tilde{P}$-$\tilde{T}$ diagram. The regions of small, large and supercritical black hole phases are clearly displayed, and the behaviour   is similar to that of a VdW fluid and the  four-dimensional charged AdS black hole. All  large black hole spinodal curves start at the origin, and terminate at the critical points, whereas   small black hole spinodal curves start at $\tilde{T}$=0.658037, 0.629961, 0.611422, 0.598116, 0.588024, 0.580065 for $d$=5-10, with this value approaching  1/2 as $d\rightarrow\infty$. Note that the large black hole spinodal curves get closer to the coexistence curves with   increasing spacetime dimension.

\begin{figure*}
\begin{center}
\subfigure[]{\label{Hctv5}
\includegraphics[width=4.5cm]{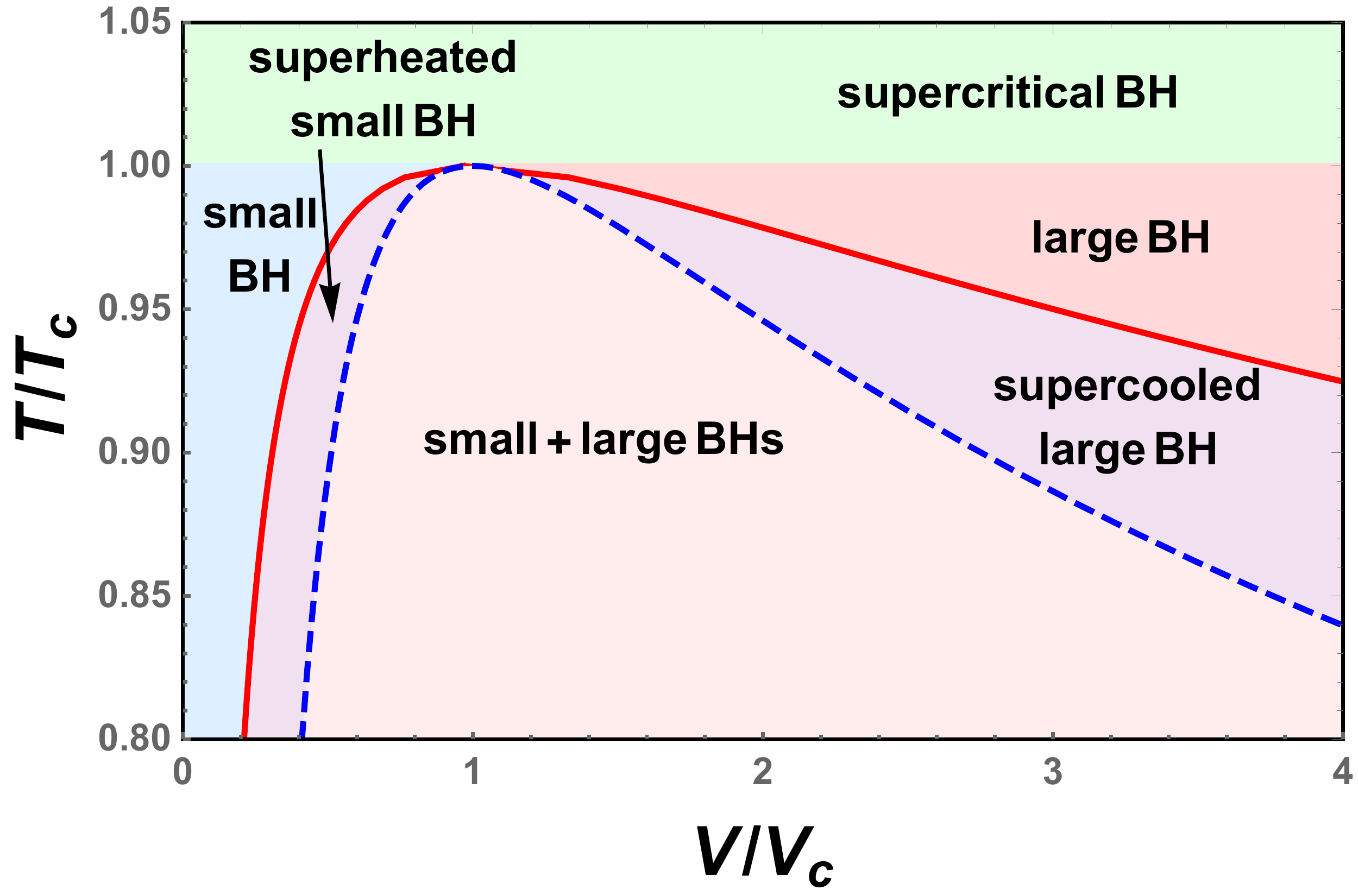}}
\subfigure[]{\label{Hctv6}
\includegraphics[width=4.5cm]{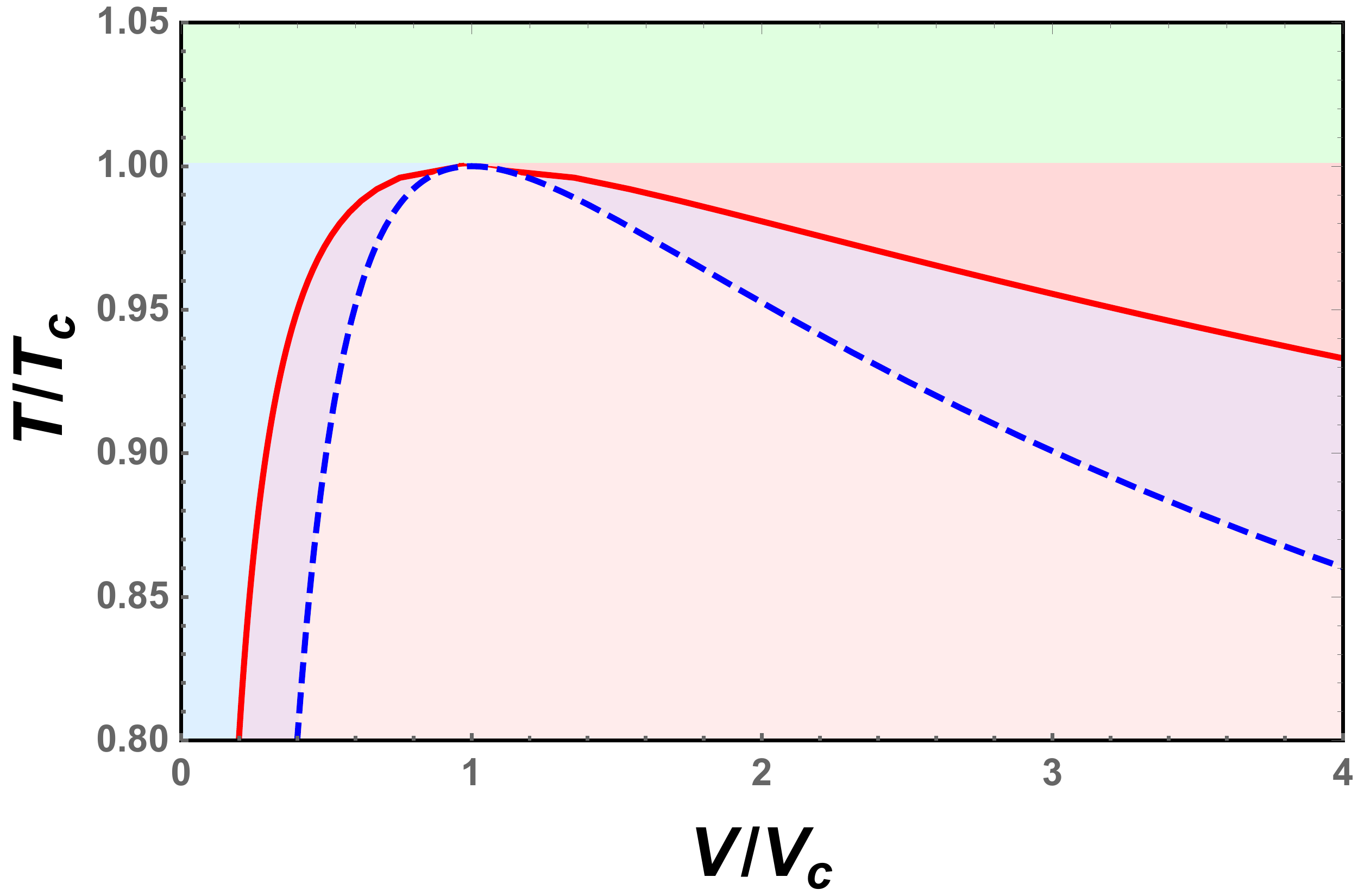}}
\subfigure[]{\label{Hctv7}
\includegraphics[width=4.5cm]{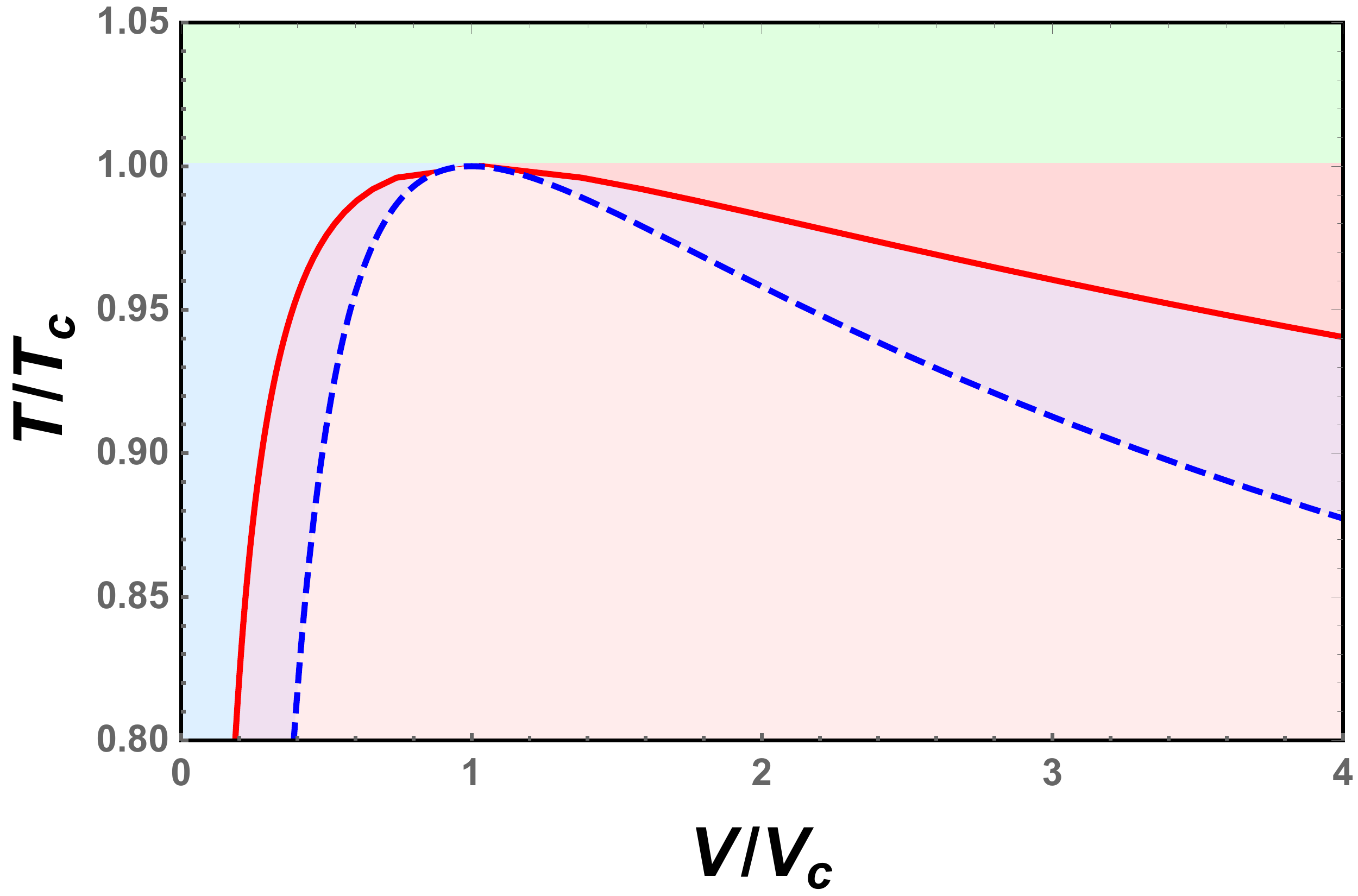}}\\
\subfigure[]{\label{Hctv8}
\includegraphics[width=4.5cm]{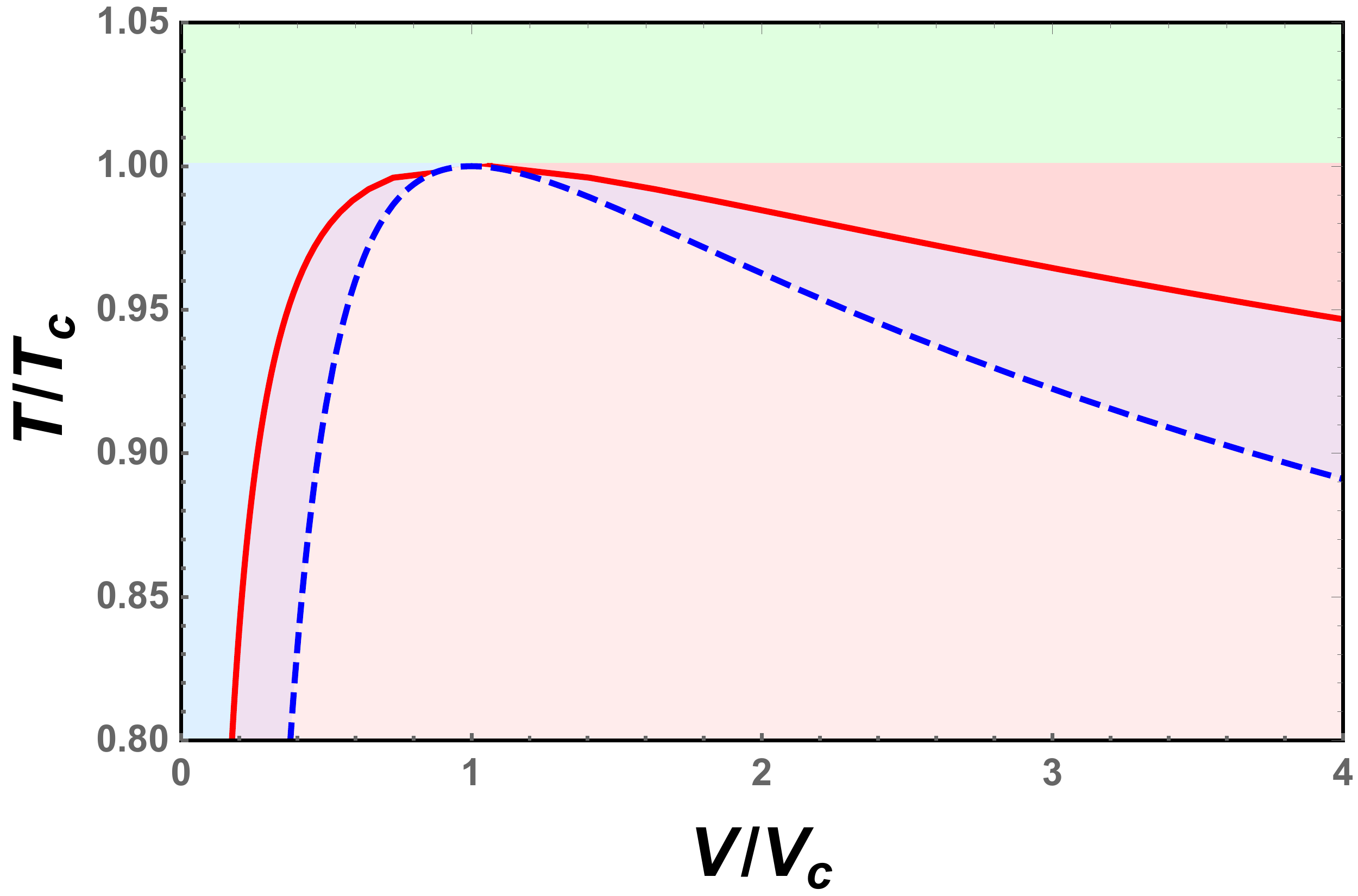}}
\subfigure[]{\label{Hctv9}
\includegraphics[width=4.5cm]{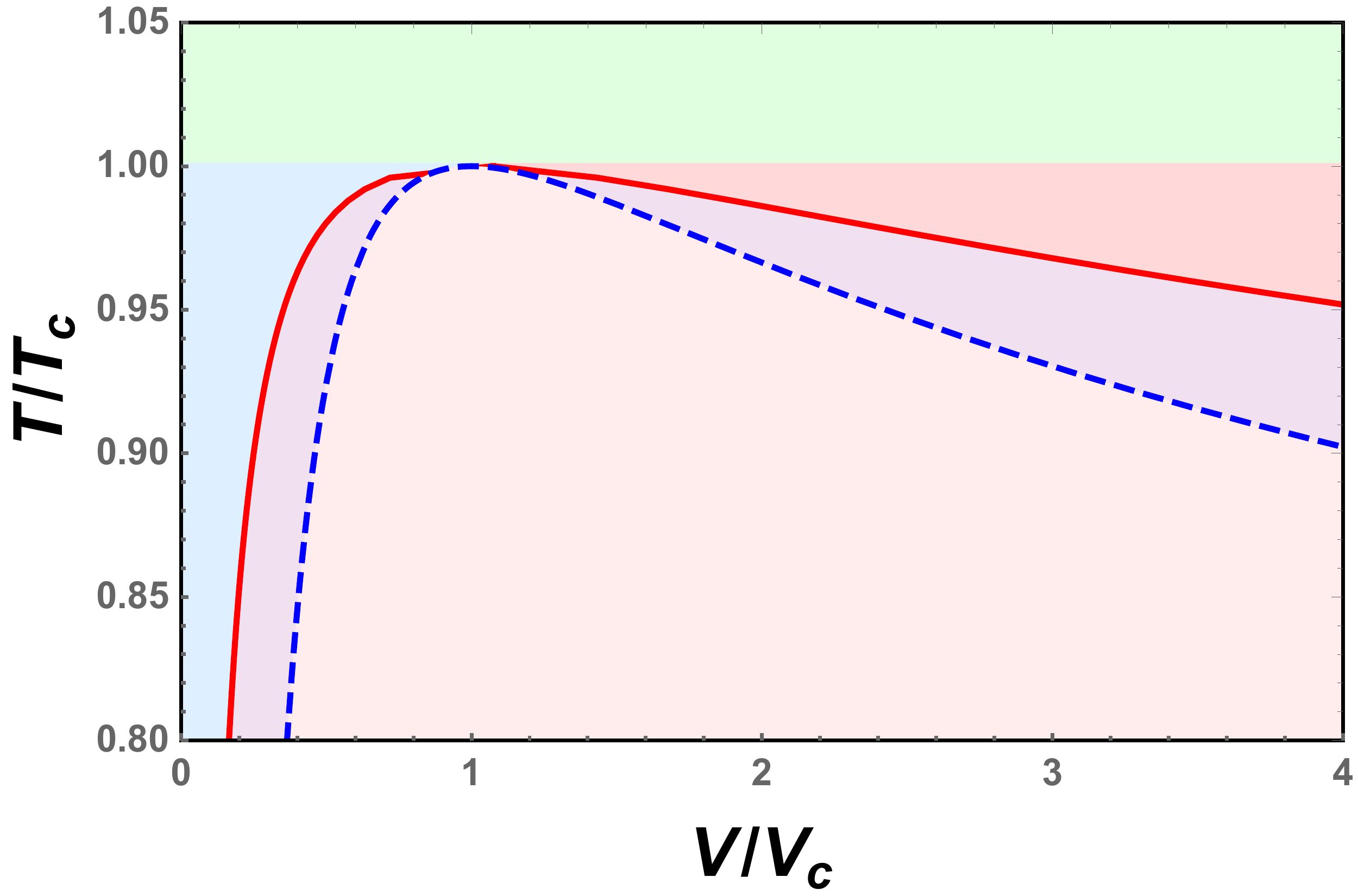}}
\subfigure[]{\label{Hctv10}
\includegraphics[width=4.5cm]{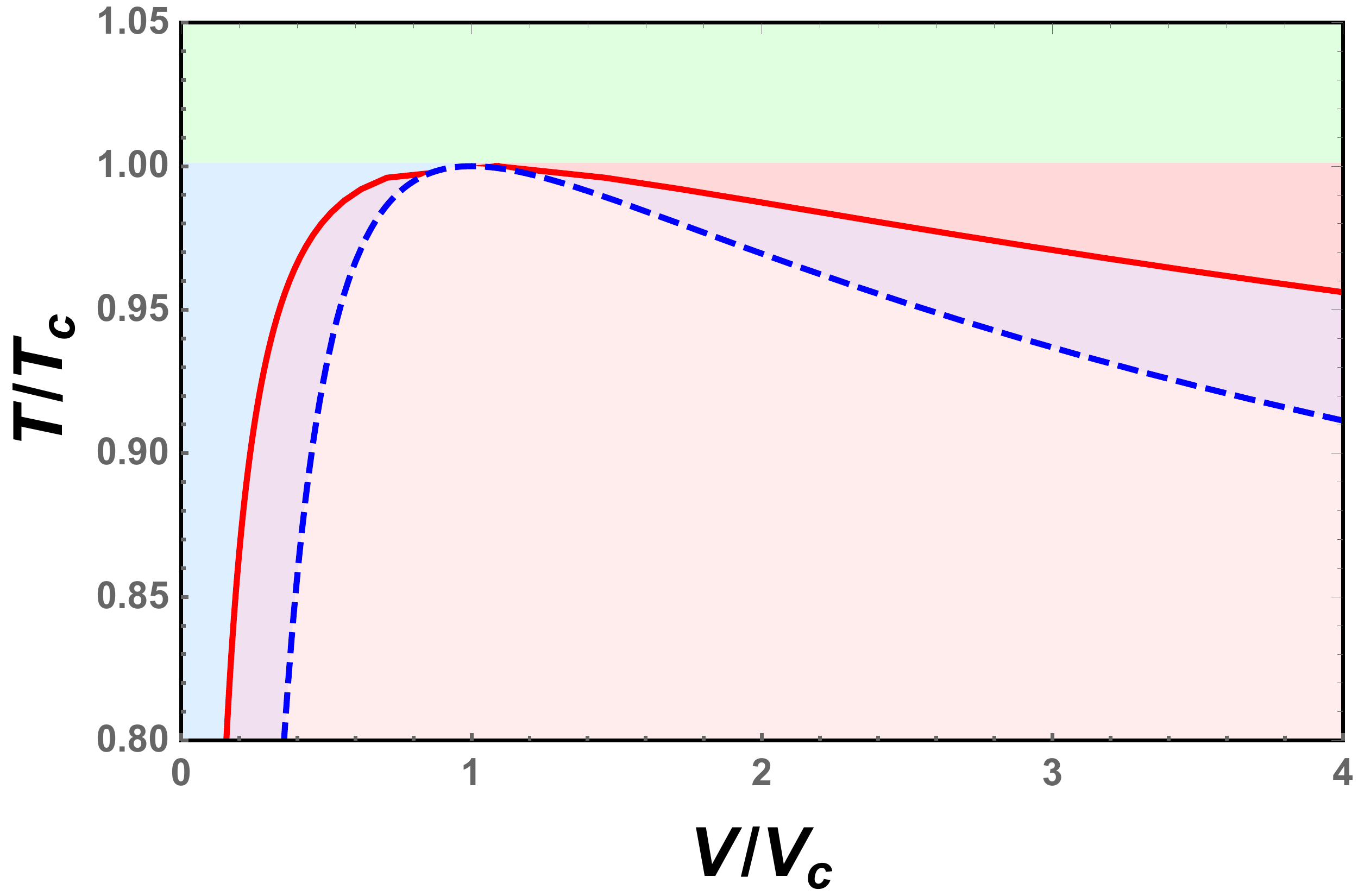}}
\end{center}
\caption{Phase diagrams and spinodal curves for   higher-dimensional charged AdS black  holes in $\tilde{T}$-$\tilde{v}$ space for $d$=5-10. (a) $d$=5. (b) $d$=6. (c) $d$=7. (d) $d$=8. (e) $d$=9. (f) $d$=10.
The red solid curves and blue dashed curves are the respective coexistence curves and spinodal curves.  The small black hole spinodal  curves have a reduced volume $\tilde{V}=(\frac{1}{2d-5})^{\frac{d-1}{2(d-3)}}$ when $\tilde{T}=0$. Note that we only mark the regions of different black hole phases for the five-dimensional  charged AdS black hole. Other cases are similar.}
\label{pHctv5}
\end{figure*}

Higher-dimensional charged AdS black holes  also have metastable   superheated small black hole and  supercooled larger black hole phases.  We illustrate these phase structures in Fig.~\ref{pHctv5}  in  $\tilde{T}$-$\tilde{V}$ space for $d$=5-10;  obviously all the phase structures are similar. Moreover, we find that for $d$=5-10, the small coexistence curves respectively start at $\tilde{V}$=0.066667, 0.062235, 0.057556, 0.053250, 0.049428, and 0.046066, while the small black hole spinodal curves respectively start at $\tilde{V}$=0.2, 0.197584, 0.192450, 0.186649, 0.180872, and 0.175364;   the spinodal curves are always below the coexistence curves.

%%%%%%%%%%%%%%%
%%%%%%%%%%%%%%%

\subsection{Ruppeiner geometry}

As in the  four-dimensional case, the heat capacity at constant volume $C_{V}$ also vanishes. Here we adopt the method given in Sec. \ref{rgr} by treating $C_{V}$ as a constant and take its value infinitely close to zero. Employing (\ref{ppppq}), we obtain
\begin{equation}
 R_{\rm N}=\frac{1}{2}-\frac{2 \pi ^2 (d-1)^4 T^2 V^4
   \left(\frac{(d-1) V}{\omega_{d-2}}\right)^{\frac{2}{d-1}}}{\left[(d-1)^2 V^2
   \left(-2 \pi  T \left(\frac{(d-1) V}{\omega_{d-2}}\right)^{\frac{1}{d-1}}+d-3\right)-32 \pi
   ^2 Q^2 \left(\frac{(d-1) V}{\omega_{d-2}}\right)^{\frac{4}{d-1}}\right]^2}
\end{equation}
for the normalized scalar curvature, which becomes
\begin{equation}
 R_{\rm N}=\frac{1}{2}-\frac{2 (d-3)^2 \tilde{T}^2
   \tilde{V}^{\frac{2}{d-1}}}{\left(2 (d-3) \tilde{T}
   \tilde{V}^{\frac{1}{d-1}}+\tilde{V}^{-\frac{2
   (d-3)}{d-1}}-2 d+5\right)^2}
   \label{rrrrrn}
\end{equation}
in the reduced parameter space and is also free of the charge.  Its dependence on $\tilde{T}$ and $\tilde{V}$ is qualitatively similar to that of the four-dimensional case shown in Figs. \ref{pCBHR}, \ref{pRNa}, and \ref{pCsign} and so we will not plot the corresponding diagrams.

However the sign-changing curve indicates the microstructure interactions and so
is very important. Setting $R_{\rm N}=0$ we get two sign-changing curves
\begin{eqnarray}
 T_{0}&=&\frac{\tilde{T}_{\rm sp}}{2}=\frac{2d-5-\tilde{V}^{-\frac{2(d-3)}{d-1}}}{4(d-3)\tilde{V}^{\frac{1}{d-1}}},\\
 \tilde{V}&=&(2 d-5)^{-\frac{d-1}{2 (d-3)}}
\end{eqnarray}
where again $T_{0}$ is half of $\tilde{T}_{\rm sp}$. The second one is a vertical line in the $\tilde{T}$-$\tilde{V}$ diagram (see Fig. \ref{pCsign} for example), which gets closer to the vertical axis with increasing   dimension  $d$.

\subsection{Coexistence curves and critical phenomena}

\begin{figure*}
\begin{center}
\subfigure[]{\label{HRNT5}
\includegraphics[width=4.5cm]{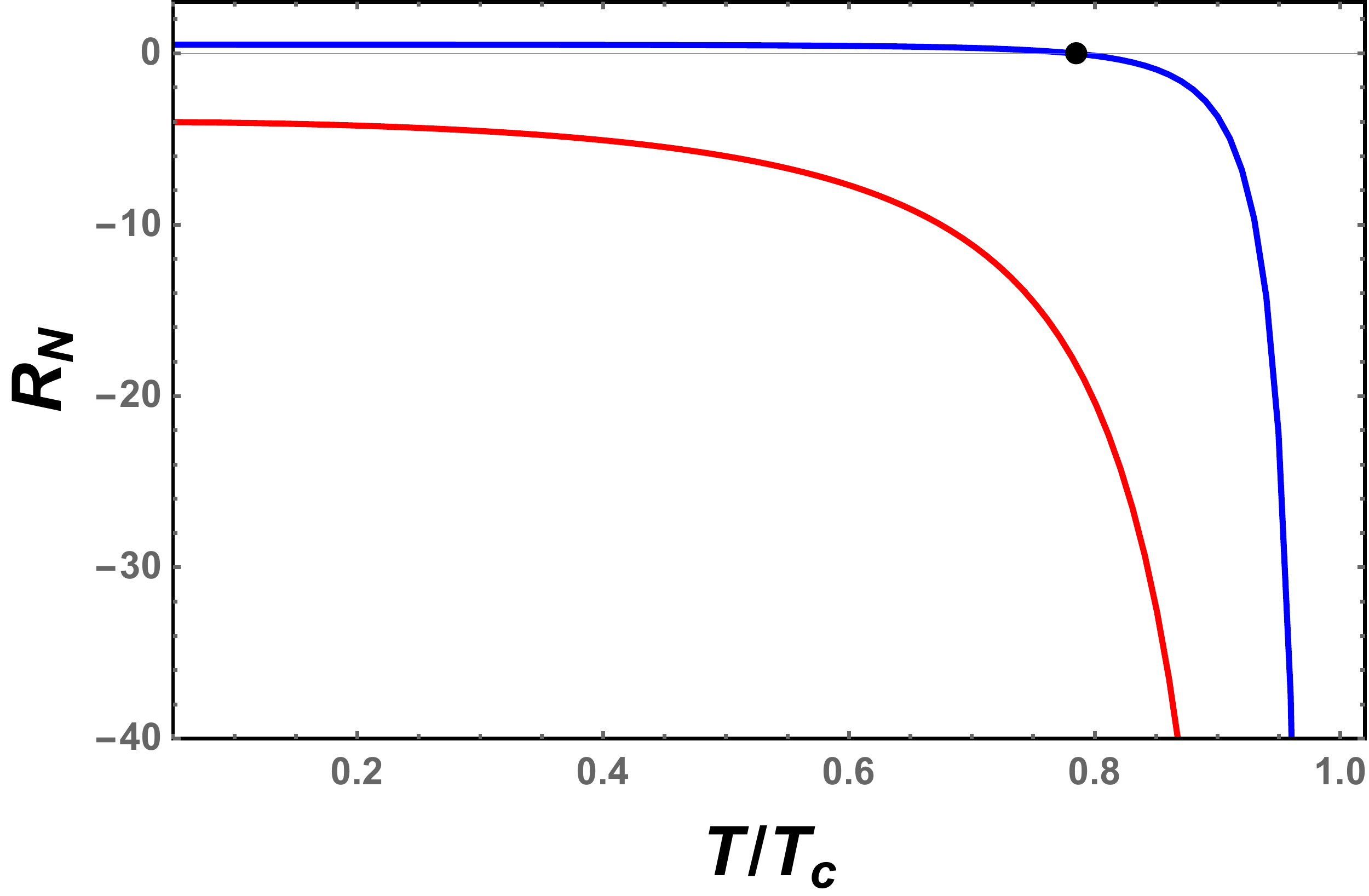}}
\subfigure[]{\label{HRNT6}
\includegraphics[width=4.5cm]{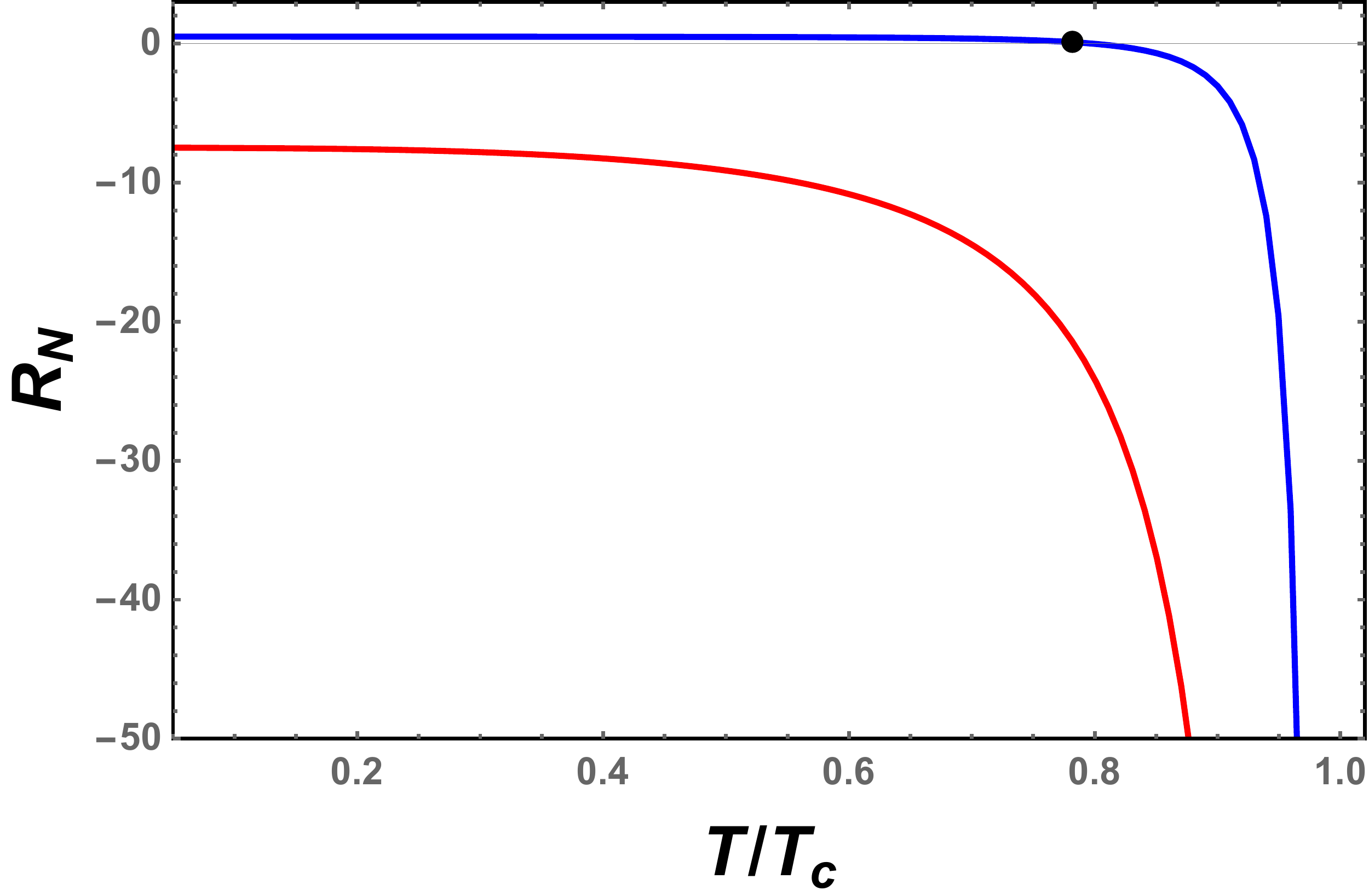}}
\subfigure[]{\label{HRNT7}
\includegraphics[width=4.5cm]{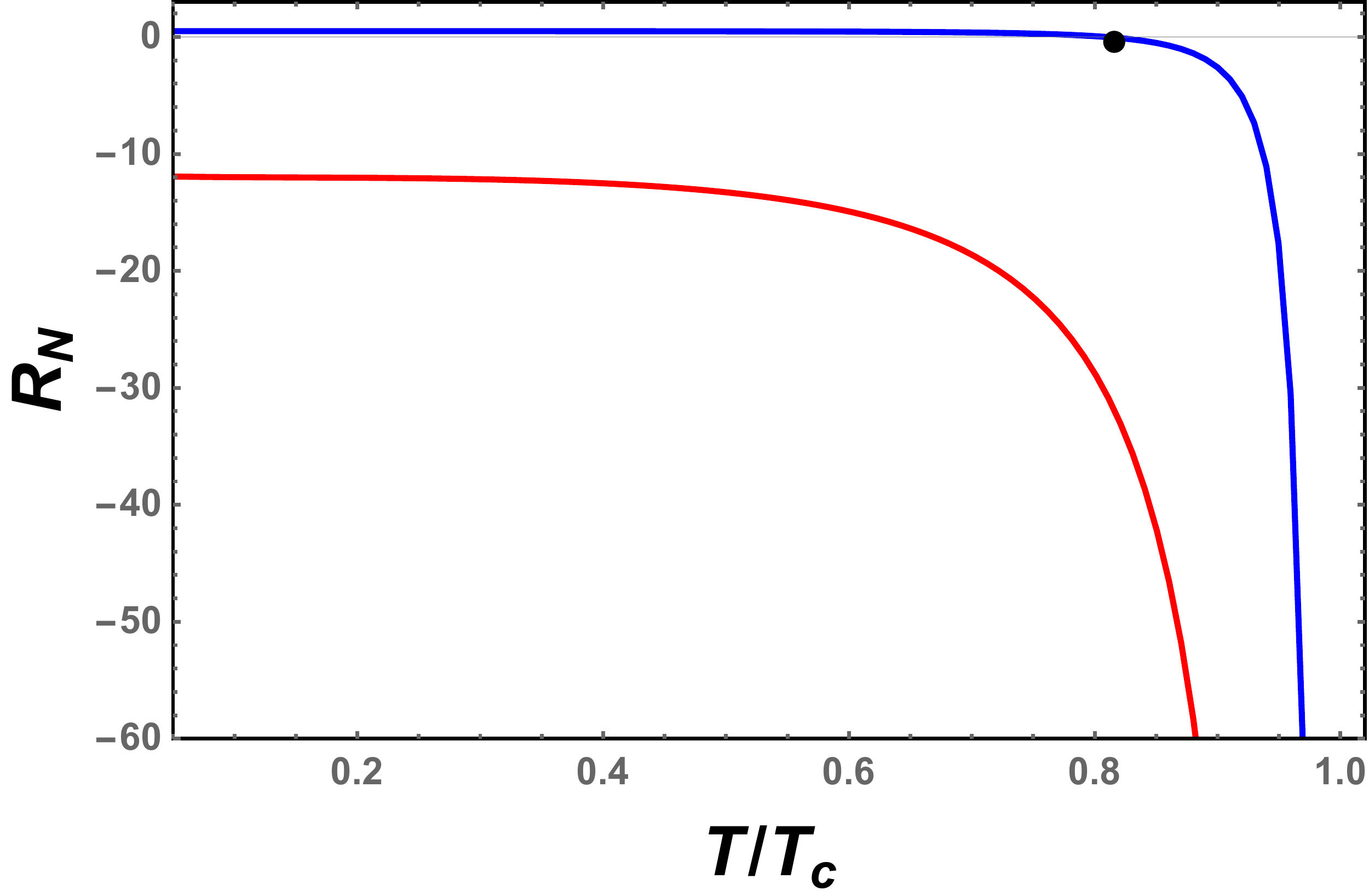}}\\
\subfigure[]{\label{HRNT8}
\includegraphics[width=4.5cm]{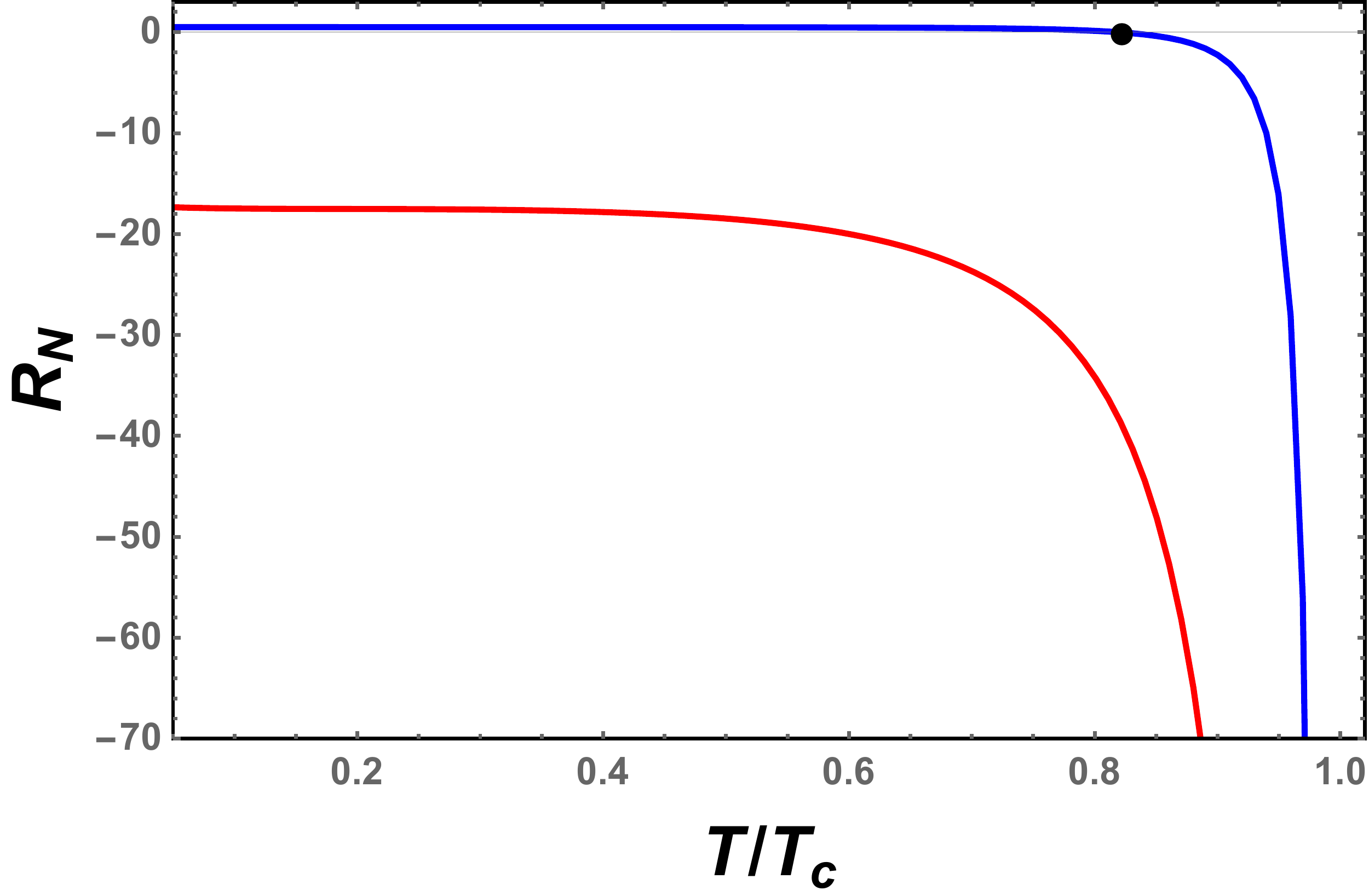}}
\subfigure[]{\label{HRNT9}
\includegraphics[width=4.5cm]{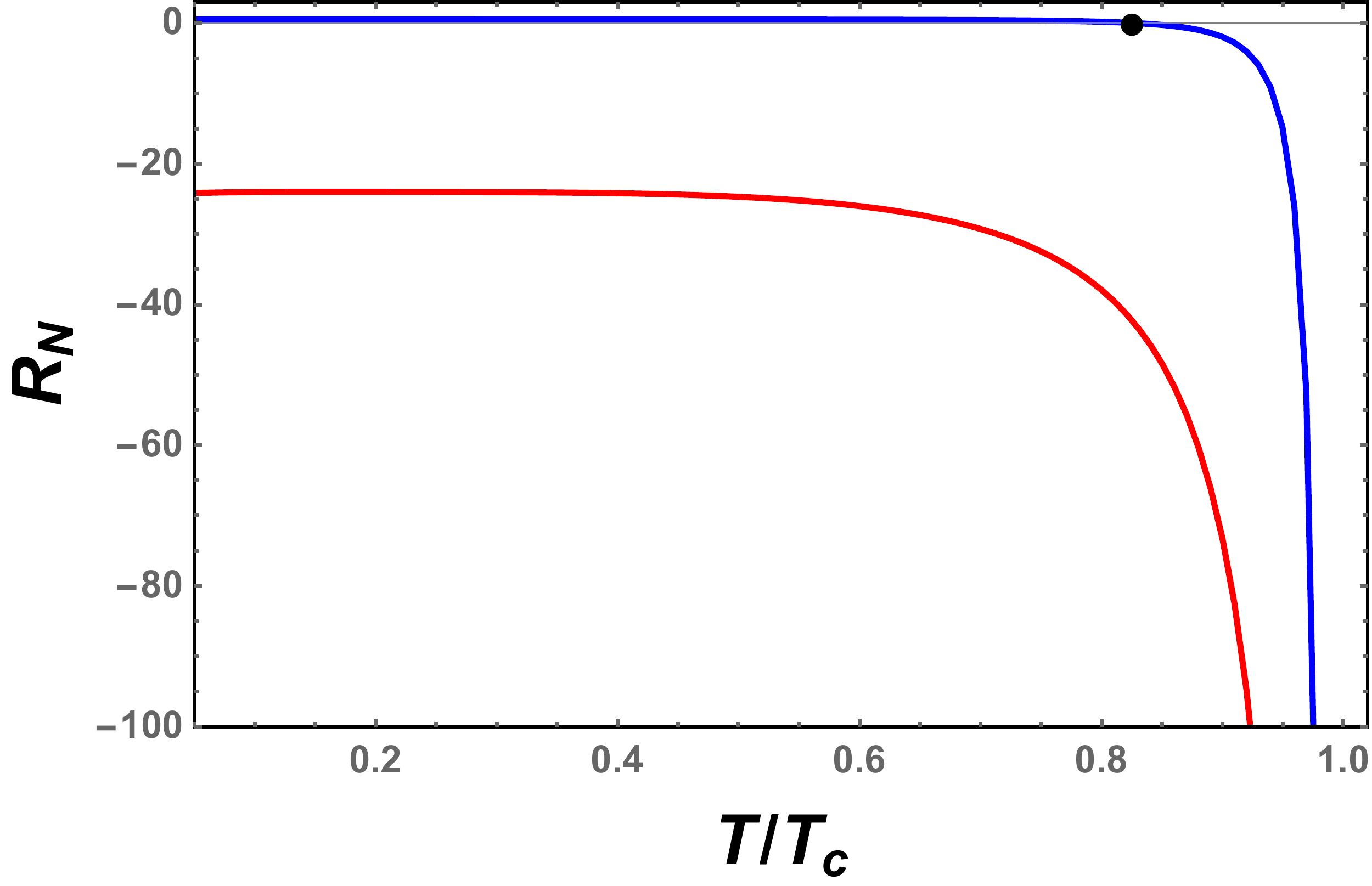}}
\subfigure[]{\label{HRNT10}
\includegraphics[width=4.5cm]{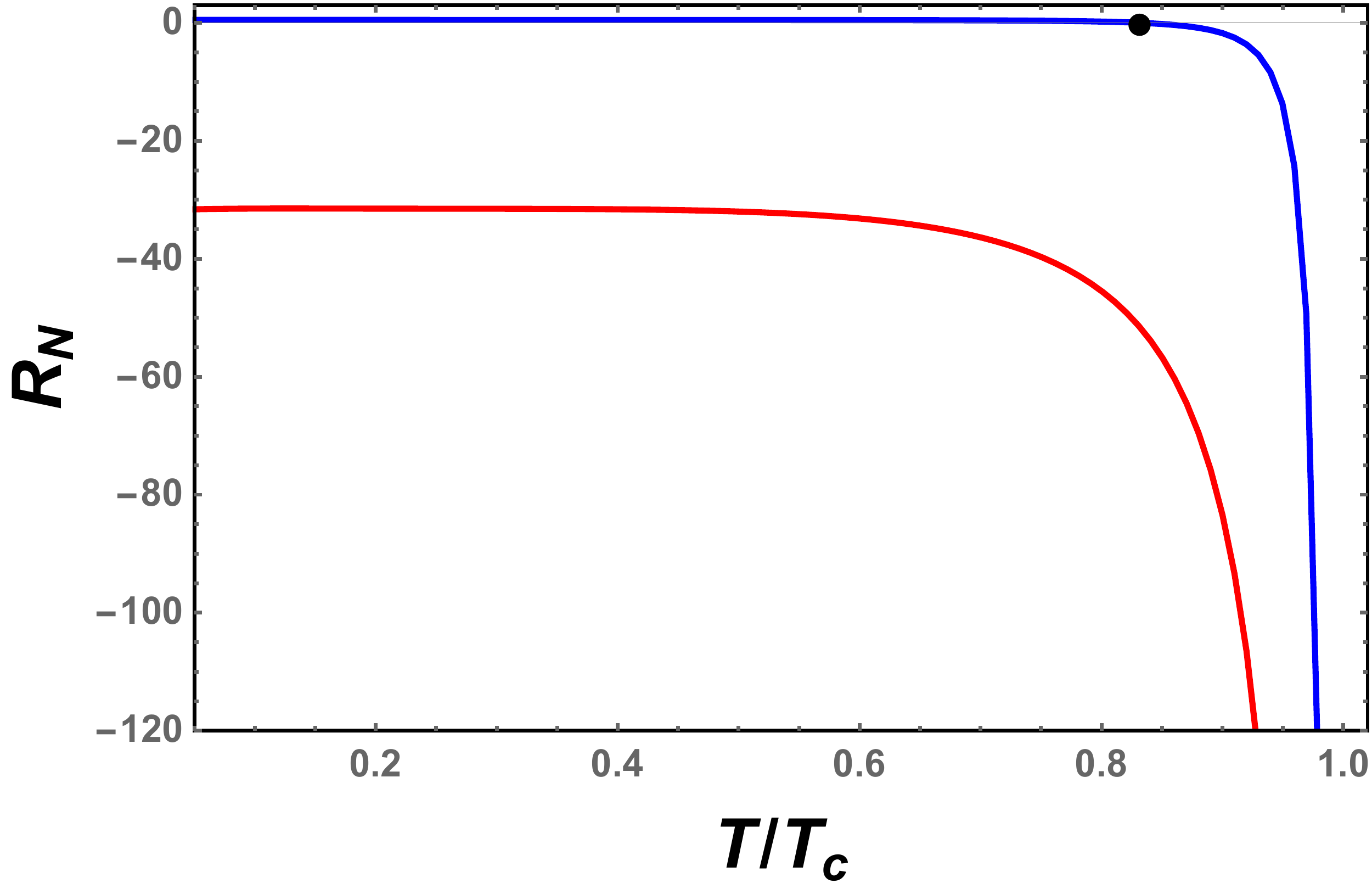}}
\end{center}
\caption{Behaviour  of the normalized scalar curvature $R_{\rm N}$ along the coexistence saturated large black hole curve (bottom red line) and small black hole curve (top blue line) for the  higher-dimensional charged AdS black hole. Black dots denote the vanishing points of $R_{\rm N}$. (a) $d$=5. (b) $d$=6. (c) $d$=7. (d) $d$=8. (e) $d$=9. (f) $d$=10.}
\label{pHRNT5}
\end{figure*}

Employing the formula (\ref{rrrrrn})  and the highly accurate fitting formulae of the coexistence curves given in Ref.~\cite{Wei5}, we display the behaviour of $R_{\rm N}$ along the coexistence curves for $d$=5-10 in Fig.~\ref{pHRNT5}, where the top blue line and the bottom red line are for the coexistence saturated small and large black holes, respectively. Black dots denote the vanishing points of $R_{\rm N}$, which are at $\tilde{T}=0.782073, 0.797955, 0.809779, 0.819143, 0.826857$, and $0.83339$, respectively for $d$=5-10 -- increasing dimension marginally increases the temperature at which $R_N$ vanishes.

From these figures, we can also find that for fixed $\tilde{T}$, the  large black hole at the coexistence curve always has smaller $R_{\rm N}$ than the small black hole at this curve. It is only for the  small black hole that $R_{\rm N}$ can be positive at low temperature. Hence according to the physical interpretation of the Ruppeiner geometry, only the small black hole with low temperature can have repulsive interaction among its microscopic structure. More importantly, both the scalar curvatures of the  small and large black holes at the coexistence curves go to negative infinity at the critical points.

%%%%%%%%%%%%%%%
%%%%%%%%%%%%%%%

In order to examine the critical behaviour of the normalized scalar curvature, we numerically calculate the coexistence curves very near the critical point, and then obtain the values of $R_{\rm N}$ along the coexistence curves. We then fit these numerical results using (\ref{423}). The fitting values of the coefficients $\alpha$ and $\beta$ are listed in Table \ref{tab1}. Combining these results for coexistence saturated small and large black holes, we find that the values of the coefficient $\alpha$ are around 2 with the error no more than 7\%.  We conclude that the critical exponent is 2,  the same as that of the  four-dimensional charged AdS black hole.

%%%%%%%%%%%%%%%%%%%%%%%%%%%%%
\begin{table}[h]
\begin{center}
\begin{tabular}{ccccccc}
  \hline\hline
                & $d$=5 & $d$=6 & $d$=7 & $d$=8 & $d$=9 & $d$=10 \\\hline
 $\alpha$ (CSSBH) & 1.93688 & 1.93127 & 1.92883 & 1.92809 & 1.92878 & 1.93012 \\
 -$\beta$ (CSSBH) & 1.73749 & 1.74037 & 1.76085 & 1.78940 & 1.82517 & 1.86293 \\\hline
 $\alpha$ (CSLBH) & 2.08393 & 2.09947 & 2.11159 & 2.12236 & 2.12976 & 2.13678 \\
 -$\beta$ (CSLBH) & 2.57094 & 2.63522 & 2.67879 & 2.71628 & 2.73099 & 2.74554 \\ \hline\hline
\end{tabular}
\caption{Fitting values of $\alpha$ and $\beta$ for coexistence saturated small black holes (CSSBH) and coexistence saturated large black hole (CSLBH).}\label{tab1}
\end{center}
\end{table}
%%%%%%%%%%%%%%%%%%%%%%%%%%%%%%%%

Furthermore, with these numerical results given in Table \ref{tab1}, we obtain
\begin{equation}
 R_{\rm N}(1-\tilde{T})^{2}=-(0.138889, 0.145833, 0.15, 0.152778, 0.154762, 0.15625),
\end{equation}
for $d$=5-10. These values are more negative than that of the VdW fluid and the four-dimensional black hole. We see a slight increase in the magnitude  of this quantity with increasing dimension.

\section{The Microstructures of Van der Waals fluids and Black Holes}

We now consider what our comparison of Ruppeiner curvatures between VdW fluids and black holes might imply about their microstructures.

Despite many years of study, we still do not know the nature and properties of the underlying degrees of freedom -- the microstructure -- of black holes. Following Boltzmann's dictum -- ``if you can heat it, it has microstructure" --
we are in the situation of obtaining what information we can from the relationship between their thermodynamic properties and their gravitational ones.

Thus far the black hole chemistry program has indicated that black holes have a rich and varied phase behaviour, indicative of a molecular kind of microstructure for their underlying degrees of freedom. For the particular case of charged AdS black holes, so far their thermodynamic behaviour has been found to be similar in every respect to that of VdW fluids. Our analysis and results provide the first indications that this is not the case for their microstructures. While there remains much to be understood, we provide here  a preliminary discussion on the implications of our research for black hole microstructure.

\begin{figure*}
\begin{center}
\includegraphics[width=9cm]{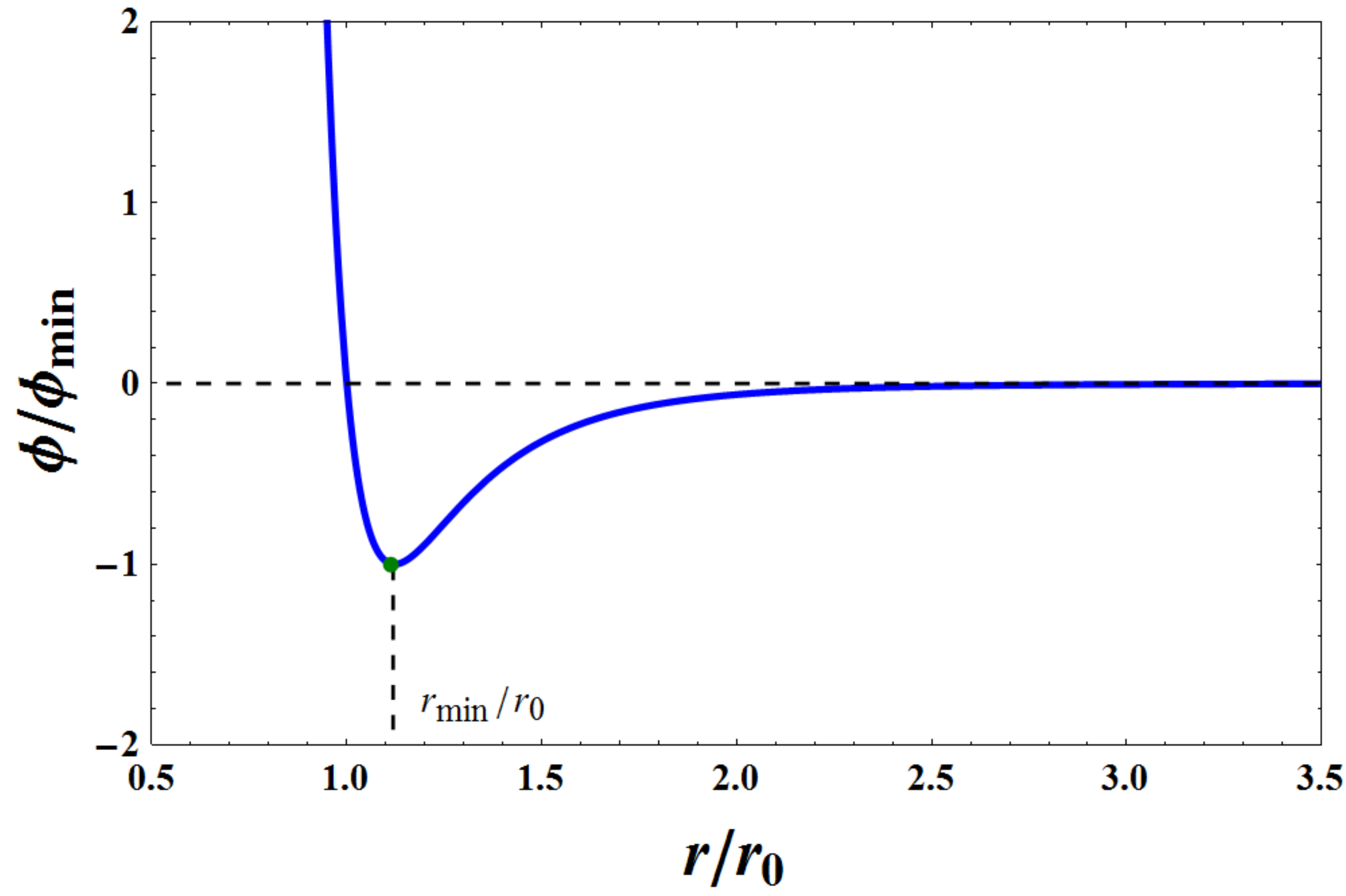}
\end{center}
\caption{Behavior of the Lennard-Jones potential energy.}\label{PhiLJ}
\end{figure*}

Let us start with the VdW fluid. It is well-known that the interaction between two fluid molecules can be modelled by the Lennard-Jones potential,
\begin{equation}
 \phi(r)=4\phi_{min}\left(\left(\frac{r_{0}}{r}\right)^{12}-\left(\frac{r_{0}}{r}\right)^{6}\right),
\end{equation}
where $\phi_{min}$ depends on the parameters $a$ and $b$ of the VdW fluid. This potential is a combination of a short-range repulsive interaction and a long-range attractive interaction. The thermodynamic quantities given in Sec. \ref{svdw} were calculated under the ``hard-core" model, where the diameter of a molecule is chosen $d=r_{min}=2^{1/6}r_{0}$, located at the extremal point of the potential, as shown in Fig. \ref{PhiLJ}. Thus only attractive interactions are allowed between two molecules. However due to collisions or mean field effects, repulsive interactions are still possible. Excluding the coexistence range, the information in the Ruppeiner geometry is exactly consistent with this microscopic model of the fluid.

%%%%%%%%%%%%%%%
%%%%%%%%%%%%%%%

The phase behaviour of large charged black holes suggests that the interactions between two underlying `black hole molecules' can still be modelled by a Lennard-Jones like potential.  This in turn implies that the diameter  $d$ of a such a molecule must have some maximal value, $d<r_{min}$. However for  small black holes at high temperature -- outside the coexistence range --   we observe that the Ruppeiner geometry is positive, indicating a repulsive interaction. At the least this implies that the microstructure of black holes is more complex than that of a VdW fluid, and that any attempt to provide a deeper explanation of the degrees of freedom of a black hole will have to take this feature into account.

There are a number of ways to explore this, but an important next step is to understand the nature of the black holes from the gravitational side in light of these results. Such an approach is based on finding potential links between black hole thermodynamics and gravity. One of the  first steps in this direction has already been made \cite{Wei01522}, in an investigation of the behavior of the photon sphere around the first and second order black hole phase transition for   charged AdS black holes, with subsequent extensions to rotating Kerr-AdS black holes and charged Born-Infeld-AdS black holes \cite{Wei03455,Wei03334}. We argued that differences in the radius and critical angular momentum of the photon sphere can be treated as order parameters to describe black hole thermodynamic phase transitions. In particular, at the critical point, these two quantities exhibit  a critical exponent of $\frac{1}{2}$. This relationship between the photon sphere and thermodynamic phase transitions not only holds for VdW-like phase transitions, but also for reentrant phase transitions \cite{Altamirano}. This further underscores the expectation that gravity and thermodynamics of a black hole is linked with each other. Further pursuit of such links will help us understand the nature of gravity and black holes.

\section{Conclusions and discussions}
 \label{discussions}

We have made use of the Ruppeiner geometry and phase behaviour of $d$-dimensional charged AdS black holes to probe their microstructures.   Our novel approach  entails using the coordinates $(T,V)$ in thermodynamic phase space, and then examining the behaviour of the Ruppeiner scalar curvature at the  first-order and second-order phase transitions.

Our investigation has shown that while clear similarities exist between a van der Waals' fluid and a charged AdS black hole, there are also considerable qualitative differences. A comparison of figures \eqref{pVdWPTP} and \ref{pCbhpt} indicate that both have gas, liquid, supercritical fluid, and coexistence phases, in addition to two metastable phases, the superheated liquid phase and supercooled gas phase. These features likewise occur in higher-dimensions for the charged AdS black hole. Likewise in both cases the quantity  $R_{\rm N}\to -\infty$ at the critical points, and its critical exponent is equal 2 and $R_{\rm N}(1-\tilde{T})^{2} \to -1/8$ as $\tilde{T}\to 1$. We believe these results are closely related to the correlation length of the VdW fluid and are of significant import regarding the microstructure of thermodynamic systems. That they are the same for the charged black hole suggests some similarity in their microstructure.

However there is a significant difference between a van der Waals fluid and a charged AdS black hole: the Ruppeiner curvature changes sign at in the latter at sufficiently low temperature $\tilde{T} < \tilde{T}_0 = 0.7581$, but not in the former. The scalar curvature $R$ is always negative for the VdW fluid, which indicates that among the microstructures of the VdW fluid there are only attractive interactions. But for the charged black hole, the positivity of $R$ for $\tilde{T} < \tilde{T}_0$, indicated in Fig. \ref{pCsign}, indicates a phase region for small black holes where the microstructure interactions are attractive.

These features are robust, appearing for higher dimensional black holes. Although no analytic formulae for coexistence curves are available, our numerical study indicated clear similarities with the four-dimensional case. The relation
\begin{equation}
 \tilde{T}_{0}=\frac{1}{2}\tilde{T}_{\rm sp}.\label{tttsp}
\end{equation}
between the sign-changing temperature $\tilde{T}_0$ and the spinodal temperature $\tilde{T}_{\rm sp}$ exactly holds, and along the coexistence curves $R_{\rm N}$ goes to negative infinity at the critical points, and its critical exponent is equal 2.  There is a   weak dependence of both $\tilde{T}_s$ and $\lim_{\tilde{T}\to 1}R_{\rm N}(1-\tilde{T})^{2}$ on the spacetime dimension. The former quantity becomes slightly larger and the latter becomes slightly more negative as the dimension increases.

In conclusion we have obtained a universal formula of the scalar curvature of the Ruppeiner geometry, see (\ref{RR}), and presented a novel approach for probing black hole microstructures.  It would
be interesting to extend our approach to other black hole systems that exhibit different phenomena,
such as reentrant phase transitions and triple points.  We believe that studying the behaviour of the scalar curvature for these systems and comparing it with the recent observations \cite{Ruppeinerg} for six representative fluids (hydrogen, helium, argon, methane, oxygen, and water) may  shine  further light on the internal microstructure of black holes.

\acknowledgments
We would like to thank Dr. Hai-Shan Liu for useful discussions. This work was supported by the National Natural Science Foundation of China (Grants No. 11675064, No. 11875151, and No. 11522541) and the Natural Sciences and Engineering Research Council of Canada. S.-W. Wei was also supported by the Chinese Scholarship Council (CSC) Scholarship (201806185016) to visit the University of Waterloo.

\end{document}